\documentclass[submission,Phys]{SciPost}

\usepackage[utf8]{inputenc} 
\usepackage[T1]{fontenc} 	
\usepackage[english]{babel} 

\usepackage[bitstream-charter]{mathdesign}
\usepackage{geometry} 		
\usepackage{amsmath} 		
\usepackage{amsthm} 		
\usepackage{mathtools} 		
\usepackage{float} 			
\usepackage{graphicx} 		
\usepackage{tabularx} 		
\usepackage{booktabs} 		
\usepackage{color, xcolor} 	
\usepackage{pdfpages} 		
\usepackage{extarrows} 		
\usepackage{multirow} 		
\usepackage{multicol} 		
\usepackage{caption} 		
\usepackage{subcaption} 	
\usepackage{enumitem} 		
\usepackage{setspace} 		
\usepackage{xspace} 		
\usepackage{ragged2e} 		
\usepackage{stackrel} 		
\usepackage{tikz} 			
\usepackage{braket} 		
\usepackage{bm} 			
\usepackage{tensor} 		
\usepackage{slashed} 		
\usepackage{siunitx} 		
\usepackage{lastpage} 		
\usepackage{cite} 			
\usepackage[normalem]{ulem} 
\usepackage{fontawesome} 	
\usepackage{tocloft} 		
\usepackage{titlesec} 		
\usepackage{doi} 			
\usepackage{hyperref} 		
\usepackage[most]{tcolorbox} 					
\usepackage[nameinlink, capitalize]{cleveref} 	
\usepackage[nottoc, notlot, notlof]{tocbibind} 	
\usepackage[ruled, vlined]{algorithm2e} 		

\hypersetup{
	pdftitle={How to Understand Limitations of Generative Networks},
	pdfauthor={Ranit Das, Luigi Favaro, Theo Heimel, Claudius Krause, Tilman Plehn, and David Shih},
	colorlinks=true, 			
	linkcolor={red!50!black}, 	
	citecolor={blue!50!black}, 	
	urlcolor={blue!80!black} 	
} 

\DeclareSymbolFont{usualmathcal}{OMS}{cmsy}{m}{n}
\DeclareSymbolFontAlphabet{\mathcal}{usualmathcal}

\SetArgSty{textnormal}
\SetKwComment{Comment}{{\small\#}~}{}
\SetCommentSty{mycommfont}

\setitemize{itemsep=0pt, parsep=0pt} 				
\setenumerate{itemsep=0pt, parsep=0pt} 				
\setitemize{itemsep=2pt,topsep=2pt,parsep=0pt,partopsep=0pt,leftmargin=*}
\setenumerate{itemsep=0pt,topsep=2pt,parsep=0pt,partopsep=0pt,labelindent=3pt,leftmargin=*}
\newcommand{\pd}{p_\text{data}}

\newcommand{\pmd}{p_\text{model}}

\newcommand\one{\leavevmode\hbox{\small1\normalsize\kern-.33em1}}

\def\slashchar#1{\setbox0=\hbox{$#1$}           
   \dimen0=\wd0                                 
   \setbox1=\hbox{/} \dimen1=\wd1               
   \ifdim\dimen0>\dimen1                        
      \rlap{\hbox to \dimen0{\hfil/\hfil}}      
      #1                                        
   \else                                        
      \rlap{\hbox to \dimen1{\hfil$#1$\hfil}}   
      /                                         
   \fi}

\def\ie{{\sl i.e.} \,}

\graphicspath{{./figs/}}                

\begin{document}

\begin{center}{\Large \textbf{
How to Understand Limitations of Generative Networks
}}\end{center}

\begin{center}
  Ranit Das\textsuperscript{1},
  Luigi Favaro\textsuperscript{2},
  Theo Heimel\textsuperscript{2},
  Claudius Krause\textsuperscript{2},
  Tilman Plehn\textsuperscript{2}, and 
  David Shih\textsuperscript{1}
  
\end{center}

\begin{center}
  {\bf 1} NHETC, Department of Physics \& Astronomy, Rutgers University, Piscataway, NJ USA \\
  {\bf 2} Institut f\"ur Theoretische Physik, Universit\"at Heidelberg, Germany
\end{center}

\begin{center}
\today
\end{center}

\section*{Abstract}
         {\bf Well-trained classifiers and their complete weight
           distributions provide us with a well-moti\-vated and
           practicable method to test generative networks in particle
           physics. We illustrate their benefits for distribution-shifted jets,
           calorimeter showers, 
           and reconstruction-level events. In all cases, the classifier weights make for a
           powerful test of the generative network, identify potential
           problems in the density estimation, relate them to the underlying physics, and tie in 
           with a comprehensive precision and uncertainty treatment for
           generative networks.}

\vspace{10pt}
\noindent\rule{\textwidth}{1pt}
\tableofcontents\thispagestyle{fancy}
\noindent\rule{\textwidth}{1pt}
\vspace{10pt}

\clearpage
\section{Introduction}
\label{sec:intro}

Like all of society, LHC physics is currently undergoing a
transformation driven by modern data science. The experimental and
theoretical methods of LHC physics have always been numerical in
nature, with the goal to quantitatively, systematically, and
comprehensively understand data in terms of fundamental
theory. Generative networks are an exciting concept of modern machine
learning (ML), combining unsupervised density estimation in an
interpretable phase space with fast and flexible sampling and
simulations~\cite{Butter:2022rso}. Currently, the most promising architectures
for precision generation are normalizing flows and their invertible network (INN) variants,
but we will see that diffusion models and generative transformers might offer an even
better balance of precision and expressivity.

The range of tasks for generative networks in LHC simulations and
analysis is extensive. Given the modular structure of LHC simulations,
it starts with phase space integration and
sampling~\cite{Chen:2020nfb,Bothmann:2020ywa,Gao:2020vdv,Gao:2020zvv,Danziger:2021eeg,Heimel:2022wyj},
for instance of ML-encoded transition amplitudes. More LHC-specific
tasks include event subtraction~\cite{Butter:2019eyo}, event
unweighting~\cite{Verheyen:2020bjw,Backes:2020vka}, or
super-resolution
enhancement~\cite{DiBello:2020bas,Baldi:2020hjm}. Generative networks
working on physics phase spaces have been developed and tested as
event
generators~\cite{dutch,gan_datasets,DijetGAN2,Butter:2019cae,Alanazi:2020klf,Butter:2021csz},
parton
showers~\cite{locationGAN,Andreassen:2018apy,Bothmann:2018trh,Dohi:2020eda,Buhmann:2023pmh},
and detector
simulations~\cite{Paganini:2017hrr,deOliveira:2017rwa,Paganini:2017dwg,Erdmann:2018kuh,Erdmann:2018jxd,Belayneh:2019vyx,Buhmann:2020pmy,ATL-SOFT-PUB-2020-006,Buhmann:2021lxj,Krause:2021ilc,
ATLAS:2021pzo,Krause:2021wez,Buhmann:2021caf,Chen:2021gdz,
Adelmann:2022ozp,Mikuni:2022xry,ATLAS:2022jhk,Krause:2022jna,Cresswell:2022tof,Diefenbacher:2023vsw,
Hashemi:2023ruu,Xu:2023xdc,Diefenbacher:2023prl,Buhmann:2023bwk,Buckley:2023rez
}. These
networks should be trained on first-principle simulations, easy to
handle, efficient to ship, powerful in amplifying the training
samples~\cite{Butter:2020qhk,Bieringer:2022cbs}, and --- most
importantly --- precise. Going beyond forward generation, conditional
generative networks can also be applied to probabilistic
unfolding~\cite{Datta:2018mwd,Bellagente:2019uyp,Andreassen:2019cjw,Bellagente:2020piv,Backes:2022vmn,Shmakov:2023kjj},
inference~\cite{Bieringer:2020tnw,Butter:2022vkj}, or anomaly detection~\cite{Nachman:2020lpy,Hallin:2021wme,Raine:2022hht,Hallin:2022eoq,Golling:2022nkl,Sengupta:2023xqy}, reinforcing the
precision requirements.\medskip

For all the above tasks, normalizing flows or INNs have reached the
level of precision, stability, and control required by LHC
physics. Methods to control the performance of these generative
networks include Bayesian network
setups~\cite{Bellagente:2021yyh,Butter:2021csz},
classifier-reweighting~\cite{Diefenbacher:2020rna,Winterhalder:2021ave,Butter:2021csz,Nachman:2023clf},
and conditional training on augmented
data~\cite{Butter:2021csz}. Building on these developments, LHC physics
needs methods to systematically evaluate the performance and the
precision of generative networks~\cite{Kansal:2022spb}, for example to
quantify possible gains through new
architectures~\cite{Mikuni:2022xry,Leigh:2023toe,Mikuni:2023dvk,Butter:2023fov}.\medskip

In this paper we will explore the merits of a classifier-based
evaluation of generative networks in particle physics. We will start
by defining the goals of such a systematic evaluation and then
introduce the classifier metric in Sec.~\ref{sec:general}. We will present
our approach for jet generators~\cite{Kansal:2022spb} in
Sec.~\ref{sec:jets}, and discuss it in more details for a calorimeter
simulation similar to Ref.~\cite{Krause:2021ilc} in
Sec.~\ref{sec:calo}.  Finally, we will show how to use event weights
to track progress between two versions of an ML-event
generator~\cite{Butter:2021csz} in Sec.~\ref{sec:events}. We will also
illustrate how a systematic scan over kinematic distributions of
events with anomalous weights can identify issues of a trained network
and how Bayesian networks help us identify the reason for this
discrepancy.

All three applications combined illustrate how 
the distribution of learned control weights over
phase space is a reliable measure of the quality if the generative
networks and that its shape provides a powerful ``explainable AI" (xAI) tool which allows 
us to systematically search for failure modes of generative models,  
identify the underlying physics cause, and  
improve the tested networks efficiently.

\section{Testing generative networks}
\label{sec:general}

Given a generative model trained on some reference data, we would like
to know how well it reproduces the data in the full phase space. This
includes correct reproduction of critical high-level features, such as
transverse momenta and invariant masses in the case of event
generation, or shower profiles and MIP peaks in the case of
calorimeter simulation. But it also includes all the multi-dimensional
correlations between all the features throughout phase space, which
might not be visible at the level of histograms of pre-defined
high-level features.

We know some typical failure modes of generative
networks~\cite{Butter:2021csz}, including features completely removed
by a fit-like density estimation, washed-out features with poor
resolution, underpopulated kinematic tails, or wrongly learned phase
space boundaries.  Comparing kinematic distributions of generated and
training events allows us to identify many of these issues, making use
of the fact that phase space is interpretable and we can typically
derive phase space distributions using first principles in quantum
field theory or detector design. However, looking at pre-defined phase
space distributions runs the risk that we miss a problem, for example
when it only affects complex correlations.\medskip

Clearly, sensitive metrics are needed to assess the quality of a
generative model throughout all of phase space. These metrics should
be both multi-variate (capturing all correlations) and interpretable
(offering a way to diagnose which critical high-level features are
most discrepant). Ideally, these metrics could also offer a systematic
way of improving the generative model.

An optimal binary classifier, trained to distinguish generated from
reference data in the full phase space, fits the bill in every
respect. By the Neyman-Pearson (NP) lemma, this classifier is the most
powerful discriminant between generative model and reference data. It
is already well-established that one can use the classifier to {\it
  reweight} the generative model and bring it closer to the reference
data~\cite{Diefenbacher:2020rna}.
However, simply reweighting the generated data can lead to problems, for
instance in regions with large weights. This can be remedied by using the
classifier weights to improve the generative model~\cite{Winterhalder:2021ave,Butter:2021csz,Nachman:2023clf}.
This is also the basis of training GANs, where the training objective is
to minimize the difference of the generated distribution to the reference
distribution quantified by a classifier network. However, in this case the
classifier is not trained to reach close to optimal discrimination power
to allow for an equilibrium between generator and discriminator.
In addition to these methods to directly improve the generative model with
the classifier weights, they can also be used as a diagnostic tool, which is
the focus of this work.
By examining the generated and reference data as a function of the cut on
the classifier, one can zoom in on the most anomalous regions of phase
space, \ie those that are worst-reproduced by the generative
model. This facilitates the interpretability of the classifier metric,
which could be further enhanced using recent xAI techniques developed in HEP such as Refs.~\cite{Faucett:2020vbu,Das:2022cjl}.

Studies that have used the classifier metric to judge the quality of generative models have tended to focus exclusively on single numbers~\cite{Krause:2021ilc,Krause:2021wez,Mikuni:2022xry,
Krause:2022jna,Cresswell:2022tof,Kansal:2022spb,Buckley:2023rez}, like the
AUC, the loss, or the accuracy of the classifier. While these aggregate measures certainly have their uses, there is much more useful information to be gleaned from the classifier than a single
number~\cite{Butter:2021csz,Badger:2022hwf,Diefenbacher:2023vsw}. For example, a global integral measure
such as the AUC will not detect discrepancies in tails of
distributions.  Also, the AUC becomes less and less informative the
closer the generated and reference samples become. Finally, declaring
the model with the highest AUC as the ``best" model is oversimplistic,
because the definition of the ``best" generative network depends on
what we actually require from the generative network and how we want
to use its output.\medskip

In this work, we will explore what the distribution of
classifier outputs tells us about the quality of the generative
model. We will choose to work in terms of \textit{weights} which can
be obtained from the classifier outputs $C$ as
\begin{align}\label{eq:wformula}
  w(x)
  = \frac{\pd(x)}{\pmd(x)}
  = \frac{C(x)}{1-C(x)} 
  \qquad \text{with} \qquad 
  C(x) = \frac{\pd(x)}{\pd(x)+\pmd(x)} \; .  
\end{align}
The assumption is that the NP-classifier learns the density ratio. 
For a good generative model and an optimal classifier, the weight
distribution will typically peak near one, with tails to the left ($w\ll 1$)
and right ($w\gg 1$), corresponding to regions of phase space where the
generative model is overproducing and underproducing the reference
data, respectively. On general grounds, the NP classifier should have
an excess of generated events as a small-weights tail of the
distribution, and an excess of reference events as a large-weight
tail. Indeed this is a general pattern we will observe in the different examples we consider in this work. Having it the other way around, an excess of true events on the left tail and an excess of generated events on the right, would generate a ROC curve below the diagonal, indicating an anti-classifier. A renaming of the classes would then solve the problem in principle by switching the weights of true and generated events. However, finding an anti-classifier after training would lead to a troubleshooting and retraining of the classifier in practice.

Since phase space is interpretable, we can study patterns and
clustering of anomalous weights to learn more about the generative
network. For instance, a positive feature or tail missed by the
generative training will be resurrected through large weights $w(x_i)
\gg 1$, clustered in phase space. A wrongly modelled phase space
boundary will lead to small weights $w(x_i) \ll 1$ or even $w(x_i) =
0$, also clustered in phase space. 

Local features in the weight distribution, not necessarily along the tails, also carry useful information about the performance of the generative model. A simple example is the smearing 
of a peak in phase space, at $x_\text{max}$, corrected by universal weights $w>1$ around the peak.
If the smeared phase space feature dominates the total rate, a maximum in 
the weight distribution appears at
\begin{align}
    w(x) \approx \frac{\pd(x_\text{max})}{\pmd(x_\text{max})} > 1 \; .
\end{align}
Depending on the exact shape in the training data and the kind of smearing, the weights enhancing the tails of the smeared
peak can, but do not have to produce a second maximum in the weight 
distribution. We will discuss all of these patterns in the following sections.\medskip

The practical reason why we can measure the performance of generative
networks with classifiers is the typical precision of the two
networks.  For LHC events with a relatively small number of particles
in the final state, we know that generative networks reach a precision
around
\begin{align}
 \frac{\pd(x)}{\pmd(x)} - 1 \sim  
 \begin{cases}
    1\% & \text{INN~\cite{Butter:2021csz}} \\
    10\% & \text{GAN~\cite{Butter:2019cae}}
 \end{cases} \; .
\end{align}
Classifiers are not fundamentally different from regression networks,
so we expect them to learn the density ratio at the sub-percent
level~\cite{Badger:2022hwf,Maitre:2023dqz},
\begin{align}
 w(x) - \frac{\pd}{\pmd}(x) \sim 0.1 \% \; .
\end{align}
Thus it should be possible to obtain classifiers which
are precise enough to be sensitive to the failure modes of the
generative model. This is consistent with the observation that,
while classification with generative models can be useful for small
training data sets with low complexity, it is typically outperformed
by discriminative models for higher-dimensional data with large training
statistics~\cite{fetaya2019understanding,bouchard2004tradeoff,mackowiak2021generative,ardizzone2020training,zimmermann2021score}.

Because the weights are constructed from a classifier, we can use
standard methods, such as calibration curves, to ensure the classifier
is trained properly. We can also reweight generated samples with the
learned classifier and see if they become closer to the reference
data; this will be another sign that the classifier approximates well
the likelihood ratio.\medskip

In our study of the weight distributions for generative models, we can
draw inspiration from a similar approach to supervised amplitude
regression~\cite{Aylett-Bullock:2021hmo,Maitre:2021uaa,Badger:2022hwf,Maitre:2023dqz}. There,
the weights can be constructed directly from the regression task,
because the ``generated" amplitude is learned directly from the known
theoretical calculation. There, as here, tails of the weight
distribution will be induced by stochastic training data, a lack of
expressivity of the network, or overtraining~\cite{Badger:2022hwf}.
For a well-motivated statistical test we can use the fact that many
networks are trained on likelihood losses. Those losses include an
uncertainty estimate $\sigma_i$, for instance from a Bayesian
regression network, so we can supplement the weight distributions
by a pull and analyse both~\cite{Badger:2022hwf},
\begin{align}
  w_i = \frac{A_{i,\text{data}}}{A_{i,\text{model}}} 
  \qquad \text{and} \qquad 
  t_i
  &= \frac{ A_{i,\text{model}} - A_{i,\text{data}}}{\sigma_i}
  = \frac{A_{i,\text{model}}}{\sigma_i}
   \left( 1 - w_i \right) \; .
\label{eq:pulls}
\end{align}
The pull should follow a standard Gaussian for uncorrelated stochastic
deviations.  The combination of weights and pulls it is extremely
useful for testing regression networks, so we will try to generalize
it to generative networks.\medskip

Finally, we make the (obvious) observation that the AUC of a
classifier can be extracted from weight samples $w(x_i)$ evaluated on
training and generated configurations. As a function of the signal
efficiency, the ROC curve is a step-wise, monotonically increasing
function. Its integral can be estimated by the sum of bins with width
$1/N_\text{gen}$, the inverse size of the generated dataset. The
height of each rectangle is the fraction of weights in the true, or
training dataset, which are larger than a given $w_\text{cut}$,
normalized to $N_\text{true}$. The AUC is then given by the sum,
\begin{align}
  \text{AUC} = \frac{1}{N_{\text{gen}}N_{\text{true}}}\sum_{w_i \in \text{gen}} |\{w | w \in \text{true and } w>w_i\}| \; ,
  \label{eq:auc_from_weights}
\end{align}
where $|\{S\}|$ denotes the cardinality of the set $S$. Therefore, by
focusing on the weight distribution of the classifier, we are not
missing any information otherwise contained in the AUC.

In this paper we will use three standard LHC applications of
generative models to develop a common strategy to quantify the performance
of the networks. To test the specific generative networks introduced
in the following sections, we use a very generic classifier for
Secs.~\ref{sec:calo} and~\ref{sec:events}, described in
Tab.~\ref{tab:hyperparameters}. In Sec.~\ref{sec:jets} we apply the
state-of-the-art in jet classification,
\textsc{ParticleNet-Lite}~\cite{Qu:2019gqs}.

\begin{table}[t]
    \centering
    \begin{small} \begin{tabular}[t]{l|l|l}
    \toprule
    Parameter & Calorimeter & Events $Z+\{ 1,2,3\}$~jets \\
    \midrule
    Optimizer & Adam & Adam \\
    Learning rate & 0.001 & 0.001 \\
    LR schedule & reduce on plateau & reduce on plateau\\
    Decay factor & 0.1 & 0.1 \\
    Decay patience (epochs) & 5 & 5 \\
    Batch size & 1000 & 1024\\
    Epochs & 150 & 50 \\
    Number of layers & 3 & 5 \\
    Hidden nodes & 512 & 256 \\
    Dropout & $10\%$ & $10\%$ \\
    Activation function & leaky ReLU & leaky ReLU \\
    Training samples & 60k & 2.7M / 750k / 210k \\
    Validation samples & 20k & 300k / 80k / 20k \\
    Testing samples & 20k & 3.0M / 830k / 240k \\
    \bottomrule
    \end{tabular} \end{small}
    \caption{Hyperparameters of the classifier network applied to the calorimeter simulation and event generation datasets.}
    \label{tab:hyperparameters}
\end{table}

\section{Distribution-shifted jets}
\label{sec:jets}

\begin{figure}[t]
    \includegraphics[width=0.49\textwidth,page=1]{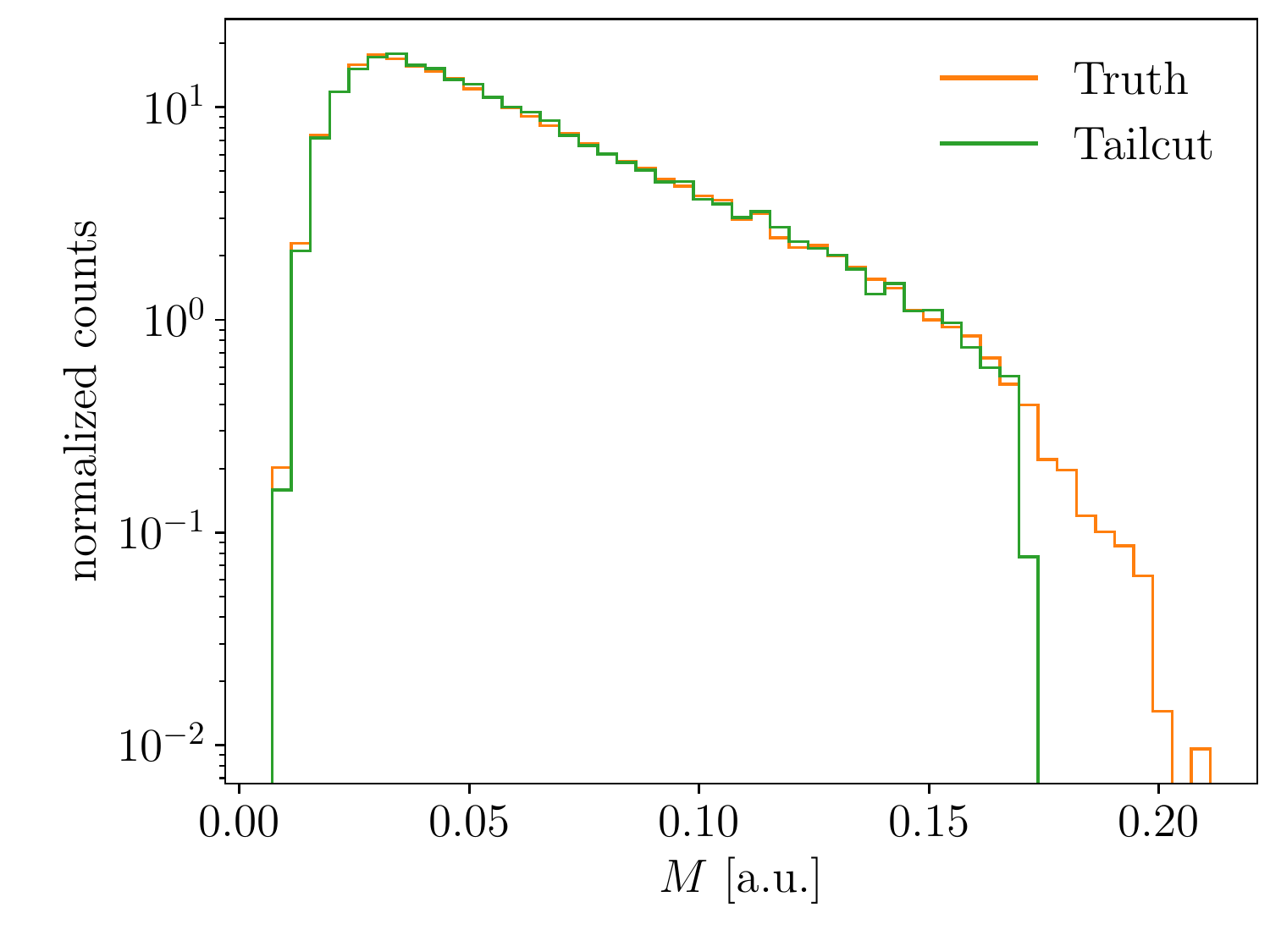}
    \includegraphics[width=0.49\textwidth,page=1]{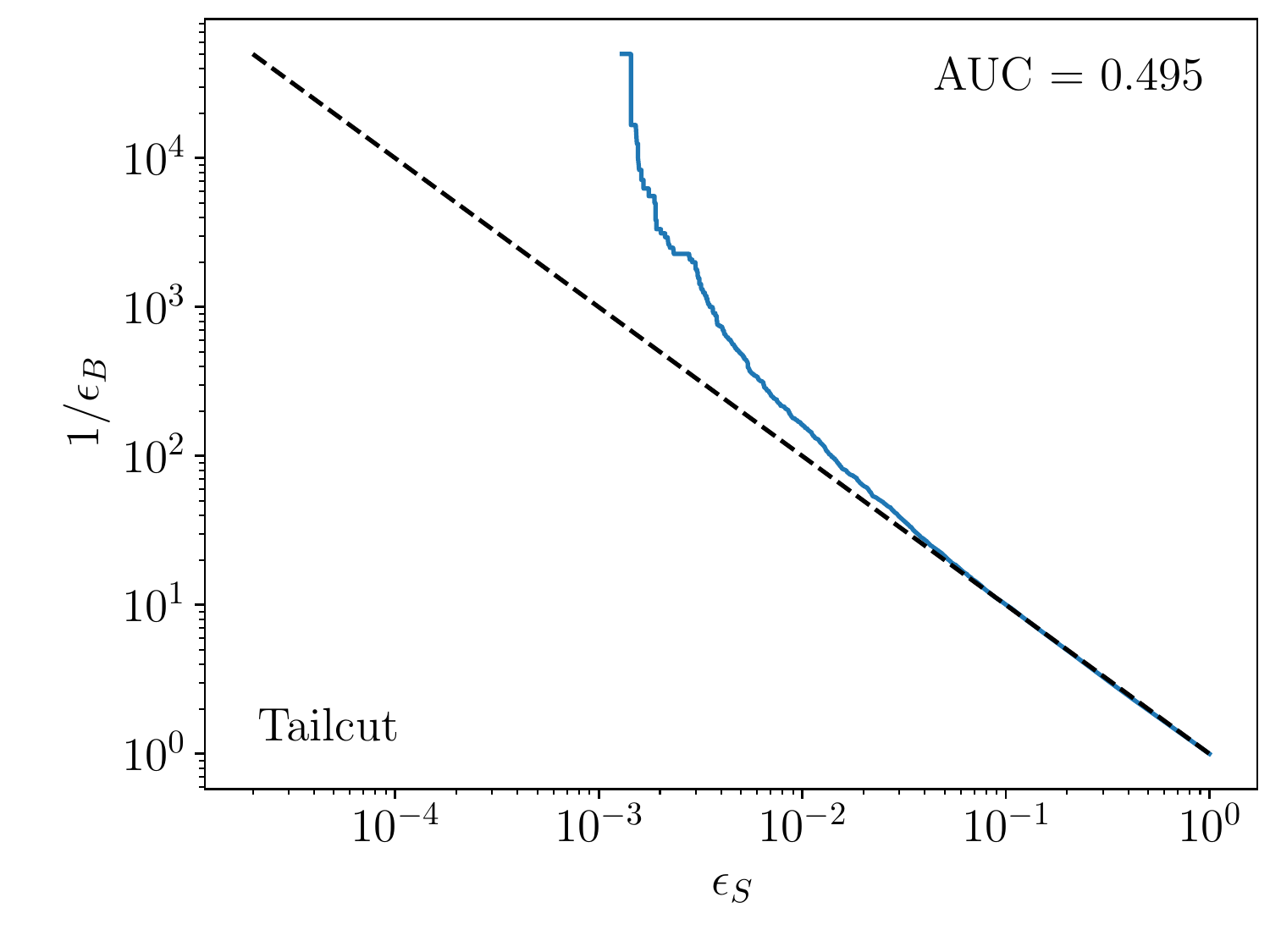} \\
    \includegraphics[width=0.49\textwidth,page=1]{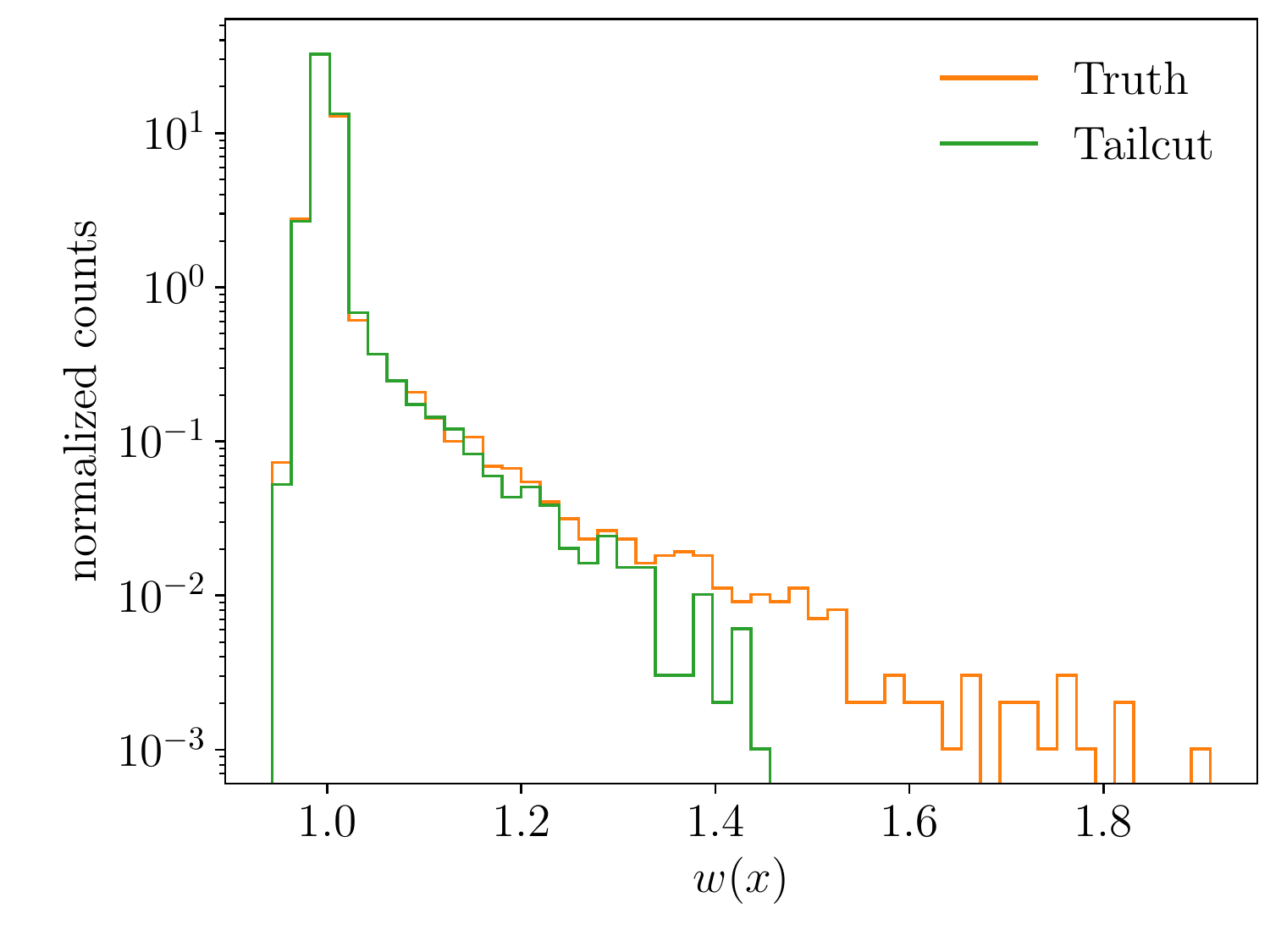}
    \includegraphics[width=0.49\textwidth,page=1]{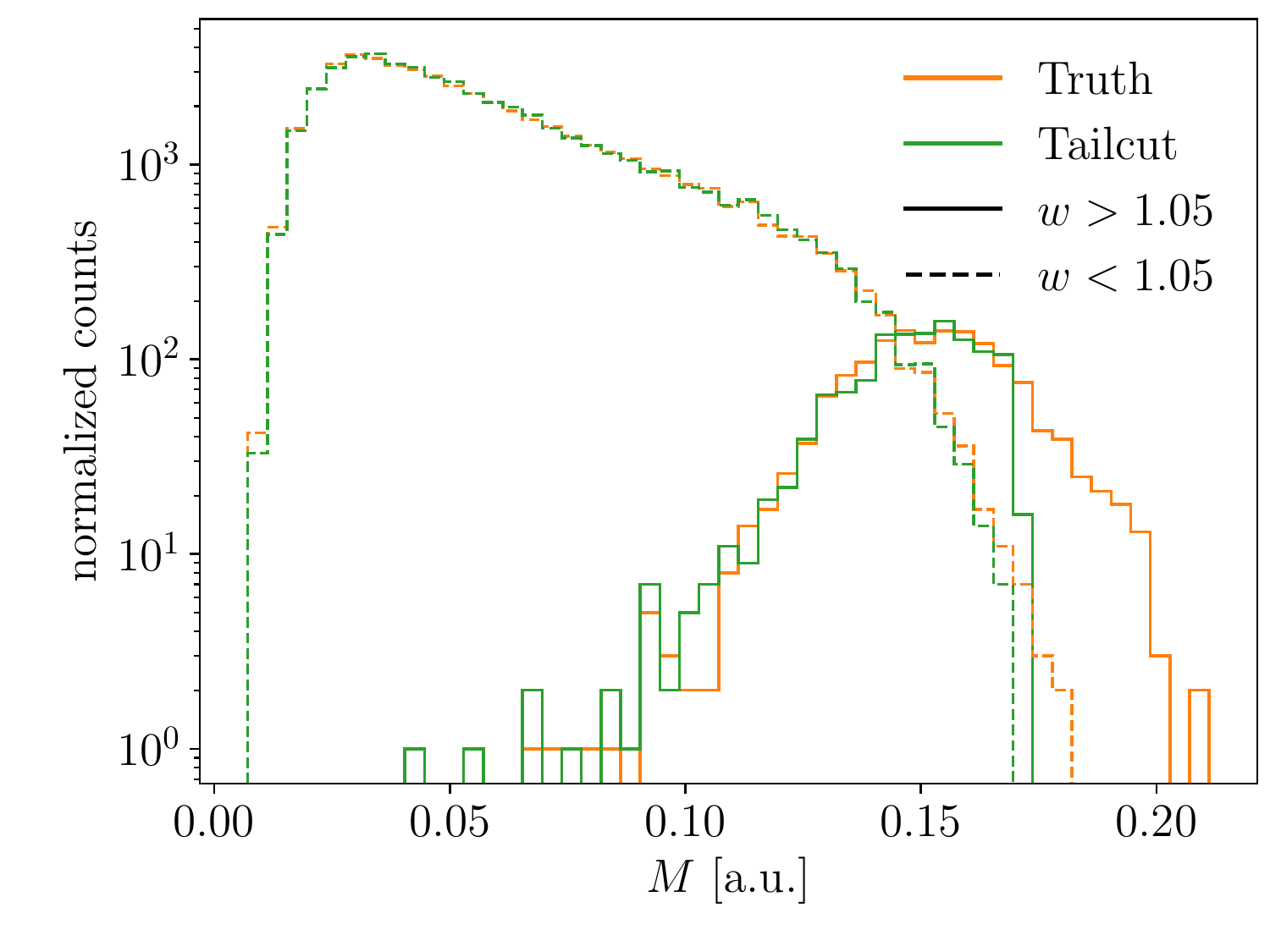}
    \caption{From top left to bottom right: jet mass distribution for
      the tailcut distortion; ROC curve from the trained ensemble of
      classifiers; learned weight distribution; jet mass distribution
      for jets in different classifier weight ranges, to identify
      clustering.}
    \label{fig:tailcut_weights}
\end{figure}

As a first application, we consider the JetNet example recently used
in Ref.~\cite{Kansal:2022spb} to illustrate different metrics for
generative models. They distort jets generated by \textsc{Pythia} at the
particle level and at the distribution level. In the particle level
distortions, each particle in the jet is altered in some way.
While this is a realistic scenario, the amount of distortion in
Ref.~\cite{Kansal:2022spb} was taken so large, that it makes the
classification task almost trivial.
For the distribution-level distortions, a single distribution like the
jet mass is modified. Jets are reweighted so that all other features
and correlations are identical to the reference data, only the one
distribution is modified. This is a highly unrealistic toy scenario,
and we would not advertize it as physics-motivated, but it provides an
interesting challenge for metrics to detect small differences.

In Ref.~\cite{Kansal:2022spb} it was pointed out that the AUC of a
classifier metric trained on distribution-level distortion versus
reference data is not very sensitive, and metrics such as FID and MMD
can detect the flaw in the generative model more sensitively. In line
with the general philosophy outlined in
Sec.~\ref{sec:general}, we argue that the AUC is indeed the wrong
metric, and examining the distribution of classifier weights,
especially the behavior on the tails, is a much more sensitive probe
and does detect all distribution-level distortions introduced into the
toy generative models.\medskip

We perform three distortions on the jet mass, extracted from the
relative polar coordinates provided in the JetNet dataset~\cite{Kansal:2022spb}:
\begin{enumerate}
    \item ``Tail cut": remove the tail with an acceptance cut $M
      <0.17$;
    \item ``Smear": smear the distribution by multiplying with a Gaussian
      with $\mu = 1.0$ and $\sigma = 0.25$.
    \item ``Shift": shift the distribution by multiplying with a
      Gaussian with $\mu = 1.1$ and $\sigma = 0.05$;
\end{enumerate} 
For each distortion, we train the classifier on 100000 distorted and the same number of undistorted jets. The validation set consists of 50000 jets each. In the interest of computation time, we use \textsc{ParticleNet-Lite} instead
of the full \textsc{ParticleNet} classifier~\cite{Qu:2019gqs} used in
Ref.~\cite{Kansal:2022spb}. This only has a minimal effect on the
results.
For extremely similar datasets and using limited training time we
expect a certain variability of the classifier output. To avoid
cherry-picking, we combine the five independent trainings of
\textsc{ParticleNet-Lite}, and for from each training select the
models with the five lowest validation losses. We ensemble these 25
classifier outputs and verify (by doing it all over again) that this
produced a stable, robust result. Evidence that the ensembled classifiers are well-calibrated, and hence learned the likelihood ratio, is
provided in Appendix~\ref{app:calibration}.\medskip

\begin{figure}[t]
    \includegraphics[width=0.49\textwidth,page=1]{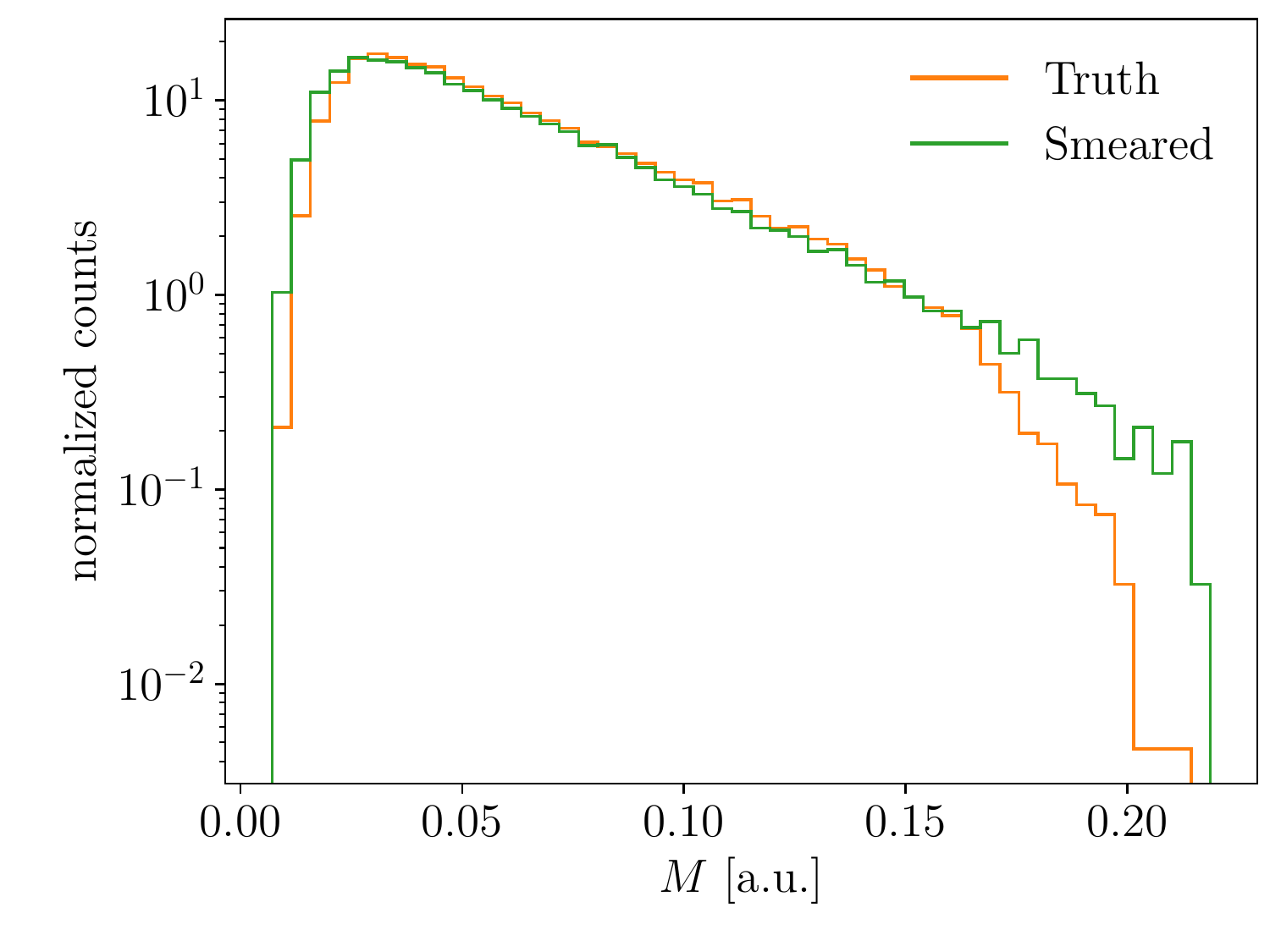}
    \includegraphics[width=0.49\textwidth,page=1]{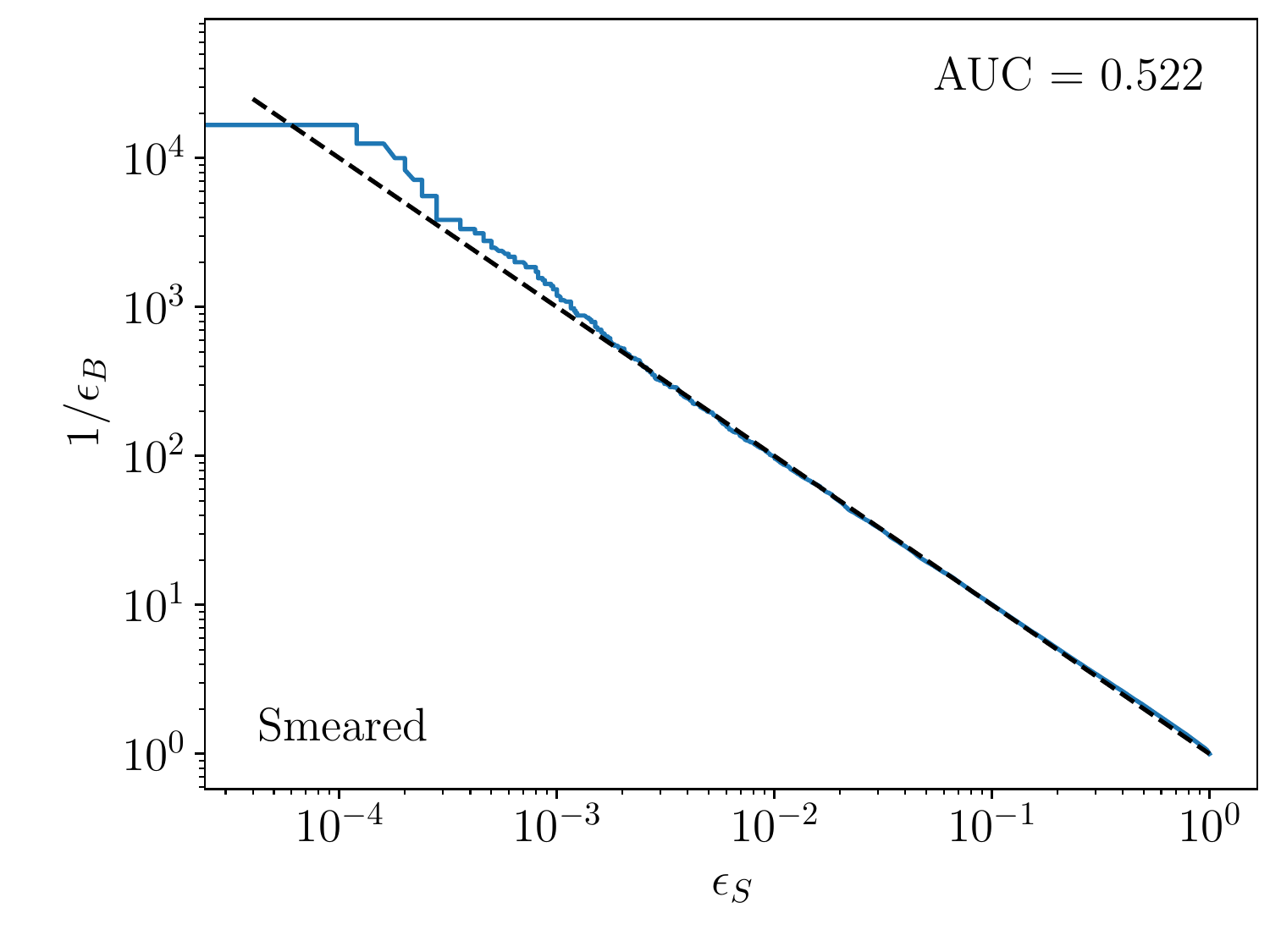} \\
    \includegraphics[width=0.49\textwidth,page=1]{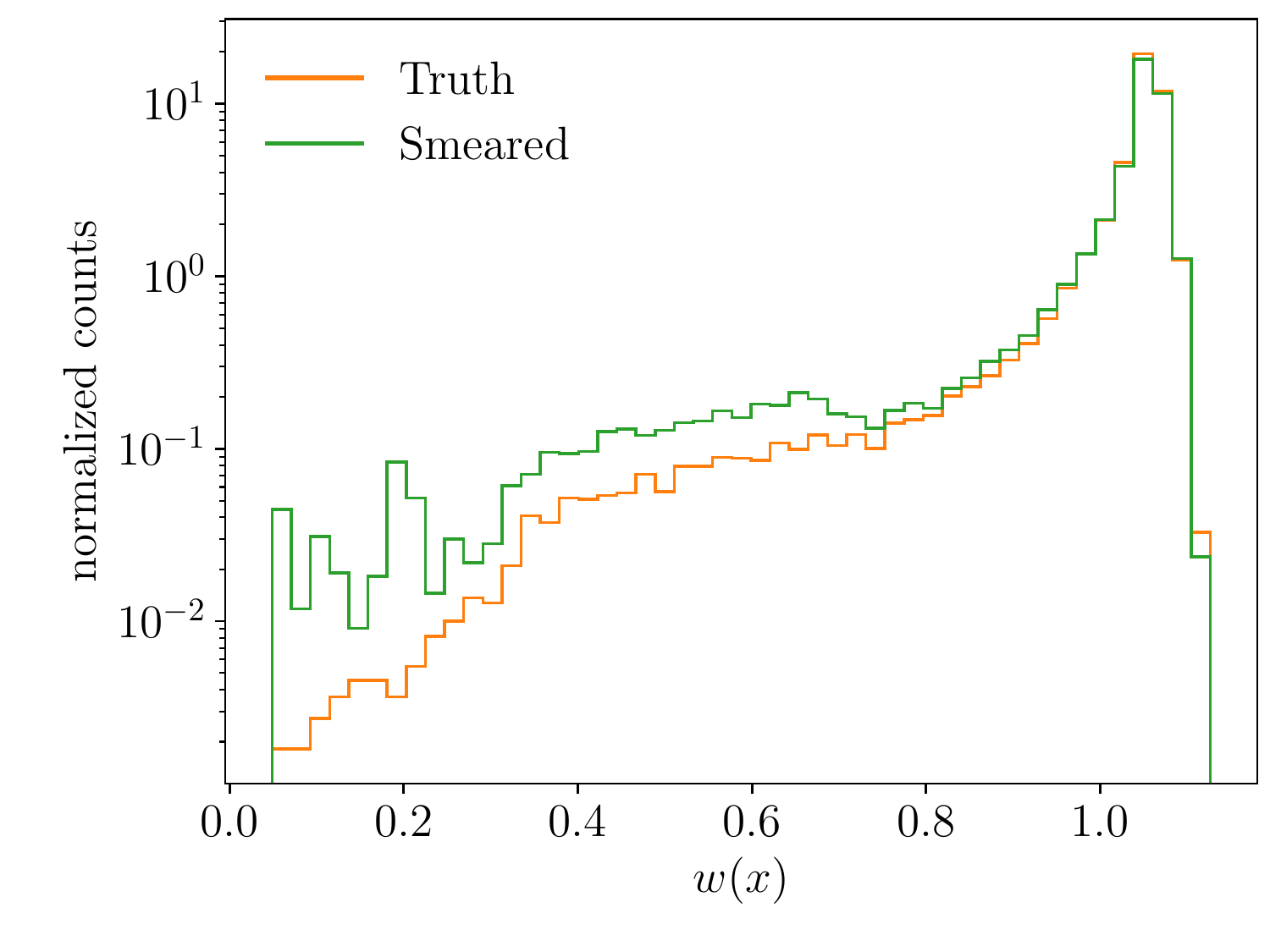}
    \includegraphics[width=0.49\textwidth,page=1]{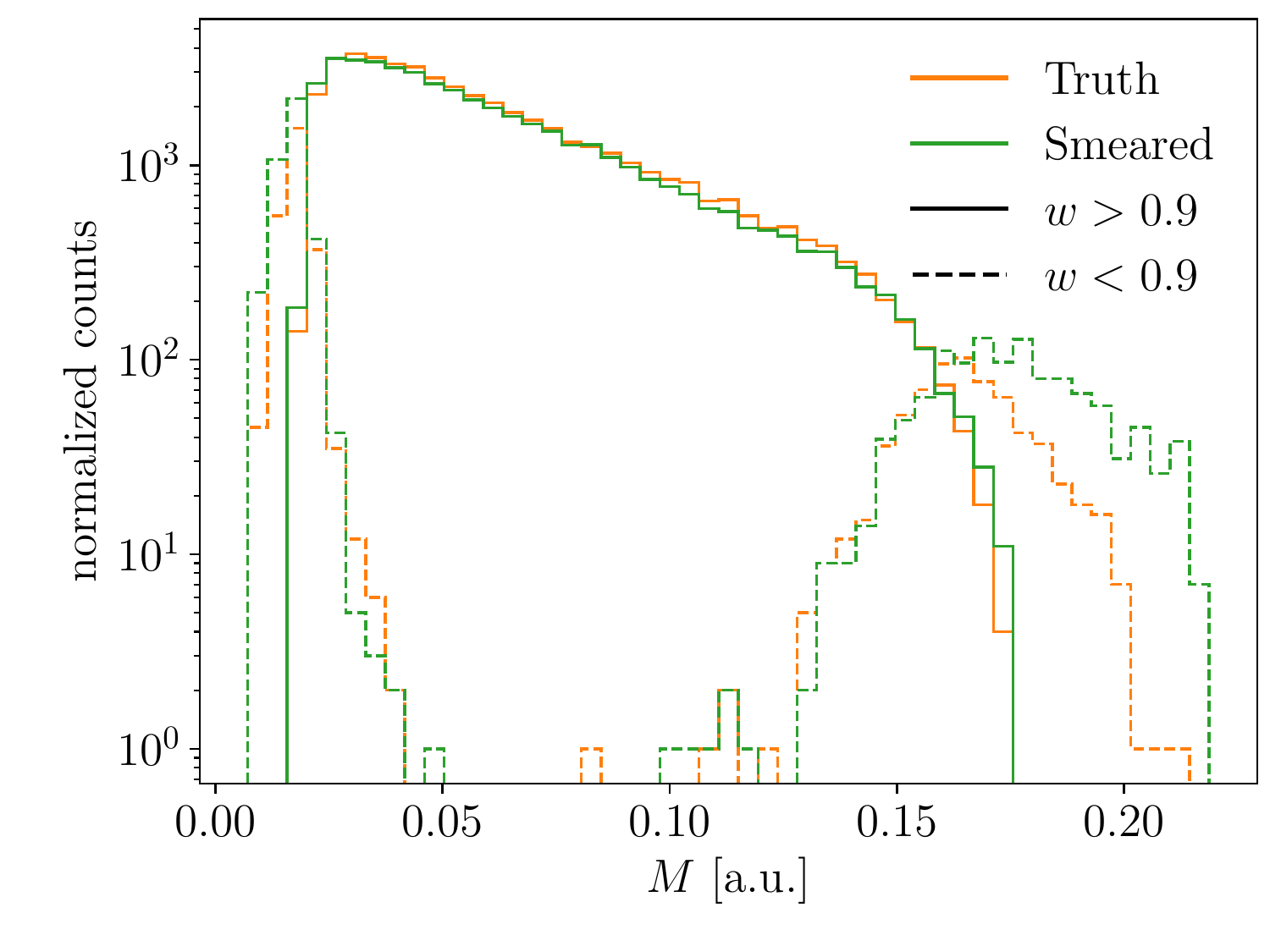}
    \caption{From top left to bottom right: jet mass distribution for
      the smeared distortion; ROC curve from the trained ensemble of
      classifiers; learned weight distribution; jet mass distribution
      for jets in different classifier weight ranges, to identify
      clustering.}
    \label{fig:smeared_weights}
\end{figure}

\begin{figure}[t]
    \includegraphics[width=0.49\textwidth,page=1]{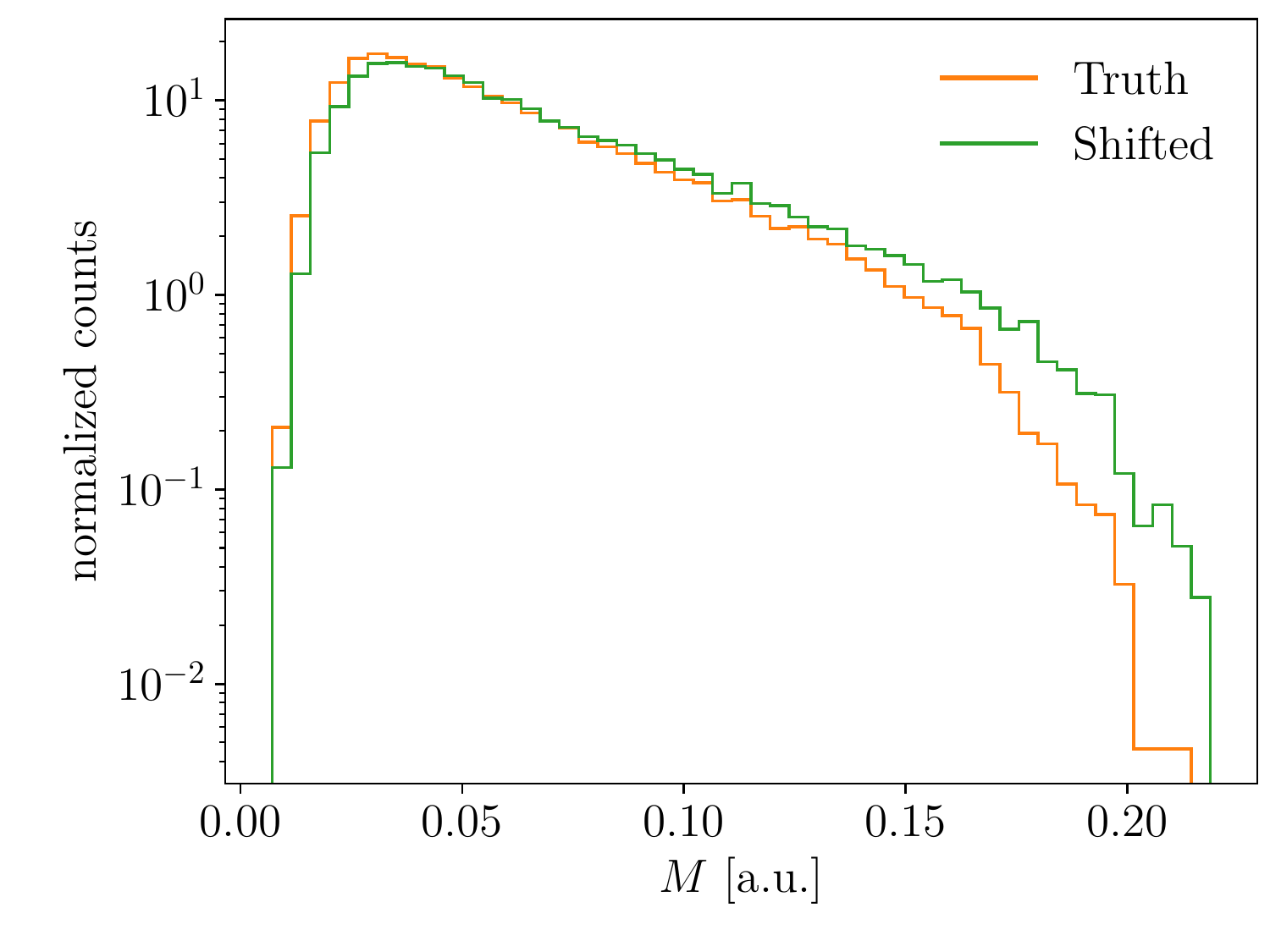}
    \includegraphics[width=0.49\textwidth,page=1]{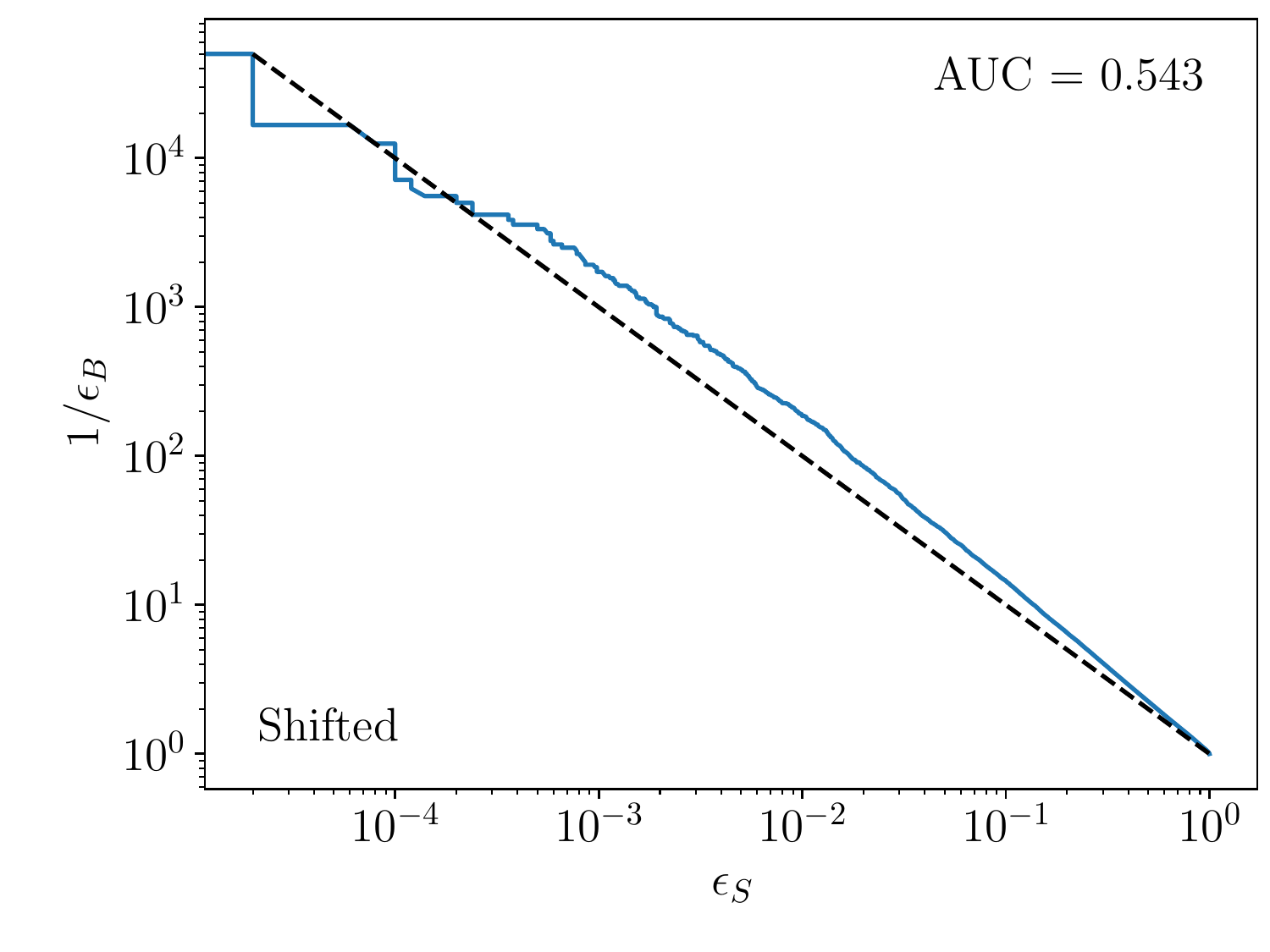} \\
    \includegraphics[width=0.49\textwidth,page=1]{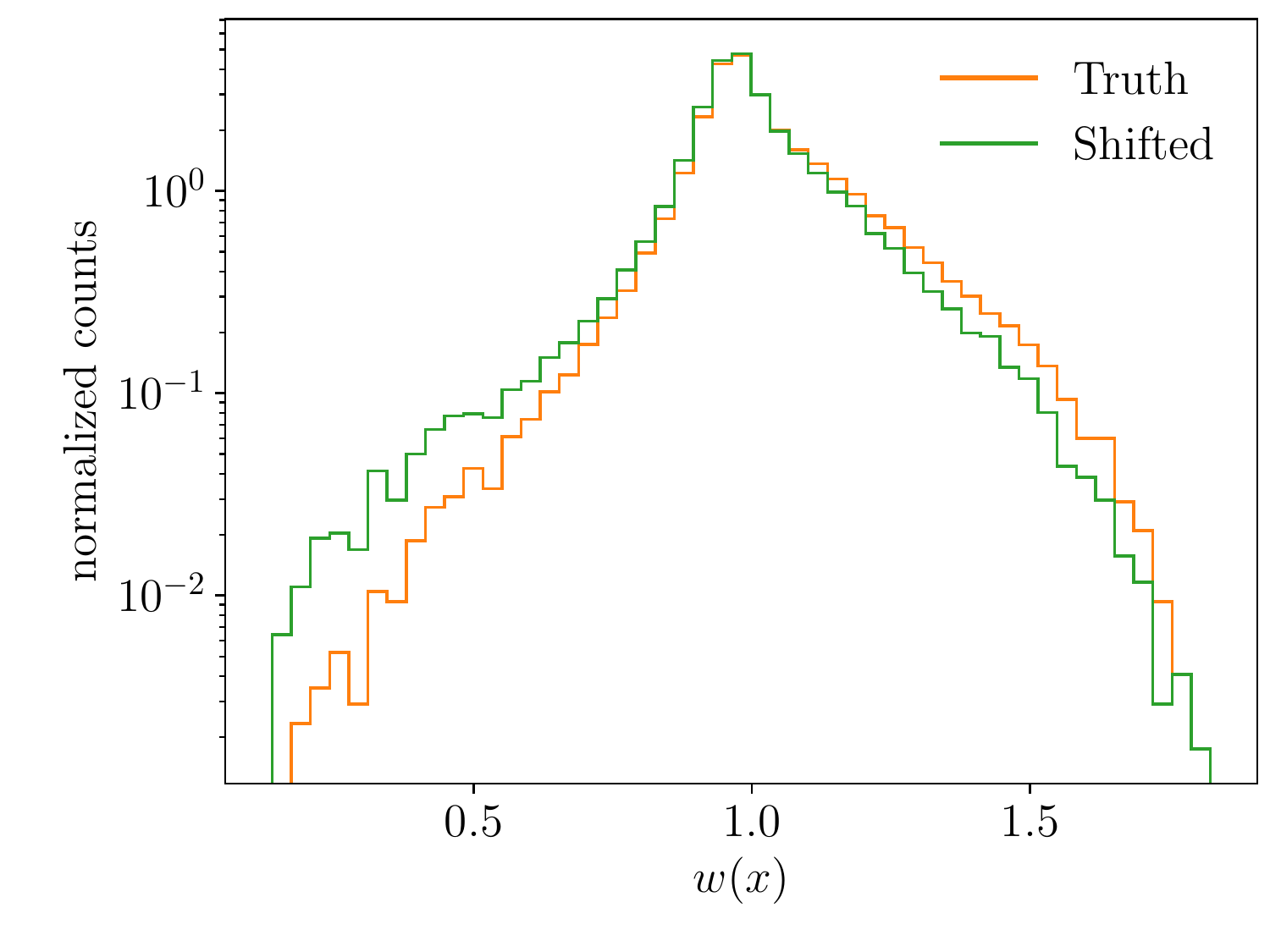}
    \includegraphics[width=0.49\textwidth,page=1]{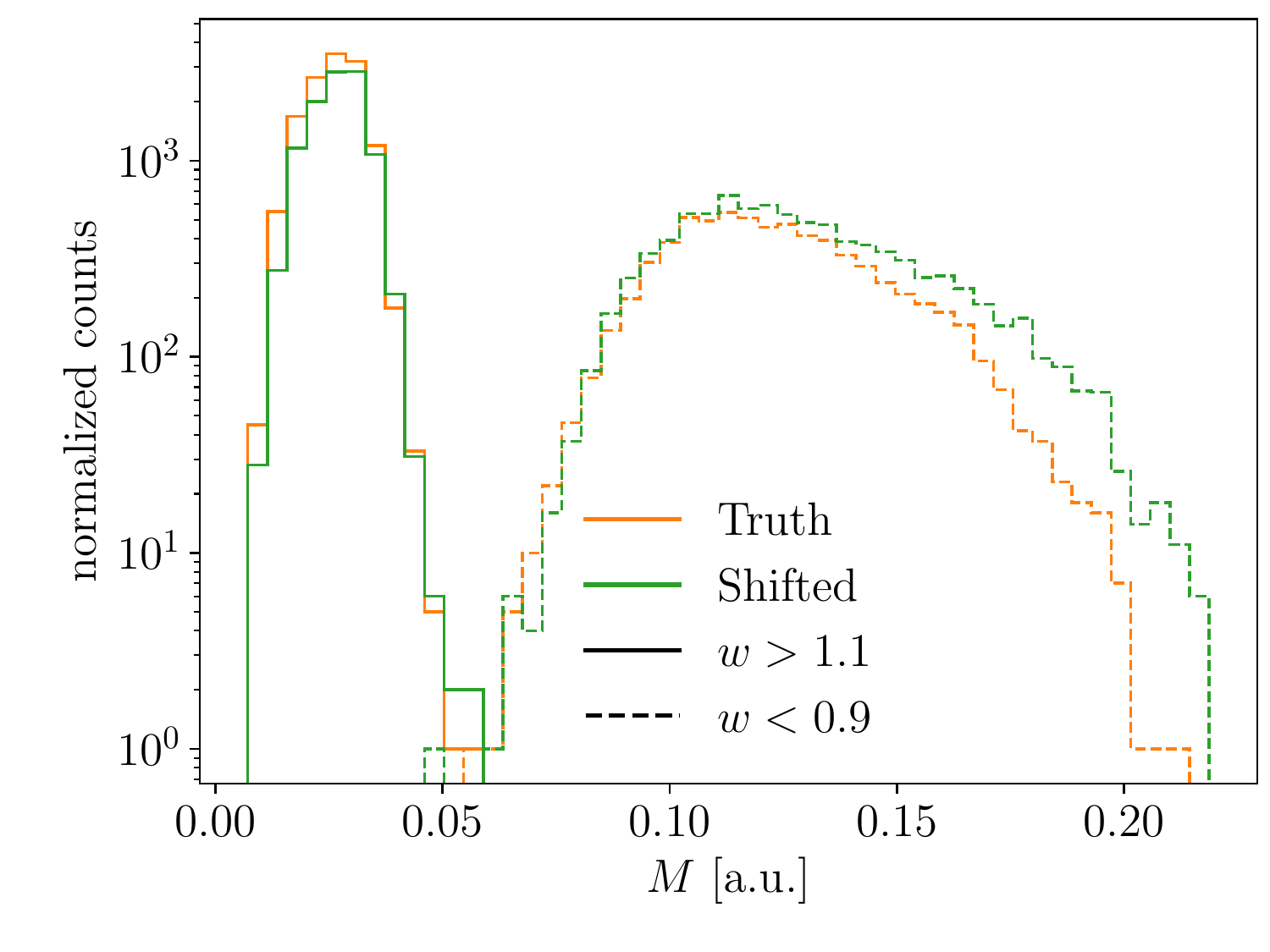}
    \caption{From top left to bottom right: jet mass distribution for
      the shifted distortion; ROC curve from the trained ensemble of
      classifiers; learned weight distribution; jet mass distribution
      for jets in different classifier weight ranges, to identify
      clustering.}
    \label{fig:shifted_weights}
\end{figure}

Figure~\ref{fig:tailcut_weights} shows the results for the tail-cut
case. Ignoring the stochastic nature of the training data, the
NP-optimal classifier should be quite singular: all jets with
$m<0.17$ should have classifier weights given by a delta function at
$w=1-\epsilon$, where $\epsilon$ is the fraction of jets removed
from the tail. In addition, there should be a delta function only for
reference jets at $w=\infty$. A realistic classifier will transform
these two features as a smooth weight distribution. In
Fig.~\ref{fig:tailcut_weights} we see the smooth weight distribution
from our classifier, with nearly all jets populating a sharp peak
near one, and a long tail extending to larger values of $w$, solely in
the reference sample.  This is an example for a large-weight tail
expected for a generative
model that is missing a tail or a feature. Since the small number of
tail jets are soaked up by the bulk, the weight distribution barely
changes and only a tail at large weights appears.

We also see that for the tail cut distortion the AUC is close to 0.5, while
even in the ROC curve itself, it is visible that the classifier is far from
a fully confused classifier. This confirms that the AUC is a terrible metric for the quality of the
generative model.  However, the tail of the weight distribution
gives us all relevant information.
Cutting on the tail of the weight distribution, we correctly identify
the discrepancy in the tail of the jet mass distribution. In Sec.~\ref{sec:events} we will see how 
we can even 
use these weights to recover such a missing feature for a 
quantitative analysis. This
example illustrates nicely how the classifier gives both a sensitive
metric of generative model quality and enables interpretability by 
allowing us to identify in which physics aspect the generative model
is wrong.\medskip

In Fig.~\ref{fig:smeared_weights}, we show the weight
distribution for the smeared distortion. The weight distribution has a maximum at $w > 1$ and is dominated by a small-weight tail. This is expected from the general discussion in Sec.~\ref{sec:general}: the smearing in this case, multiplicative in the jet mass, has a reduced effect at small jet mass and an outsize effect at large mass, and ends up heavily overpopulating the tail of the large-mass regime. 
Correspondingly, there
is hardly any tail with large weights and a very large tail with small weights. Cutting on
the small-weight tail correctly reveals the excess of generated jets,
now appearing on both ends of the jet mass distribution.

Finally, in Fig.~\ref{fig:shifted_weights}, we see the weight distribution for
the shifted distortion, again leading to an unhelpful ROC curve
and an AUC close to 0.5. Since the distortion is small enough to not
significantly overpopulate or underpopulate the tails of the jet mass
distribution, the effect on weight distribution is mild and
symmetric. We also see the characteristic tilted weight
distribution that indicates a well-calibrated classifier, with
generated jets above (below) training jets on the small-weight (large-weight)
side. Cutting on the two tails of the weight distribution
correctly reveals that the over-population and under-population of
generated jets come from the low and high ends of the jet mass
distribution, respectively.

\section{Calorimeter simulation}
\label{sec:calo}

As a second example of how to use weights over phase space to understand the
performance of a generative model, we turn to the classic calorimeter
simulation~\cite{Paganini:2017hrr,Paganini:2017dwg,Krause:2021ilc,Krause:2021wez}, but
with a slightly modified INN architecture~\cite{to_be_published}. We
study weight distributions for positron, photon, and pion showers in a
simplified calorimeter. The classifier defined in
Tab.~\ref{tab:hyperparameters} is trained on voxels, energy, and layer
energies in unnormalized shower data. We focus on the classifier with unnormalized preprocessing in this work because it appears to be better calibrated and shows less propensity for overfitting. For more discussion, see Appendix~\ref{app:calibration}. As a more realistic scenario,
learned calorimeter showers allow us to discussion some aspects of
learned weight distributions in more detail.

\subsection{Tails of weights}
\label{sec:calo_tails}



\begin{figure}[t!]
    \includegraphics[width=0.33\textwidth,page=1]{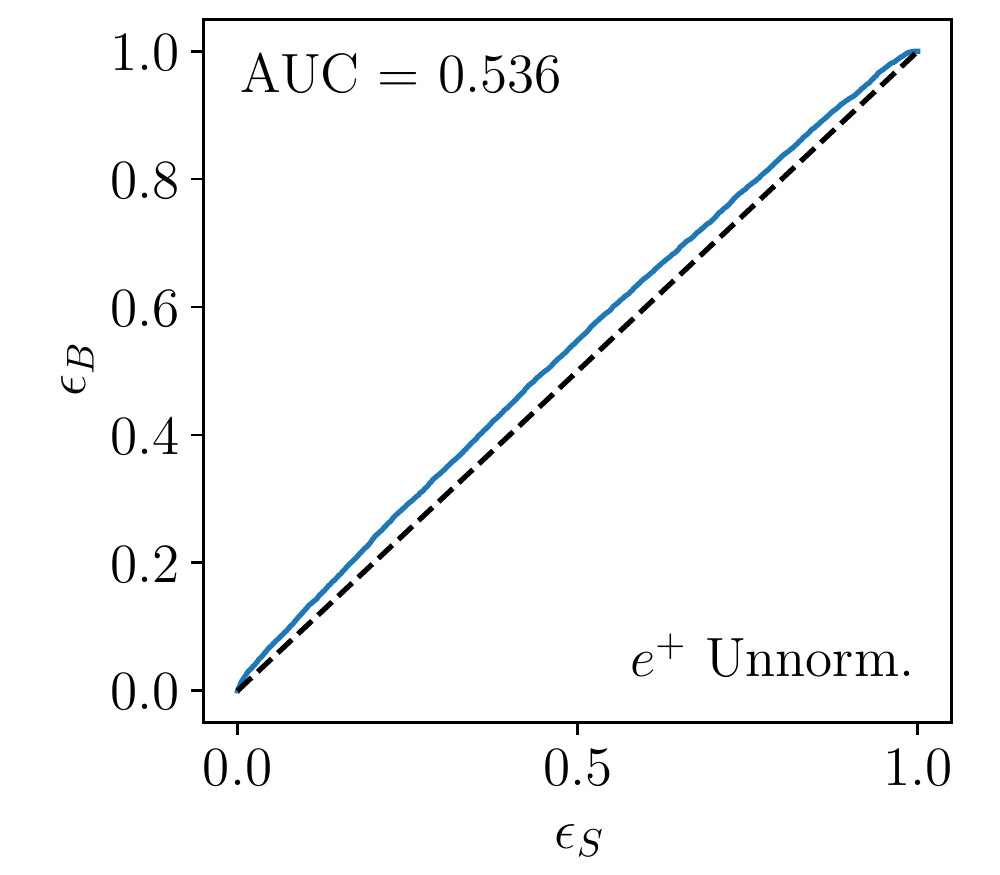}
    \includegraphics[width=0.33\textwidth,page=1]{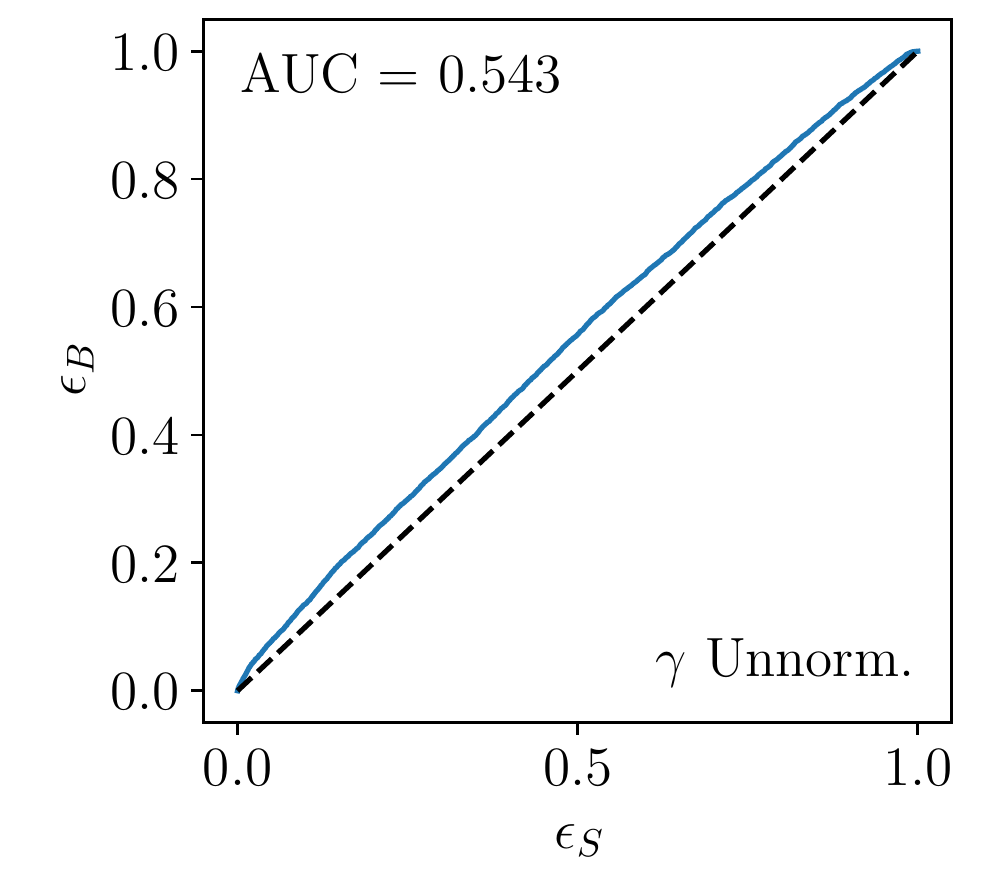} 
    \includegraphics[width=0.33\textwidth,page=1]{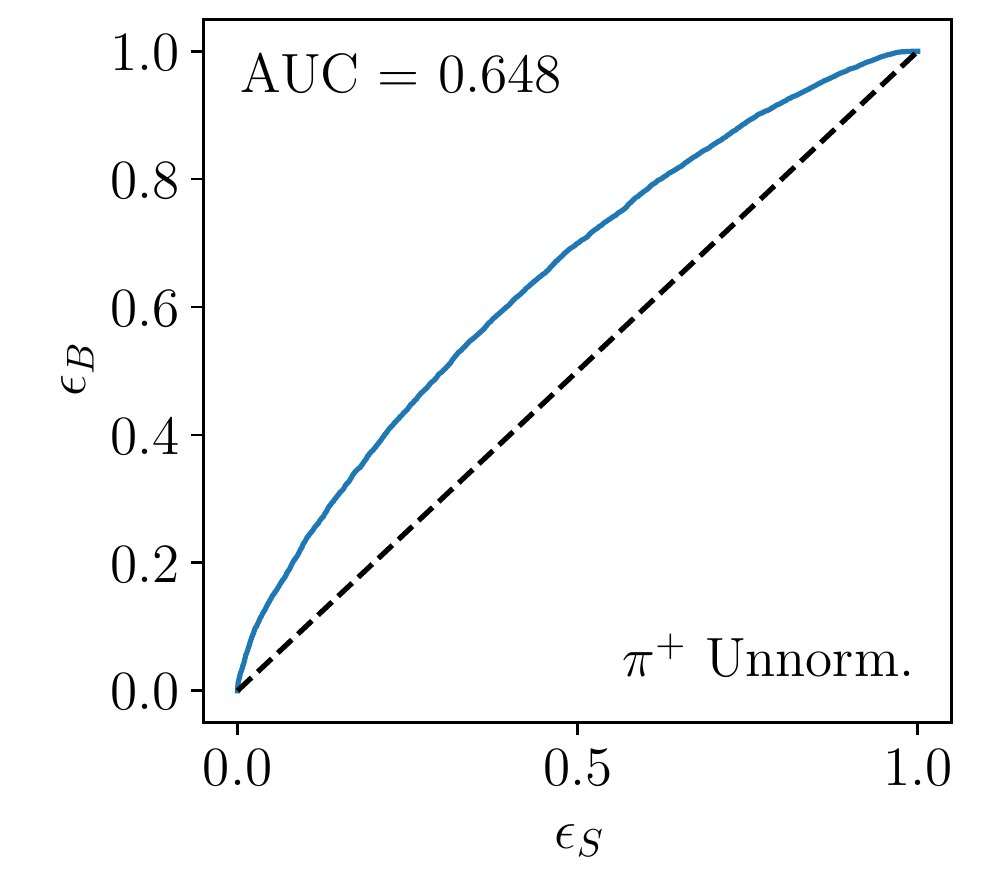} \\
    \includegraphics[width=0.33\textwidth,page=1]{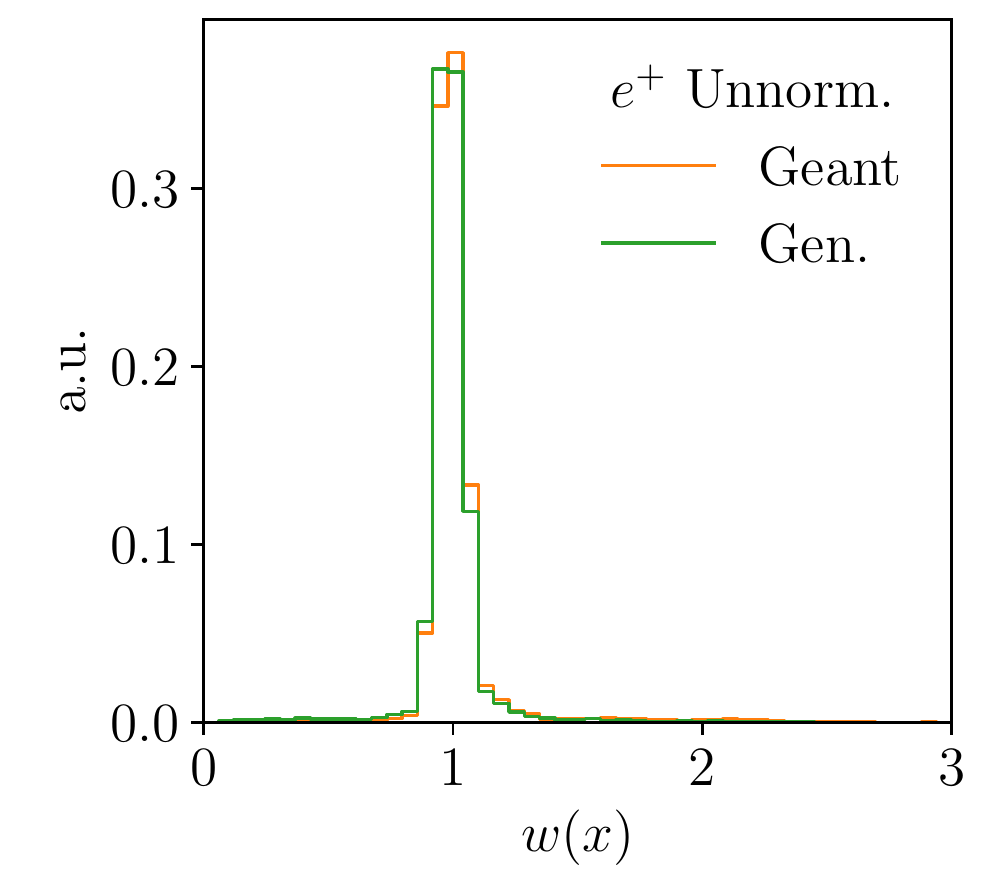} 
    \includegraphics[width=0.33\textwidth,page=1]{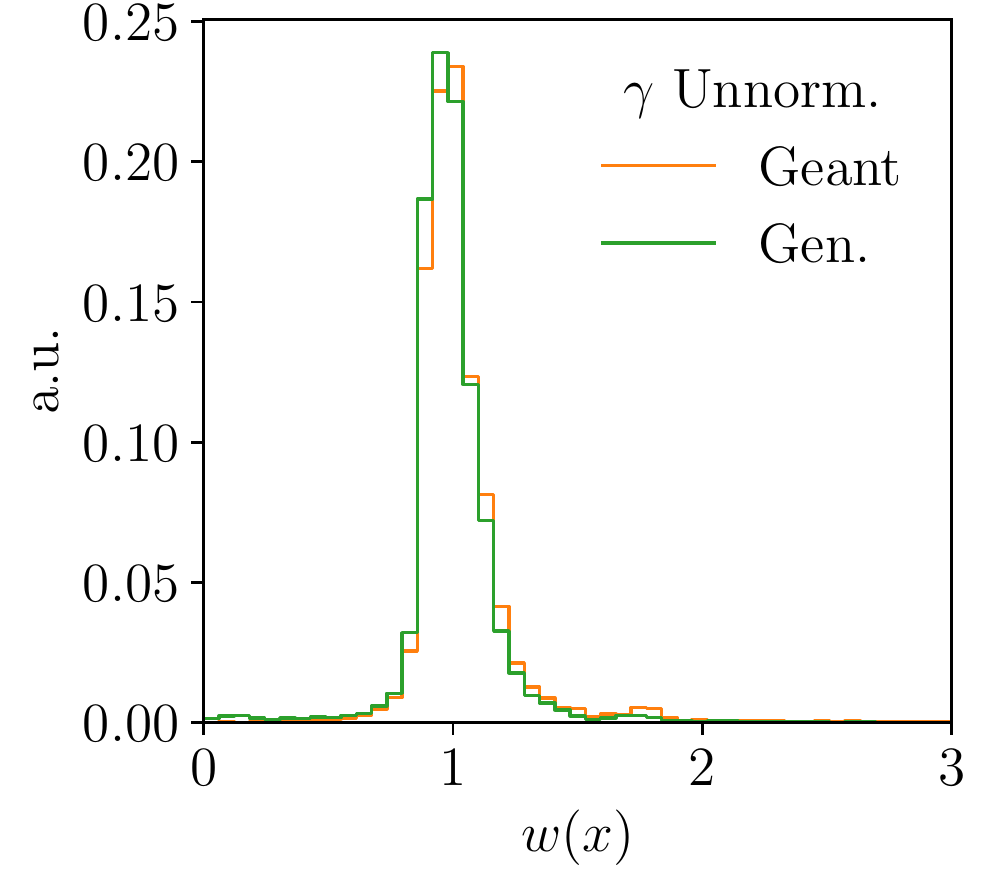} 
    \includegraphics[width=0.33\textwidth,page=1]{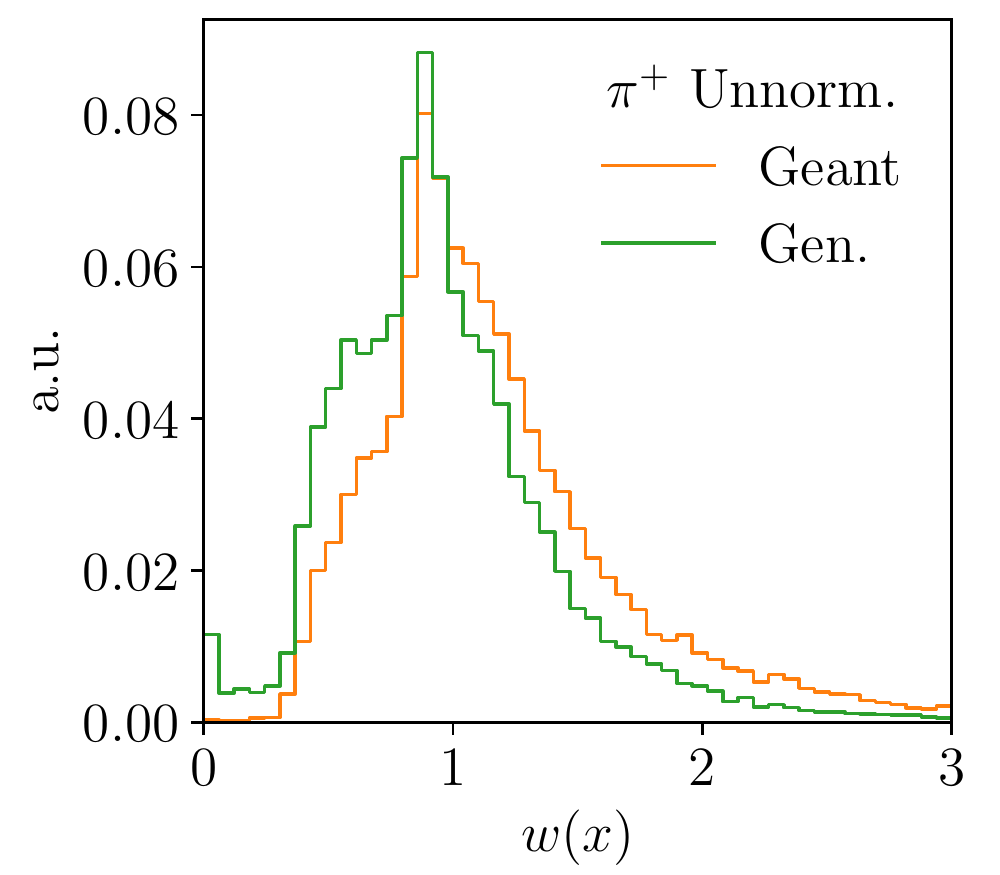} \\
    \includegraphics[width=0.33\textwidth,page=3]{figs/caloinn_plots/weights_e_unnorm}
    \includegraphics[width=0.33\textwidth,page=3]{figs/caloinn_plots/weights_g_unnorm} 
    \includegraphics[width=0.33\textwidth,page=3]{figs/caloinn_plots/weights_pi_unnorm}
    \caption{Left to right: calorimeter showers for $e^+$, $\gamma$,
      and $\pi^+$. Top to bottom: ROC curve, weight distribution on a
      linear scale, and weight distribution on a logarithmic
      scale. The weights are evaluated separately on the \textsc{Geant} dataset
      used for generator training and the generated dataset.}
    \label{fig:calo_weights}
\end{figure}

In Fig.~\ref{fig:calo_weights} we show ROC curves and weight
distributions for $e^+$, $\gamma$, and $\pi^+$ showers. The
top row confirms that positron and photon
showers are easier to generate than pion
showers. The question is which 
potential failures are related to this performance difference.


In the second row we show the weight distributions. First, we observe
that they are not symmetric, because the reweighting now compensates
features.  The limit $w(x) = 0$, most visible for the pion shower,
marks phase space points where the generator has learned a finite
density $\pmd(x)$, where the correct density is $\pd(x) = 0$, one of
the typical failure mode of generative models discussed in
Sec.~\ref{sec:general}. We will see this more clearly for LHC events in the next section, but mention here that it is not catastrophic if we can enforce
corresponding phase space boundaries during generation.

In the third row of Fig.~\ref{fig:calo_weights} we show the same
curves on a logarithmic scale to see the tails.  As expected, they are
different when evaluated on \textsc{Geant} and generated showers. Already for
positrons, the generated data includes many more showers with $w(x)
\ll 1$ than the training data. These are showers for which the
generator overpopulates phase space, so they appear preferably in the
generated dataset. This tail connects to showers with weight zero.

In contrast, showers with $w(x) \gg 1$ appear more frequently in
training dataset. These under-populated regions of phase space
correspond, for instance, to features or tails which the network does
not learn. This serious failure mode can be identified by evaluating
showers with anomalous weights on the training data.

\subsection{Phase space clustering}
\label{sec:calo_cluster}

\begin{figure}[t!]
    \begin{subfigure}{0.325\textwidth}
     \includegraphics[width=\textwidth,page= 1]{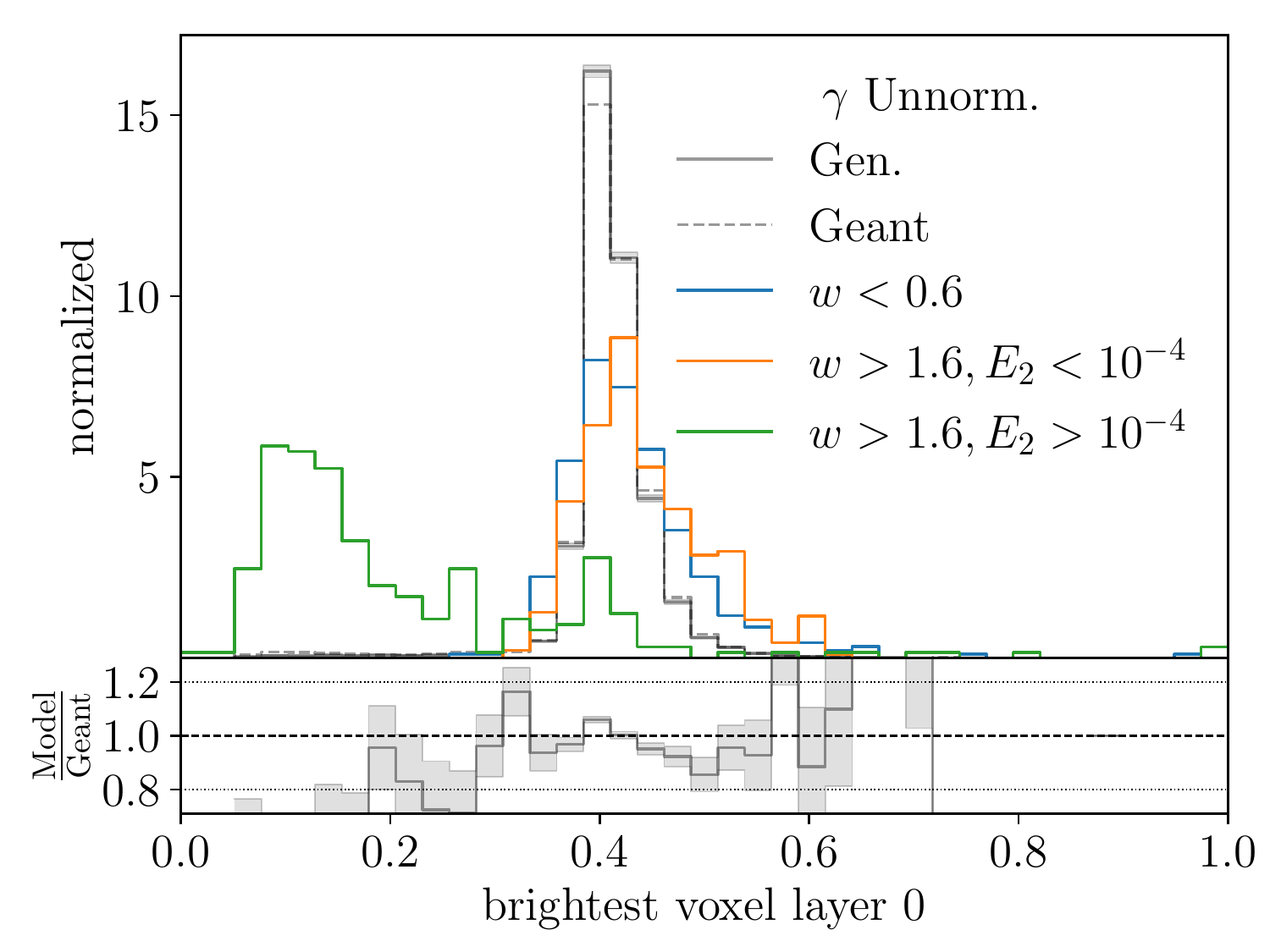} 
     \caption{}
     \end{subfigure}
     \begin{subfigure}{0.325\textwidth}
    \includegraphics[width=\textwidth,page= 2]{figs/caloinn_plots/clustering_unnorm_g.pdf}
         \caption{}
     \end{subfigure}
     \begin{subfigure}{0.325\textwidth}
    \includegraphics[width=\textwidth,page= 3]{figs/caloinn_plots/clustering_unnorm_g.pdf}
         \caption{}
     \end{subfigure}
     \begin{subfigure}{0.325\textwidth}
    \includegraphics[width=\textwidth,page= 4]{figs/caloinn_plots/clustering_unnorm_g.pdf}
         \caption{}
     \end{subfigure}
     \begin{subfigure}{0.325\textwidth}
    \includegraphics[width=\textwidth,page= 5]{figs/caloinn_plots/clustering_unnorm_g.pdf}
         \caption{}
     \end{subfigure}
     \begin{subfigure}{0.325\textwidth}
    \includegraphics[width=\textwidth,page= 6]{figs/caloinn_plots/clustering_unnorm_g.pdf}
         \caption{}
     \end{subfigure}
    \begin{subfigure}{0.325\textwidth}
    \includegraphics[width=\textwidth,page= 14]{figs/caloinn_plots/clustering_unnorm_g.pdf} 
         \caption{}
     \end{subfigure}
     \begin{subfigure}{0.325\textwidth}
    \includegraphics[width=\textwidth,page= 15]{figs/caloinn_plots/clustering_unnorm_g.pdf}
         \caption{}
     \end{subfigure}
     \begin{subfigure}{0.325\textwidth}
    \includegraphics[width=\textwidth,page= 16]{figs/caloinn_plots/clustering_unnorm_g.pdf}
         \caption{}
     \end{subfigure}
    \caption{Relevant distributions for $\gamma$ showers in the
      small-weights (blue) and large-weights regions (orange and green). We show the energy depositions, the fraction of the energy deposited in the leading voxel, and the sparsity in the three layers of the calorimeter.}
    \label{fig:calo_cluster1}
\end{figure}

The simpler structure of photon showers allows for a detailed study
of the clustered observables. By cutting on the weight values and looking at the distribution of the remaining photon showers, we identify three characteristic failure modes highlighted with different colors  Fig.~\ref{fig:calo_cluster1}.

\begin{figure}[t]
    \includegraphics[width=0.48\textwidth,page=4]{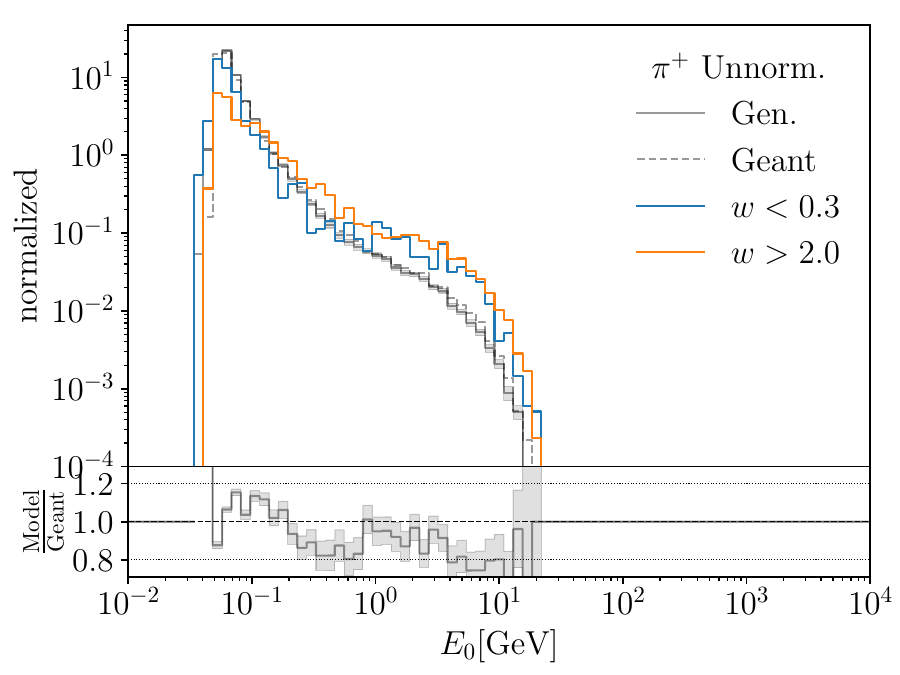}
    \includegraphics[width=0.48\textwidth,page=14]{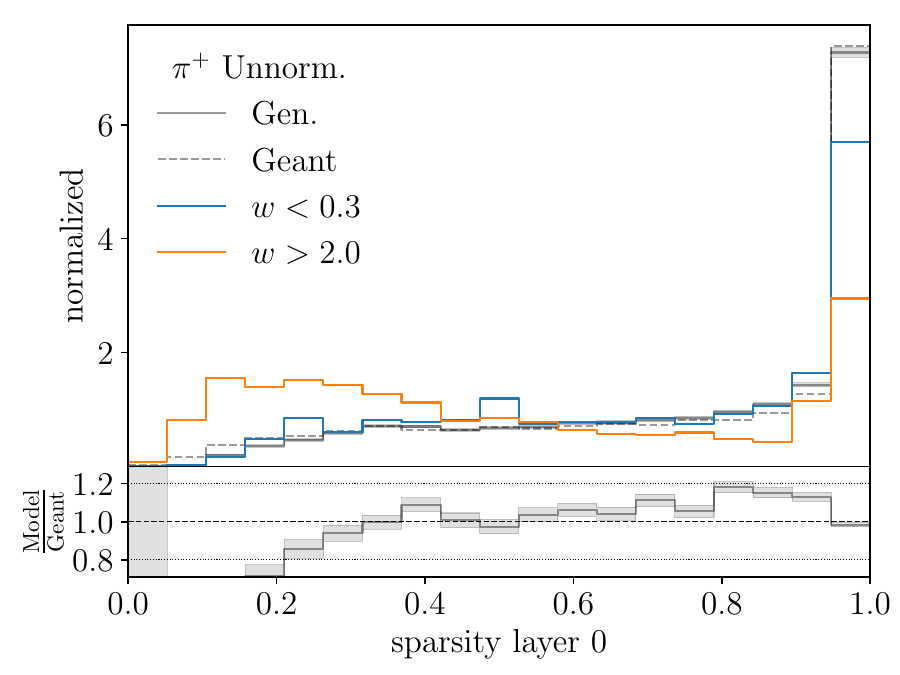} \\
    \includegraphics[width=0.48\textwidth,page=3]{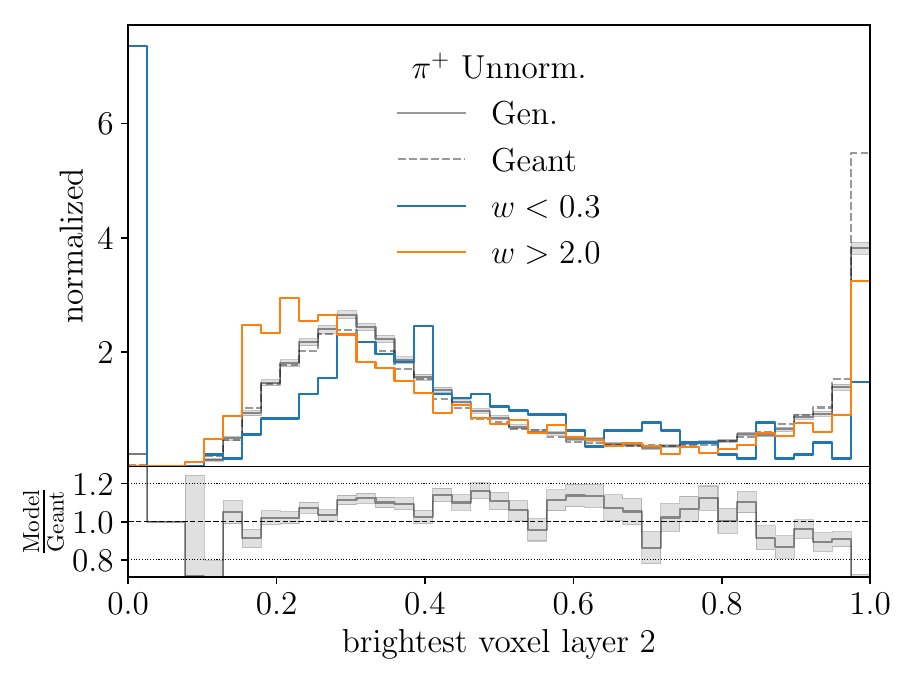} 
    \includegraphics[width=0.48\textwidth,page=16]{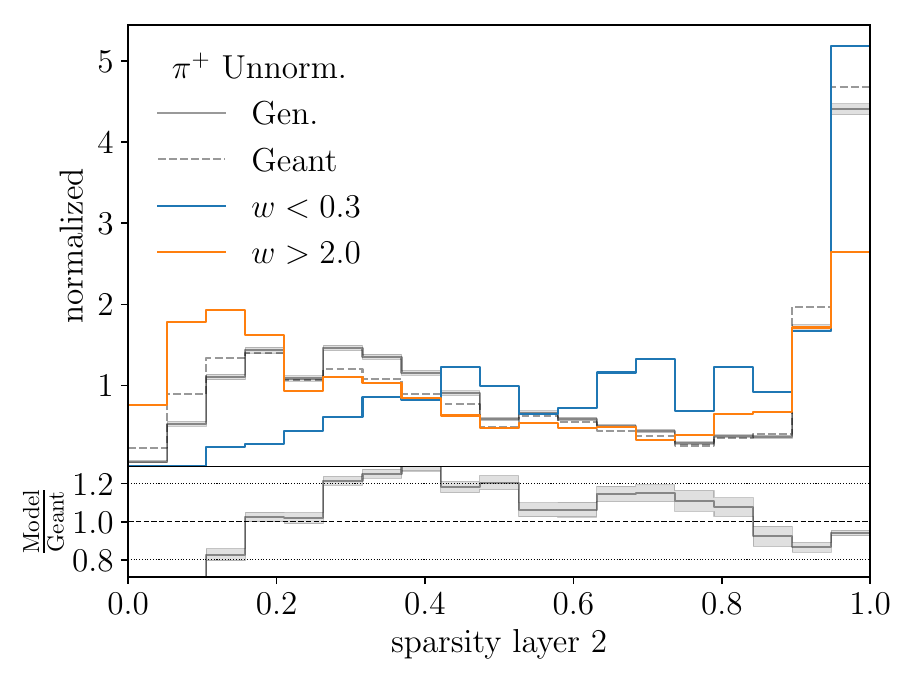}
    \caption{Relevant distributions for $\pi^{+}$ showers in the
      small-weights (blue) and large-weights regions (orange). We show the energy deposition and the sparsity in layer~0, and the brightest voxel energy and the sparsity in layer~2.}
    \label{fig:calo_cluster2}
\end{figure}

\begin{enumerate}

\item In orange, we isolate the  large-weights tail with $w>1.6$ and no energy deposited in layer~2 ($E_2 < 0.1 \, \text{MeV}$), as shown in Figs.~\ref{fig:calo_cluster1}(c) and \ref{fig:calo_cluster1}(f). As shown in Figs.~\ref{fig:calo_cluster1}(g) and \ref{fig:calo_cluster1}(h), these showers have higher sparsity\footnote{Here, we are redefining sparsity compared to previous literature \cite{Paganini:2017dwg,Krause:2021ilc}: sparsity(here)=1-sparsity(there). This way, higher sparsity means more sparse showers (i.e.\ showers with only a few voxels activated) while lower sparsity indicates less sparse showers (i.e.\ showers with many voxels activated).} in layers~0 and~1 than the typical shower. Additionally they have lower energy, shown by the $E_1$ histogram in Fig.~\ref{fig:calo_cluster1}(e), since on average most of the energy is deposited in layer~1. Overall, these showers consist of just a few activated, low-energy voxels in layers~0 and~1, and exactly none in layer~2. This sub-population of showers exists in the GEANT data, but it is not sufficiently generated by the network. 

\item In blue, we isolate the small-weights tail with $w<0.6$. Fig.~\ref{fig:calo_cluster1}(c) shows that this failure mode is characterized by a single voxel carrying all the energy in layer~2, and Fig.~\ref{fig:calo_cluster1}(e) shows that this energy is lower than the average energy deposition. Blue and orange agree in every feature that we looked at in layers~0 and~1; they only differ in layer~2. Since these are showers overproduced by the generator, we interpret this as the compensation of the generator for the underproduction of the orange showers; the compensation is only needed in layer~2. 
We think the reason for both the orange and blue failure modes is due to the low energy and the large number of zero voxels in these showers: this causes them to be especially sensitive to the noise we add during training, since a single voxel is being activated and it either falls just under or just over the minimum energy threshold. 
The vicinity of these showers to the noise threshold makes it harder for the generator to perfectly model this region of phase space.

\item Finally, in green  we isolate again the large-weights tail with $w>1.6$ that {\it does} deposit energy in layer~2 ($E_2 > 0.1 \, \text{MeV}$). These showers are also underproduced by the generator but they are distinct from the previous two classes. According to Fig.~\ref{fig:calo_cluster1}(d)-(f), these have very low energy in layer~0 (even lower energy than the orange showers), and higher-than-typical energy in layer~2. In layer~1 their energy is closer to the typical shower.  We also see in the sparsity that these photons deposit very little energy and activity in layer~0, while in layers~1 and~2 they are fairly typical.
These are showers which develop late in the calorimeter, leaving little or no energy in layer~0. Interestingly, physics tells us that these late-developing showers are possible for photons but not likely for positrons. At high energies, the latter interact  continuously with the material through Bremsstrahlung, while the former need to convert to $e^+e^-$ first\cite{Wigmans:2002kba}. This leads to showers fully absorbed deeper in the calorimeter, therefore with more energy deposited in layer 2. We see this difference in the physics clearly reflected comparing with the green showers for the positron case (see App.\ref{app:calo}). The positrons have energy deposited in layer~0, unlike the photons.

\end{enumerate}

\bigskip

The situation becomes much more complicated when looking at pions, where the more complex physics through the nuclear interaction and the poorer generative model make it harder to identify failure modes with kinematic or physics features. In line with the sobering AUC value given in Fig.~\ref{fig:calo_weights}, we see in Fig.~\ref{fig:calo_cluster2} that the generator requires correction weights essentially all over phase space. The first distinctive failure mode is corrected by small weights in the energy distributions, for instance in layer~0, which suppress the generated showers to reproduce the sharp lower edge of the energy deposition. In addition, the network produces too many showers with exactly zero energy deposition in layers~1 and~2 (see App.\ref{app:calo}). They are included in an overflow bin in the energy histograms, but appear as a failure mode in the energy fraction of the brightest voxel, for example in layer~2. Finally, we see showers with large weights cluster at low sparsities. Here the generator has a systematic bias towards simpler showers with fewer voxels. The full set of studied observables for $e^{+}$, $\gamma$, and $\pi{+}$ can be found in Appendix~\ref{app:calo}.
Given these observations, the leading improvement to the generative model concerns the low-energetic voxels. As discussed before, this can be linked to the addition of noise during training and provides us a research direction to improve the generator, e.g. sampling from a different noise distribution or the development of a noise-less training scheme.

\section{Event generation}
\label{sec:events}

\begin{figure}[b!]
    \includegraphics[width=0.33\textwidth,page=1]{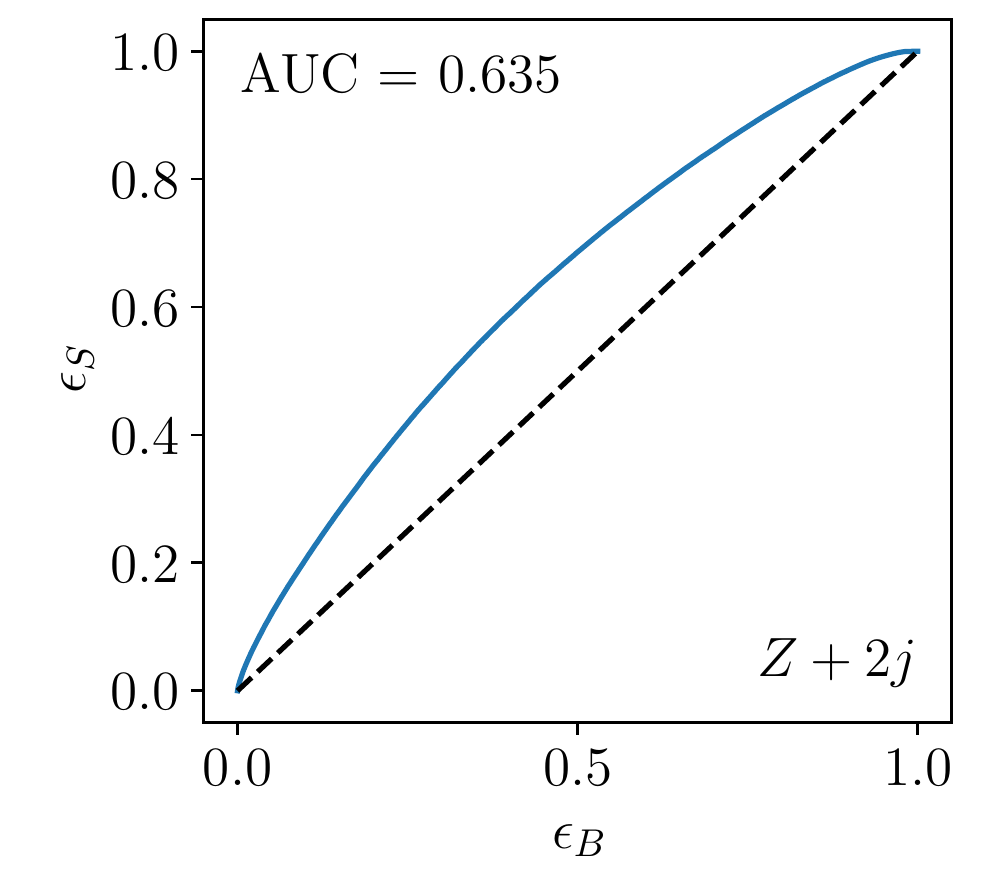}
    \includegraphics[width=0.33\textwidth,page=1]{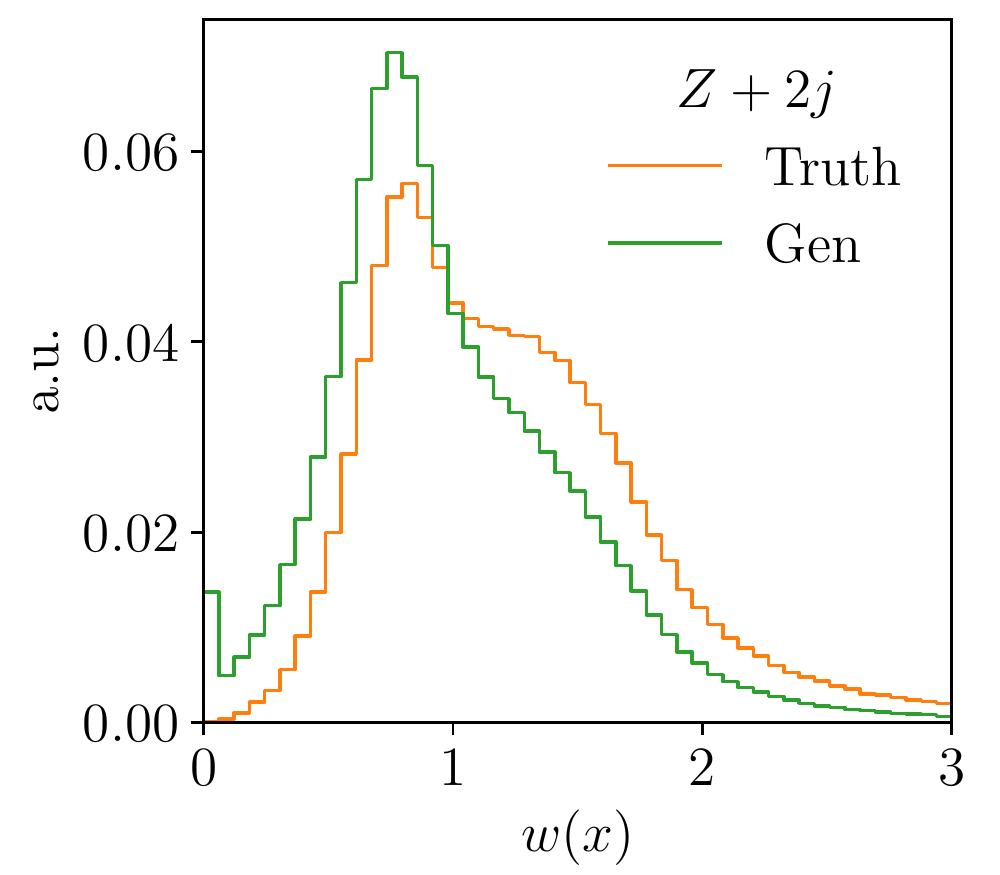}
    \includegraphics[width=0.33\textwidth,page=4]{/prec_inn_bad_generator/weights_z2j}
    \caption{Left to right: ROC curve, weight
      distribution on a linear scale, and weight distribution on a
      logarithmic scale for $Z+2$~jets events, using the outdated
      standard generator. The weights
      are evaluated separately on the true, training dataset for the generator 
      and the generated dataset.}
    \label{fig:prec_inn_weights}
\end{figure}

The third network we analyze using learned classifier weights
generates events for the process
\begin{align}
  pp \to ( Z \to \mu^+ \mu^- ) + 1,2,3~\text{jets}
\end{align}
at the reconstruction level, using the precision INN architecture
described in detail in Ref.~\cite{Butter:2021csz}. We
first use the published version and then the current state of the art,
for which we remove the PCA preprocessing, as it introduces
correlations between different jet multiplicities which make the
training harder. The convergence of the updated Bayesian version is
improved by initializing the standard deviations of the trainable
weights with a small value, bringing its performance close to the
deterministic version.

As in Ref.~\cite{Butter:2021csz}, we train a classifier on the same
observables as the generator. Because the classifier does not have an
invertibility constraint, we can add more features as network
inputs. For LHC events, the generator will wash out intermediate mass peaks
and the $\Delta R$ distribution between jets, so we provide the
classifier with
\begin{align}
    \Big\{ p_{T,i}, \eta_i, \Delta \phi_{i,i-1}, M_i \Big\} 
    \cup \Big\{ M_{\mu\mu} \Big\}
    \cup \Big\{ \Delta R_{i_1,i_2} \Big\} 
    \cup \Big\{ \Delta R_{i_2,i_3}, \Delta R_{i_1,i_3} \Big\} \; , 
\end{align}
where $M_i$ is only present for muons and there is no $\Delta \phi$
for the first particle. In addition, to help the network focus on
small $\Delta R$, we take the inverse of this observable and apply a
cutoff as a preprocessing step.

\subsection{Standard generator and mass peak}
\label{sec:events_standard}

\begin{figure}[t]
    \includegraphics[width=0.49\textwidth,page=11]{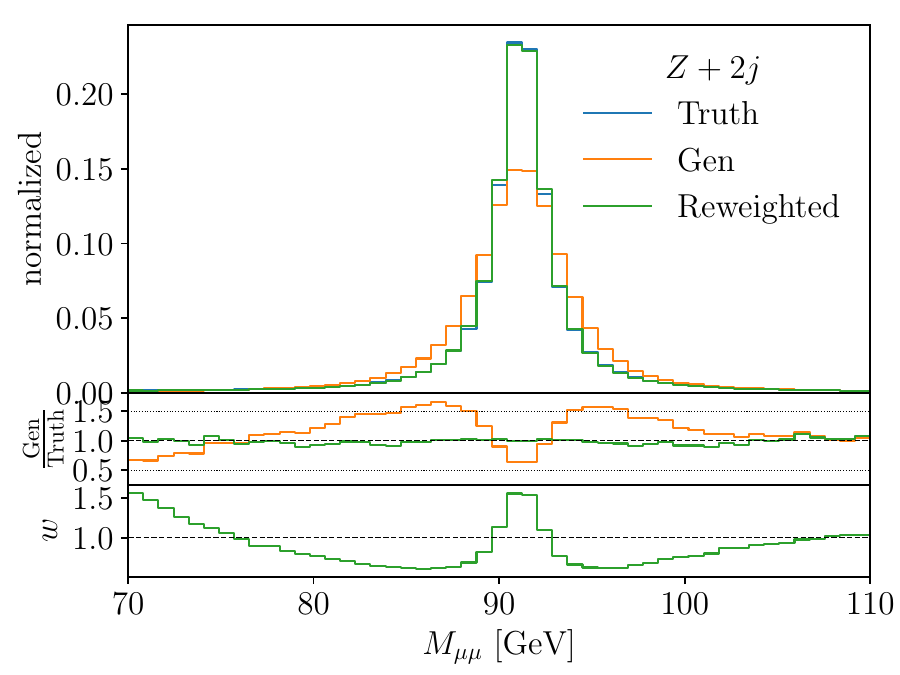}
    \includegraphics[width=0.49\textwidth,page=11]{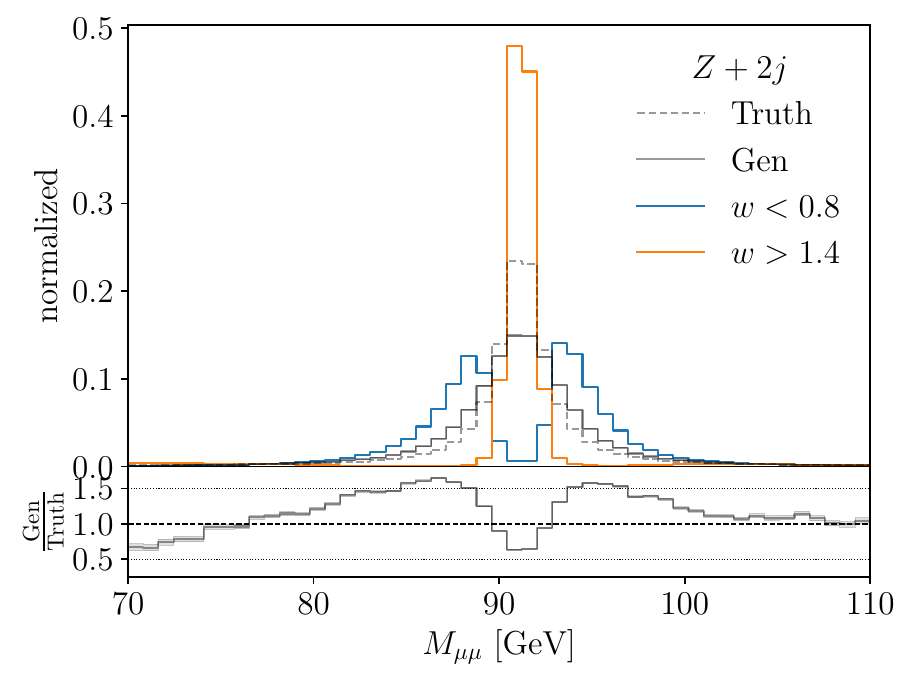}
    \caption{$Z$-peak distributions for $Z+2$~jets events from the
      outdated standard generator. We show the agreement between 
      the generated events with the truth or training data (left) and 
      evens in different weight ranges (right). The events with small 
      weights are taken from the generated distribution, the events 
      with large weights are taken from the truth distribution.}
    \label{fig:prec_inn_clustering}
\end{figure}

We start the discussion of potential failure modes of the event
generator with the old network setup from Ref.~\cite{Butter:2021csz}, and with the weight distributions shown in Fig.~\ref{fig:prec_inn_weights}. This network encounters difficulties in reproducing the $Z$-peak,
where the learned width turns out too large for two and three jets. An example of this is shown in Fig.~\ref{fig:prec_inn_clustering} for $Z+2$~jets. In the upper sub-panels we show the ratio of generated to truth density as a function of $M_{\mu\mu}$, the most discrepant distribution for this generator. We see a characteristic shape in the density ratio aligned with  the $M_{\mu\mu}$ distribution. The ratio shows a dip where the model underpopulates the true distributions due to the smearing and two massive shoulders on either side of the peak, where the smearing cause an overpopulation of generated events relative to truth. The trained classifier compensates this density ratio with values as large as $w  \sim 1.5$ on the $M_{\mu \mu}$ resonance and $w = 0.6~...~0.8$ on its shoulders.

The corresponding distribution of trained classifier weights is shown in Fig.~\ref{fig:prec_inn_weights}. In this case, the main peak is shifted to $w<1$, driven by the overpopulated wings of the smeared $M_{\mu\mu}$ distribution. A secondary peak/shoulder appears around $w \sim 1.5$, corresponding to the underpopulated $M_{\mu\mu}$ resonance. 
It is interesting to compare this weight distributions from the smeared $M_{\mu \mu}$ resonance and the smeared  distortion of the JetNet data in Fig.~\ref{fig:smeared_weights}. Although both are driven by a smearing, the weight distributions are very different. For the smeared jets the maximum of the weight distribution appears at $w > 1$, representing the actual peak configurations, while for the LHC events the maximum of the weight distribution is shifted to $w<1$, driven by the shoulders of the smeared peak. This reflects the clear differences in the form of the smeared phase space feature and the details of the actual smearing. The NP-classifier does not identify the smearing mechanism in the sense of a Wasserstein-distance, but tracks the density ratio over phase space and requires an interpretation of the entire weight distribution and the corresponding interpretable phase space.

\begin{figure}[t]
    \includegraphics[width=0.33\textwidth,page=1]{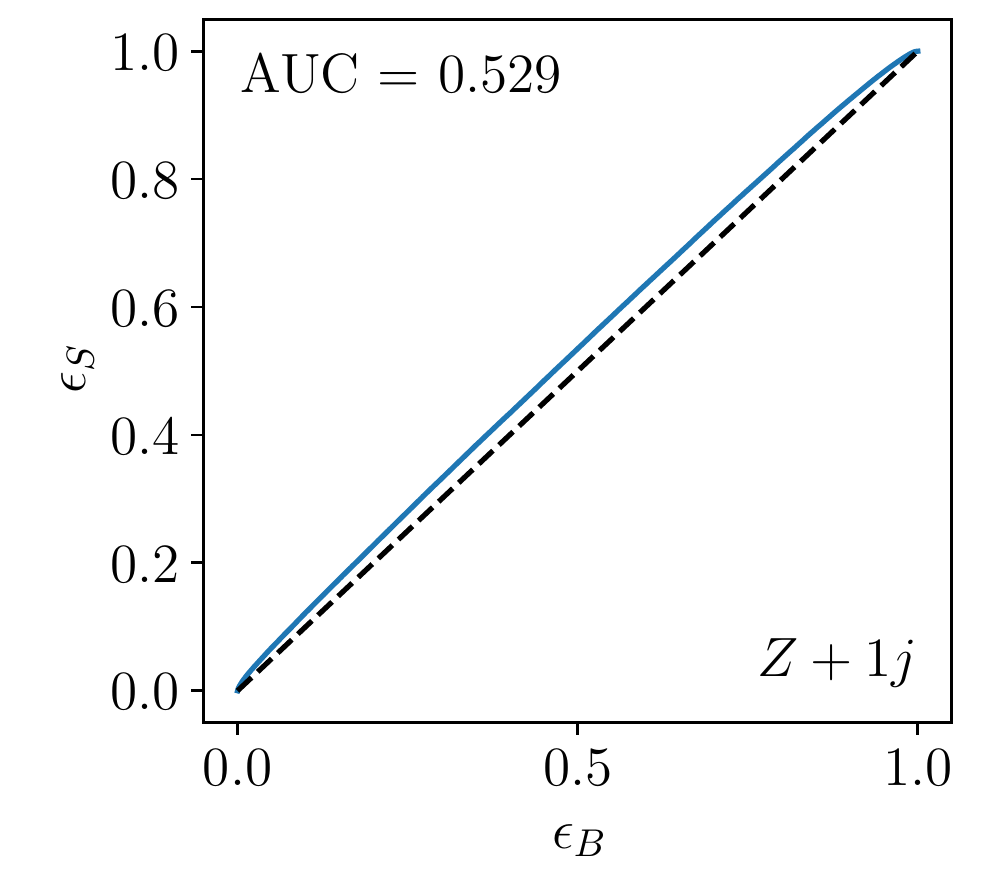}
    \includegraphics[width=0.33\textwidth,page=1]{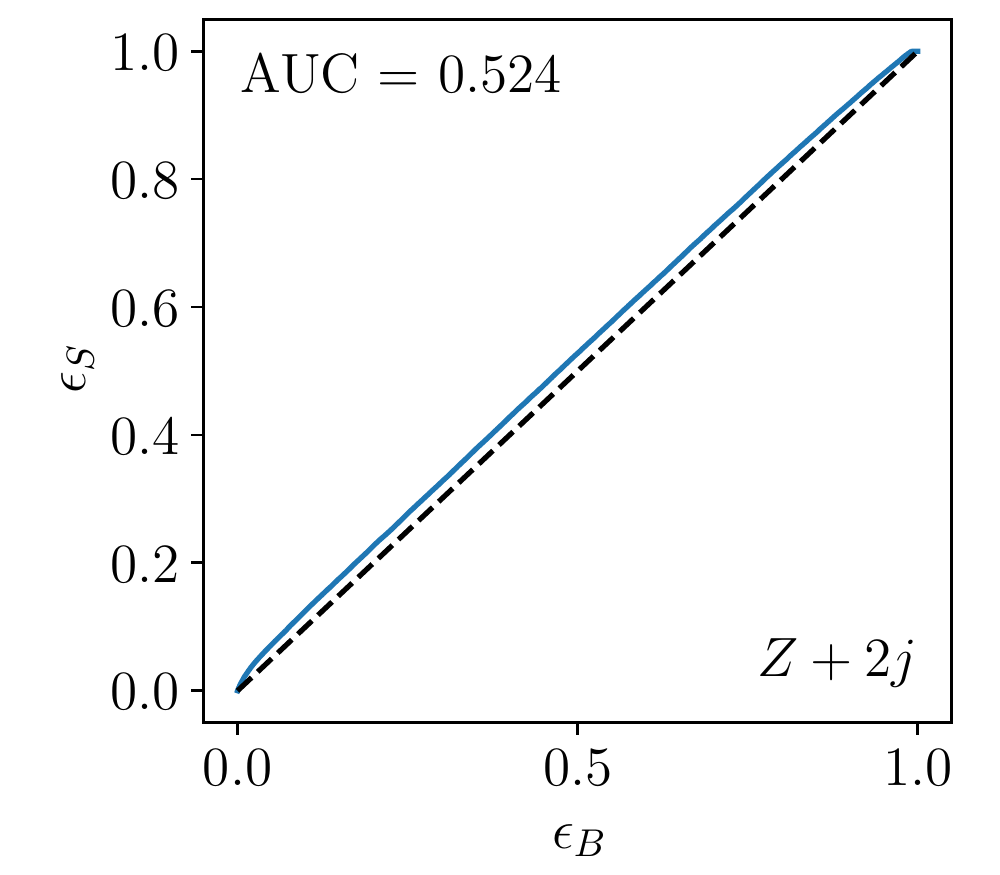}
    \includegraphics[width=0.33\textwidth,page=1]{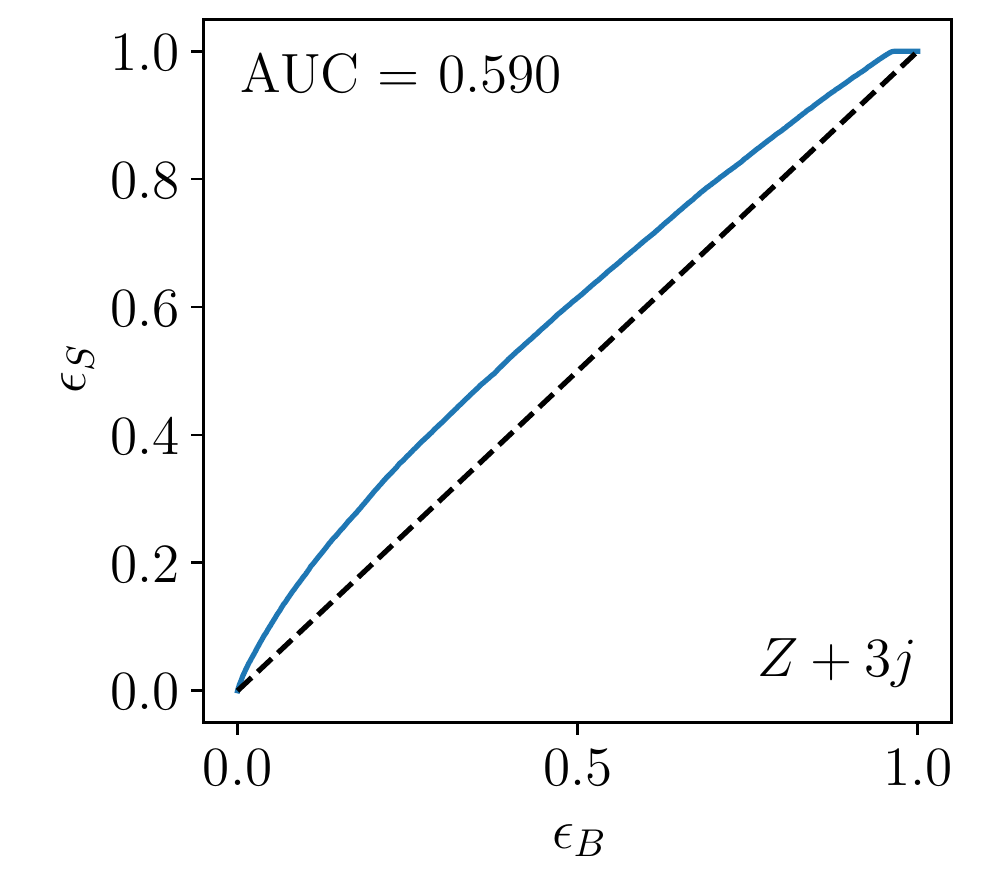}\\
    \includegraphics[width=0.33\textwidth,page=1]{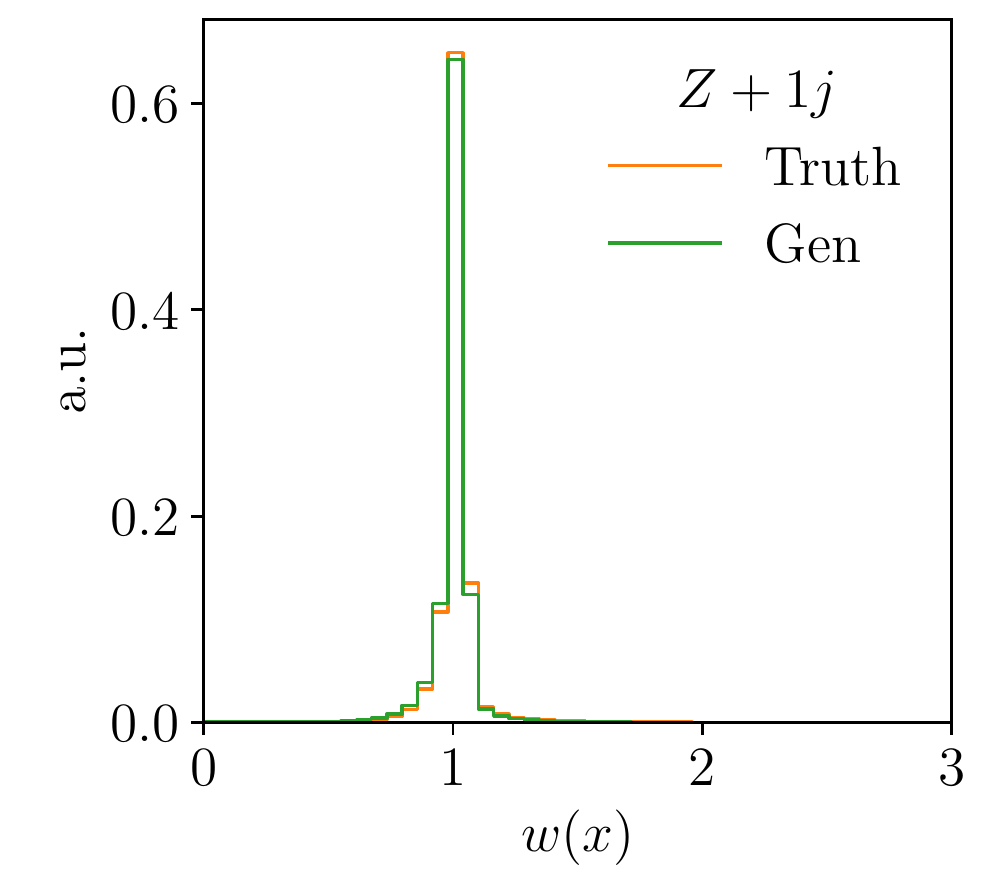}
    \includegraphics[width=0.33\textwidth,page=1]{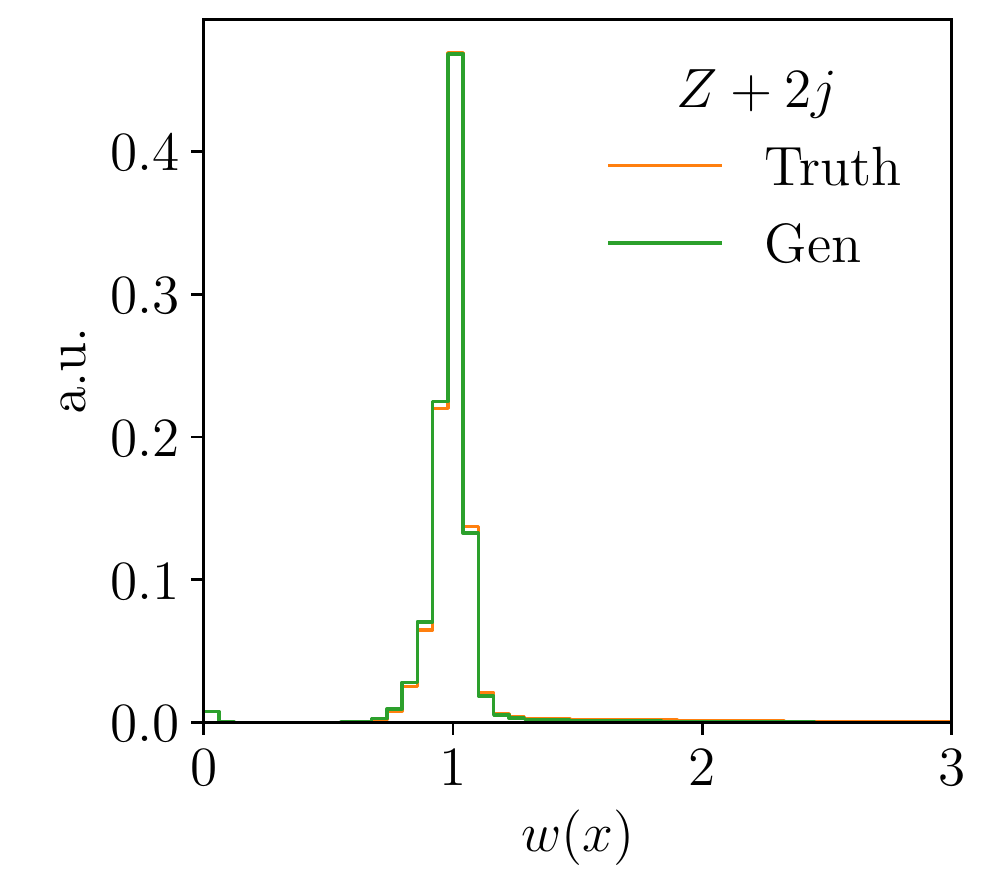}
    \includegraphics[width=0.33\textwidth,page=1]{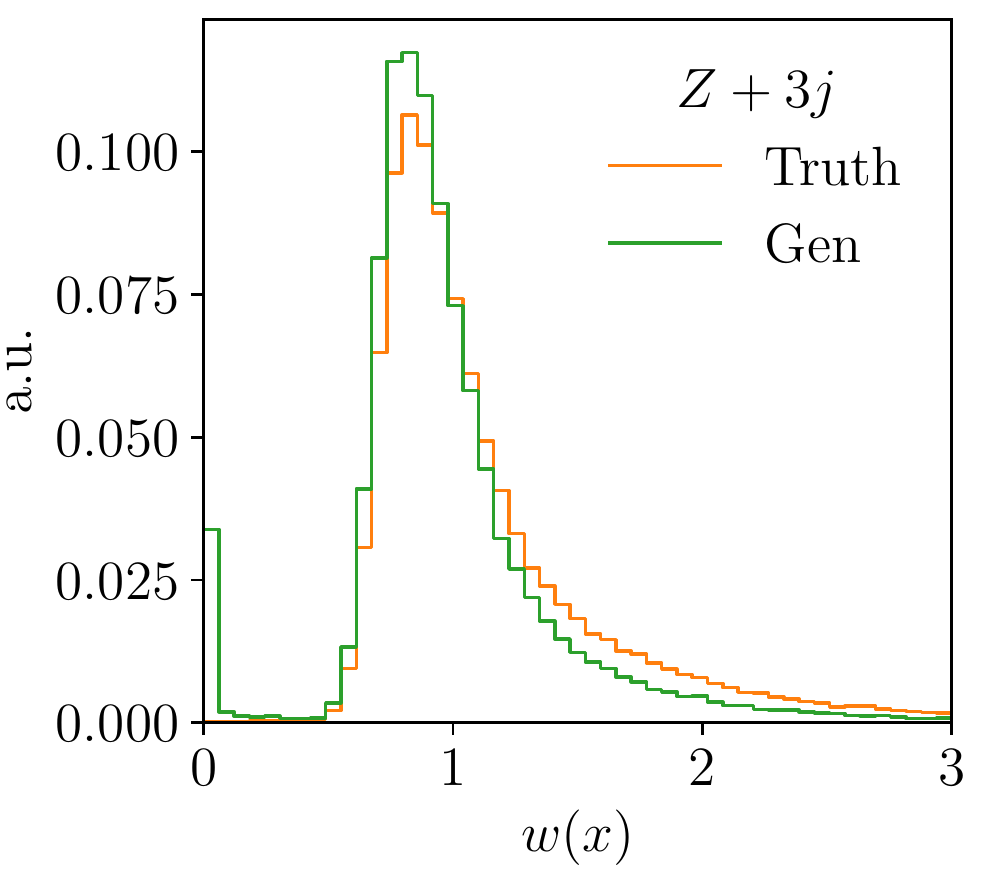}\\
    \includegraphics[width=0.33\textwidth,page=4]{/prec_inn_improved_generator/weights_z1j}
    \includegraphics[width=0.33\textwidth,page=4]{/prec_inn_improved_generator/weights_z2j}
    \includegraphics[width=0.33\textwidth,page=4]{/prec_inn_improved_generator/weights_z3j}\\
    \caption{Left to right: $Z+\{1,2,3\}$~jets using the
      state-of-the-art generator. Top to bottom: ROC curve, weight
      distribution on a linear scale, and weight distribution on a
      logarithmic scale. The weights
      are evaluated separately on the true, training dataset for the generator 
      and the generated dataset.}
    \label{fig:prec_inn_improved_weights}
\end{figure}

\begin{figure}[t]
    \includegraphics[width=0.49\textwidth,page= 4]{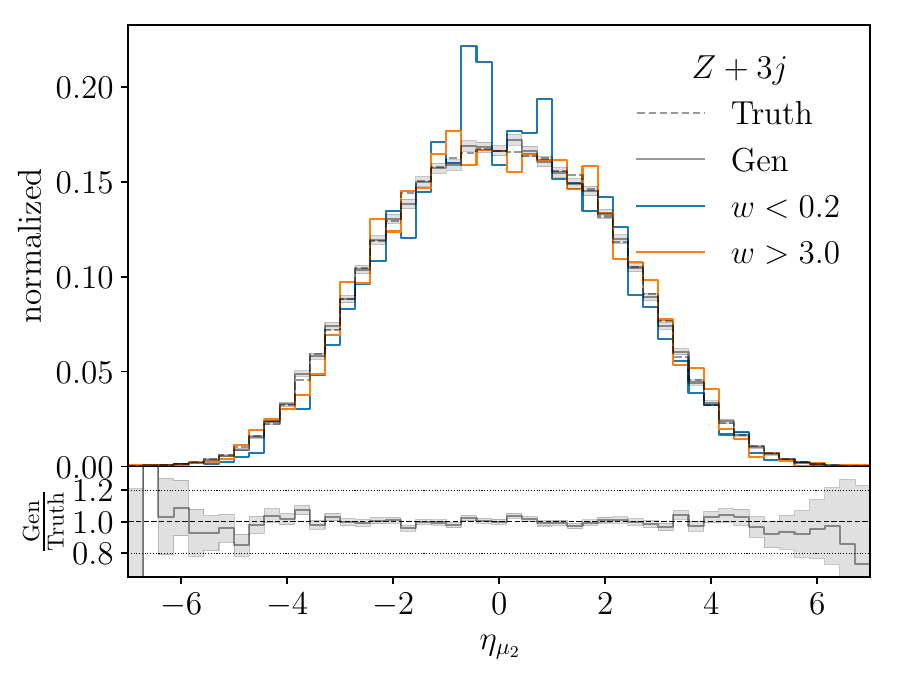}
    \includegraphics[width=0.49\textwidth,page= 5]{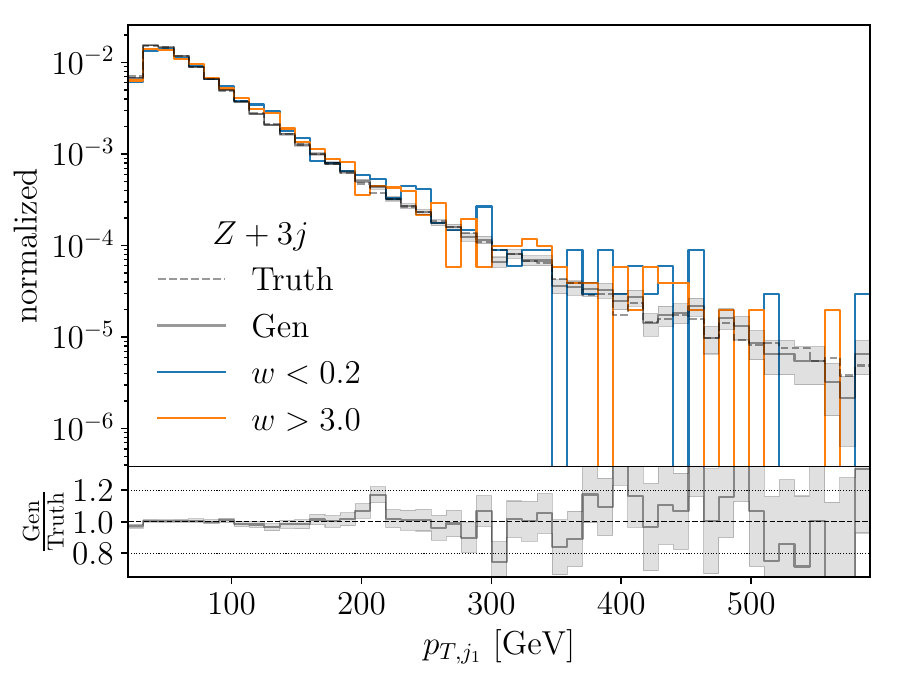}\\
    \includegraphics[width=0.49\textwidth,page=14]{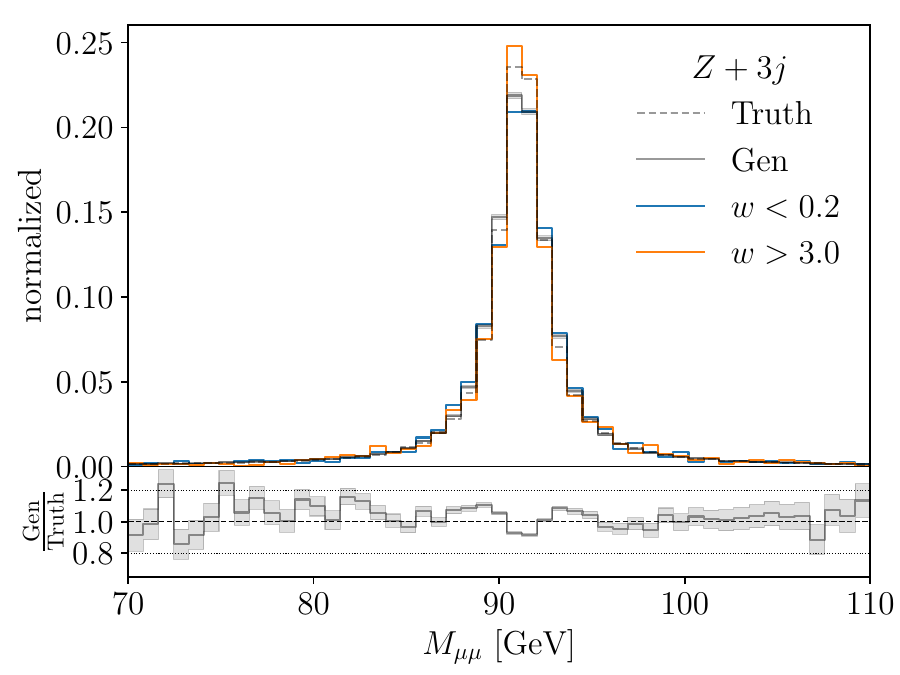}
    \includegraphics[width=0.49\textwidth,page=21]{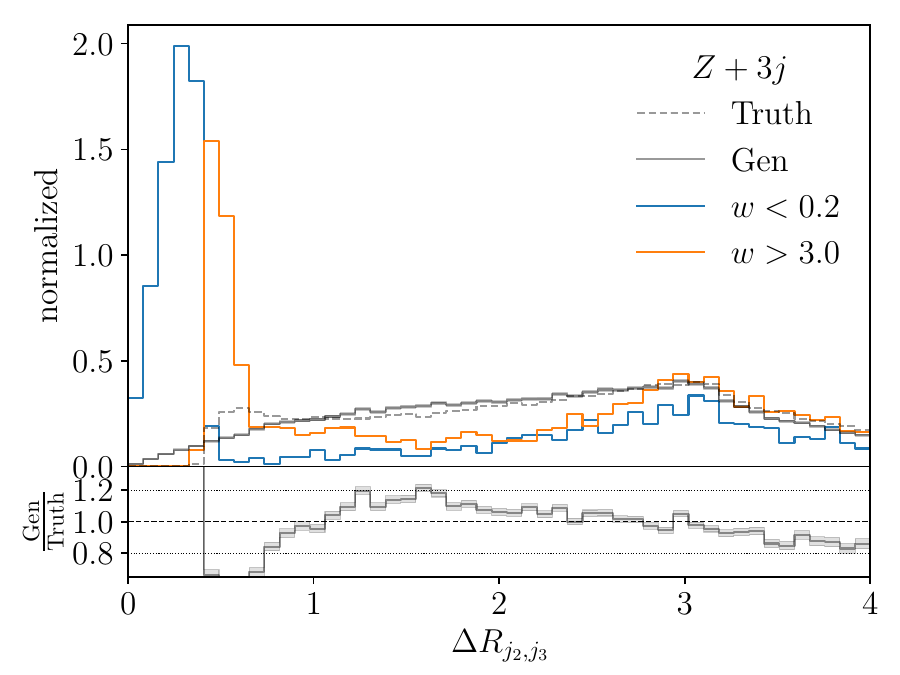}
    \caption{Kinematic distributions for $Z+3$~jets events from the
      state-of-the-art generator in different weight ranges, to see
      if events with large corrections cluster in phase space. The 
      bottom panels show two jet masses, which are not part of the 
      standard requirements testing the $Z$+jets kinematics.  The 
      events with small weights are taken from the generated 
      distribution, the events with large weights are taken from the 
      truth distribution.}
    \label{fig:prec_inn_improved_clustering}
\end{figure}

\subsection{State-of-the-art generator and feature scan}
\label{sec:events_scan}

\begin{figure}[t]
    \includegraphics[width=0.49\textwidth,page=12]{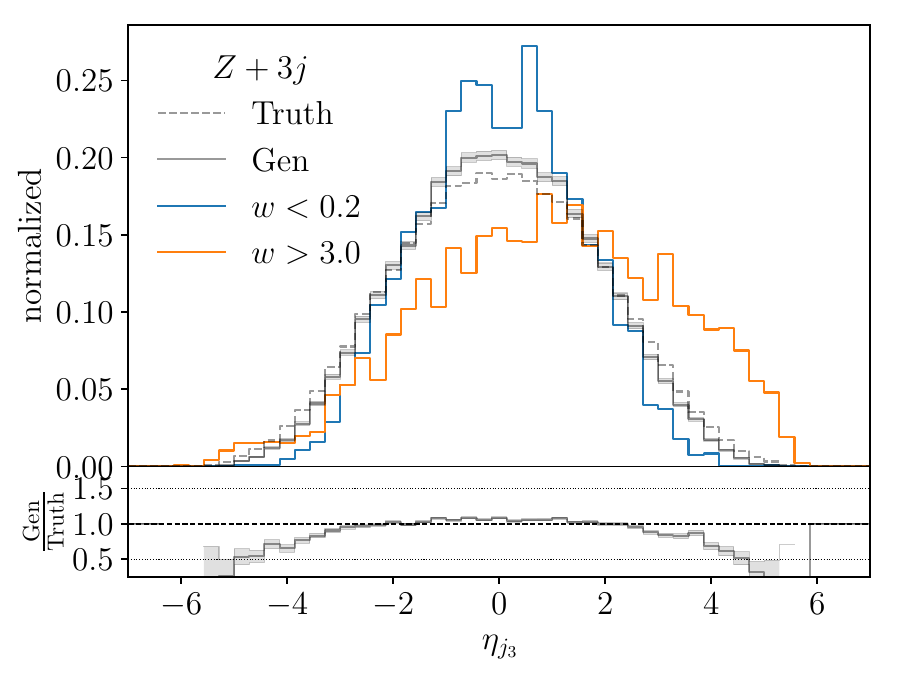}
    \includegraphics[width=0.49\textwidth,page=13]{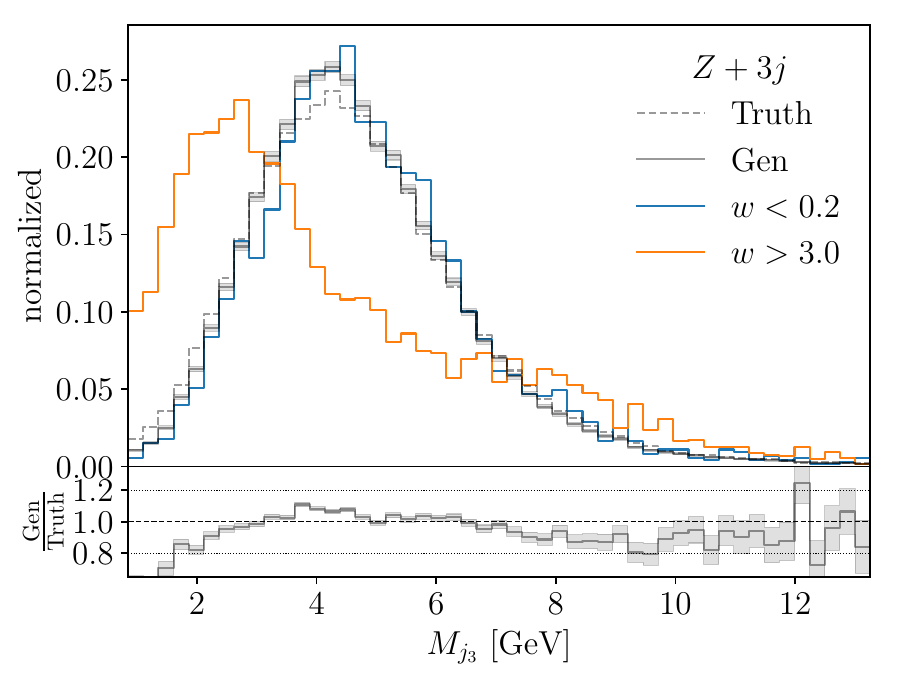}
    \includegraphics[width=0.49\textwidth,page=12]{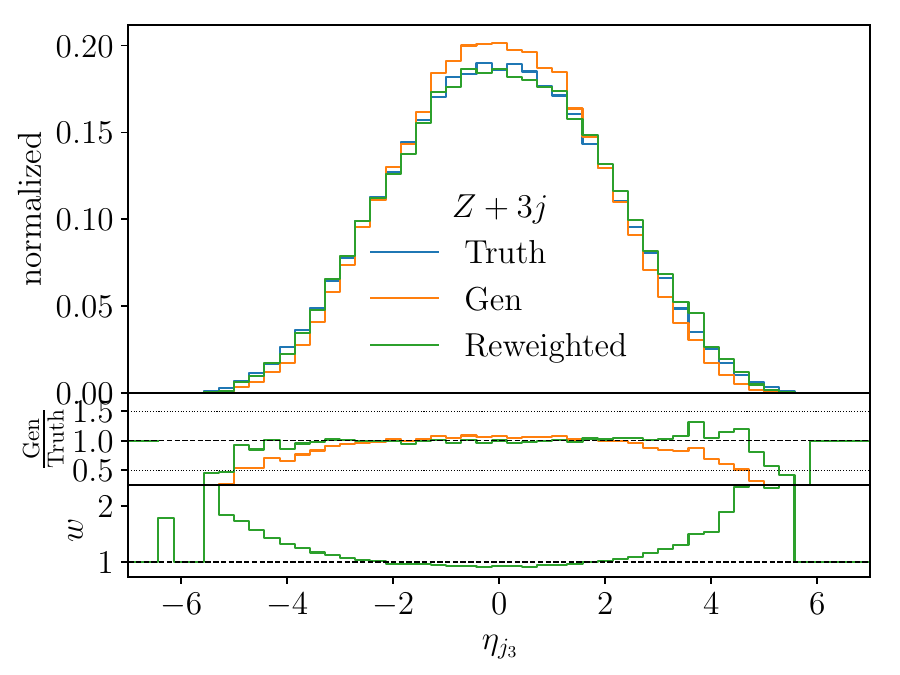}
    \includegraphics[width=0.49\textwidth,page=13]{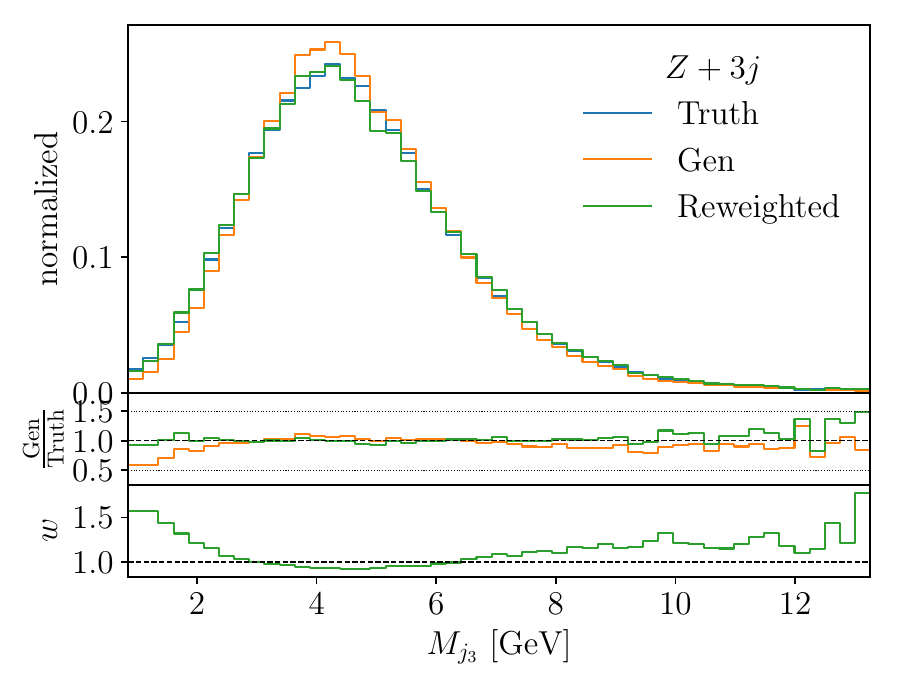}
    \caption{Critical kinematic distributions and for $Z+3$~jets events from the
      state-of-the-art generator in different weight ranges (upper)
      and comparing generated data with truth (lower).  The events 
      with small weights are taken from the generated distribution, the 
      events with large weights are taken from the truth distribution.}
    \label{fig:prec_inn_unknown}
\end{figure}

Next we turn to an improved version of the $Z$+jets event generator, where the $Z$ mass peak is much improved, and the main failure mode shifts elsewhere.  In
Fig.~\ref{fig:prec_inn_improved_weights} we show the same weight
distributions as in Fig.~\ref{fig:prec_inn_weights}, but for the
updated version of the INN event generator and one to three jets.  The
central peaks are much more narrow, and the distributions for one and
two jets are now almost identical. However, we still observe
distinctive tails of the weight distributions. They should be
evaluated on generated events, if we are interested in small weights,
and on training events, if we are interested in large weights. Even
for three jets the maximum of the weight distribution remains at one,
indicating that for the updated generator the mass peak is no longer a
serious problem. On the other hand, the tail towards large weights is
sizeable, indicating that we should look for missing sub-leading
features in the generated event sample.

Consequently, we search for phase space clustering of $Z+3$~jets
events with anomalous weights in
Fig.~\ref{fig:prec_inn_improved_clustering}, similar to
Fig.~\ref{fig:prec_inn_clustering}. We see the effect of small
statistics in the otherwise accurately learned $p_T$-tail, and the
$Z$-mass peak with hardly any reweighting required. The angular
correlations between the jets is the one distribution that is not
described well. While reweighting is not needed to describe the
maximum around $R_{jj} \sim 3$, the collinear enhancement in the range
$R_{jj} = 0.4~...~1.5$ only appears after reweighting with large
weights, while the phase space boundaries for $R_{jj} < 0.4$ requires
very small, potentially zero weights. We can confirm this by looking
at the events in the leftmost bins in the central row  of
Fig.~\ref{fig:prec_inn_improved_weights}. These correspond to
weights $0 < w < 0.06$, and we have confirmed that for two and three 
jets at least 95\% of these events have one $\Delta R_{jj} < 0.4$.

Finally, we can use event weights to identify unknown issues for a
given trained network. In App.~\ref{app:calo} we show a large set of
kinematic $Z$+jets correlations for events in the tails of the weight
distributions. Two kinematic distribution stick out as poorly
described --- the rapidity of the softest jet and its jet mass, both
shown in Fig.~\ref{fig:prec_inn_unknown}. While $\eta_{j_3}$ is part
of the standard set of distributions to check, its jet mass is not
usually used to benchmark this kind of
network~\cite{Butter:2021csz}. However, it becomes important when we
combine event-level with jet-level analysis tools.

In the lower panels of Fig~\ref{fig:prec_inn_unknown} we show the
corresponding distributions, to confirm that the reweighted generated
events reproduce the truth and the classifier output is correct.  The
reason for the poor performance on the third jet is, most likely, the
small size of the training sample. For a standard, deterministic
network the source of such a failure is hard to determine, so we
resort to a Bayesian version of the same network for this purpose.

\subsection{Bayesian generators and pull}
\label{sec:events_bayes}

\begin{figure}[t]
    \includegraphics[width=0.33\textwidth,page=1]{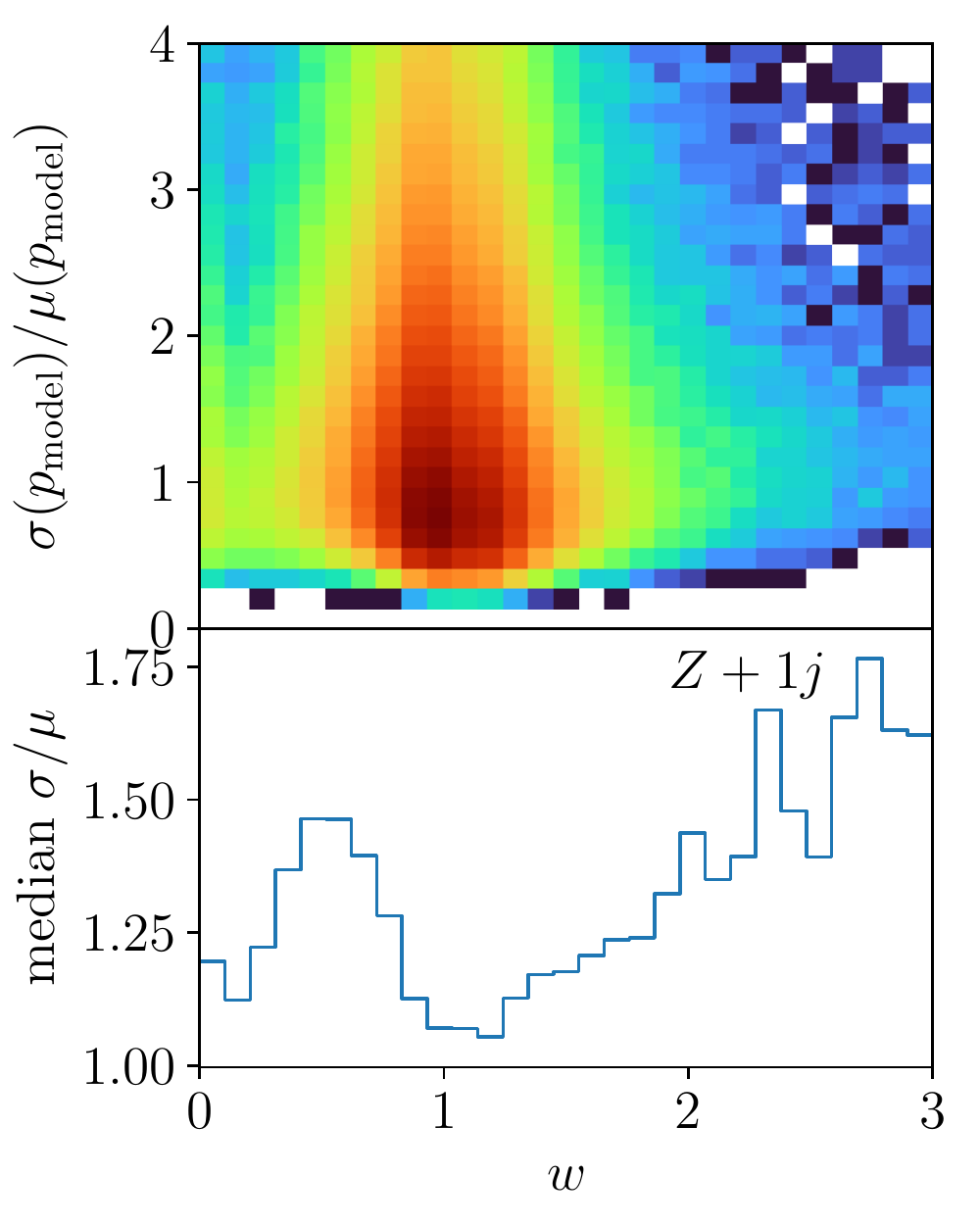}
    \includegraphics[width=0.33\textwidth,page=1]{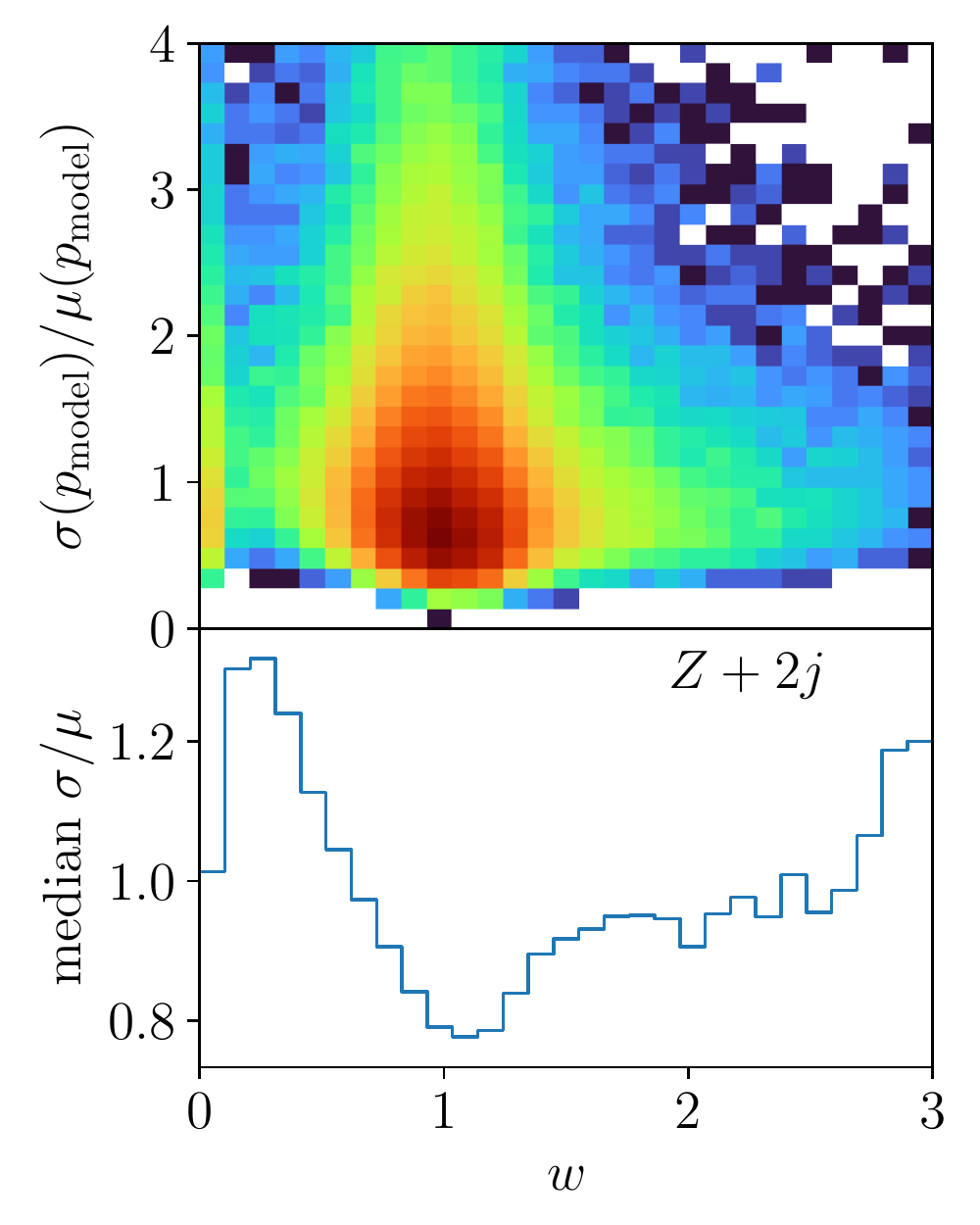}
    \includegraphics[width=0.33\textwidth,page=1]{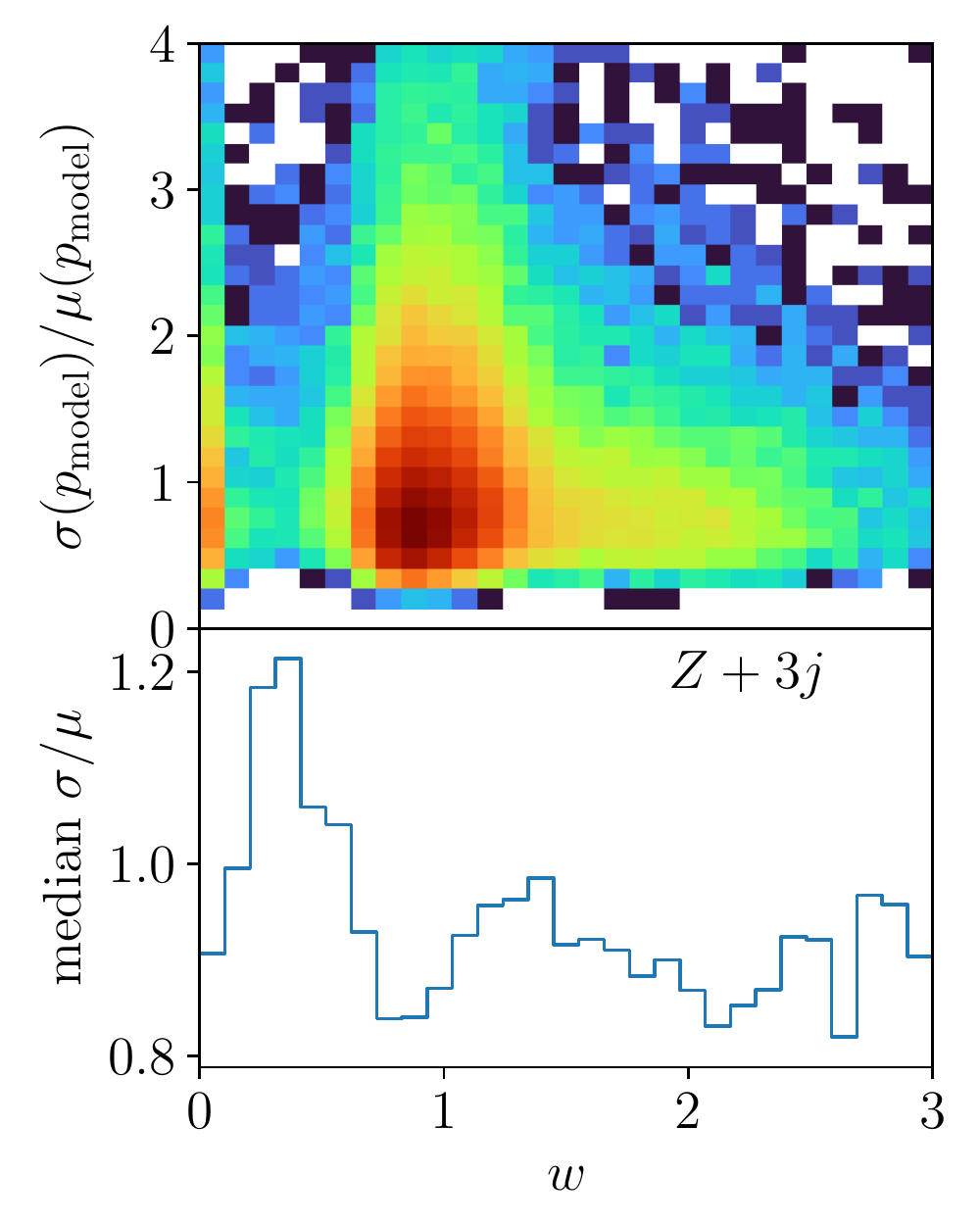}\\
    \includegraphics[width=0.33\textwidth,page=2]{/prec_inn_bayesian_generator/gen_errors_z1j}
    \includegraphics[width=0.33\textwidth,page=2]{/prec_inn_bayesian_generator/gen_errors_z2j}
    \includegraphics[width=0.33\textwidth,page=2]{/prec_inn_bayesian_generator/gen_errors_z3j}
    \caption{Top: correlation between the classifier weights and the
      relative standard deviation of the event weights from the
      Bayesian generator; Center: medians over the $w$-bins. Bottom:
      pulls combining the standard deviation of the event weight
      distribution with the error estimate from the Bayesian
      generator.}
    \label{fig:prec_inn_bayesian}
\end{figure}

\begin{figure}[t!]
    \includegraphics[width=0.49\textwidth,page= 4]{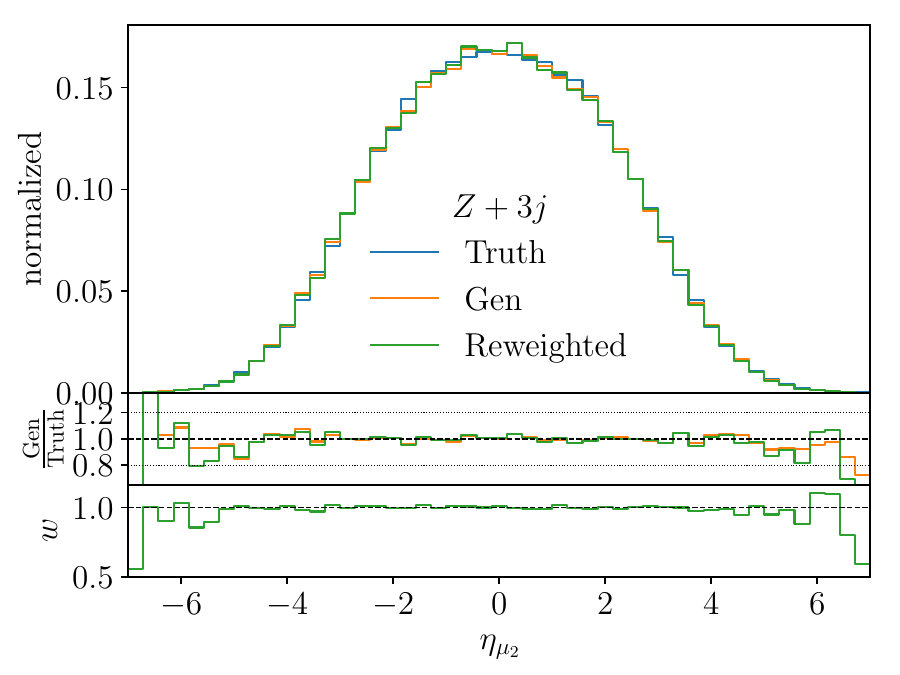}
    \includegraphics[width=0.49\textwidth,page= 5]{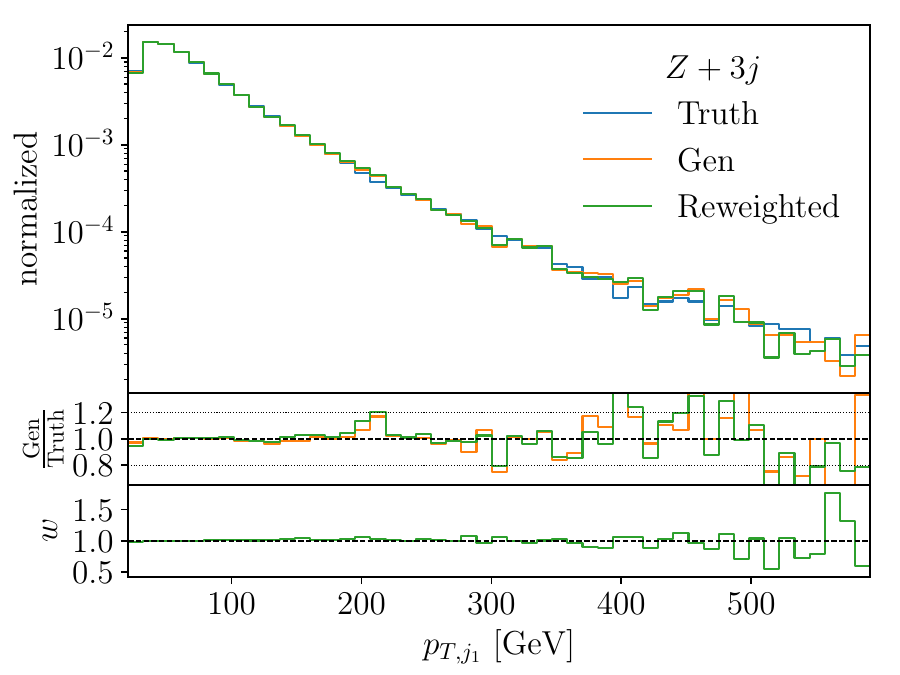} \\
    \includegraphics[width=0.49\textwidth,page=14]{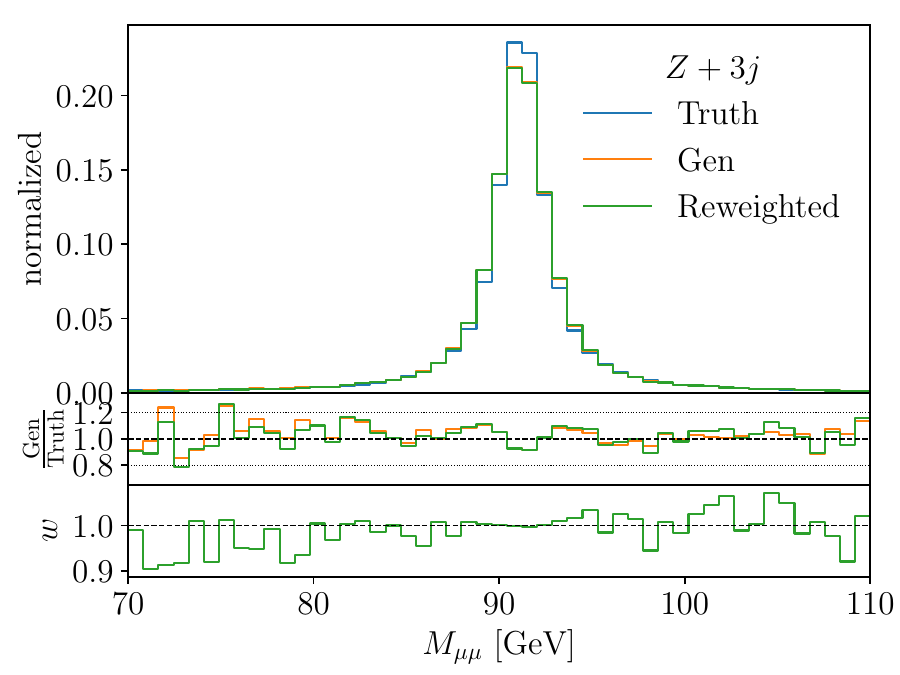}
    \includegraphics[width=0.49\textwidth,page=21]{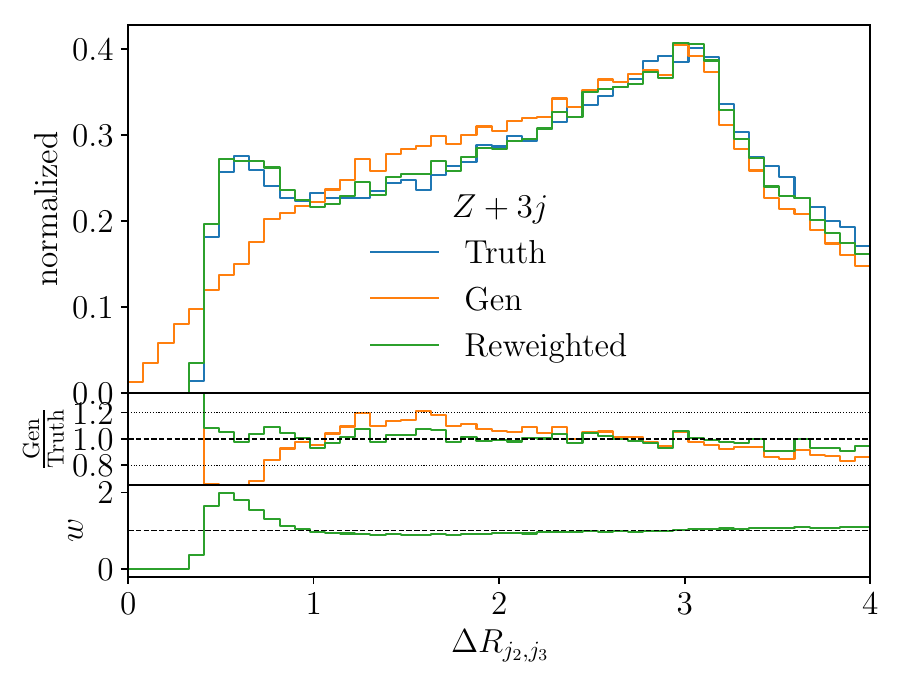} \\
    \includegraphics[width=0.49\textwidth,page=12]{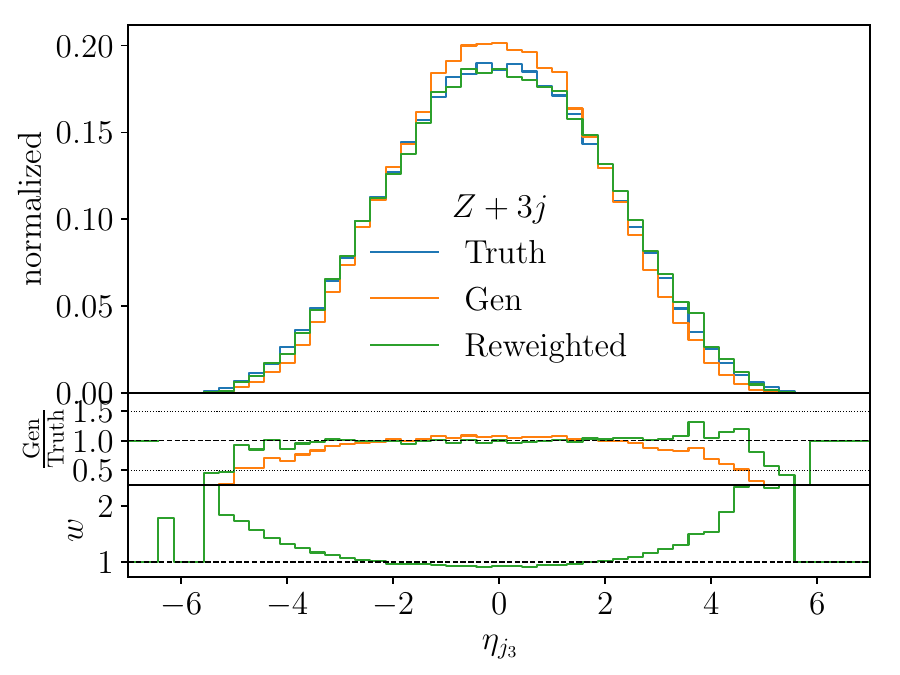}
    \includegraphics[width=0.49\textwidth,page=13]{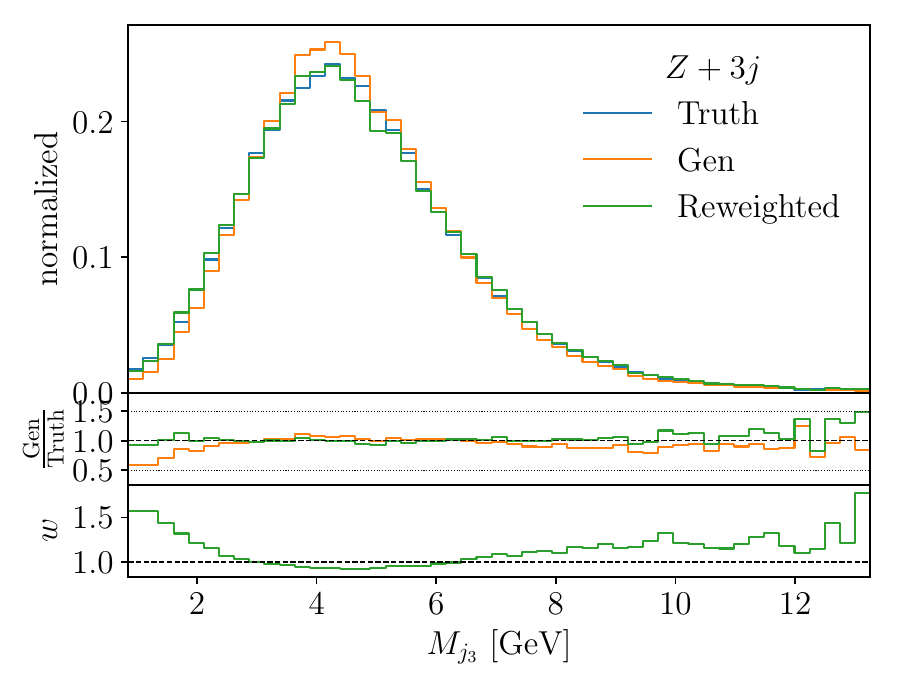}
    \caption{Kinematic distributions for $Z+3$~jets events from the
      Bayesian state-of-the-art generator, with bin-wise error bars
      on the generated events. The distribution include the set of 
      Fig.~\ref{fig:prec_inn_improved_clustering} as well as the 
      challenging distributions from Fig.~\ref{fig:prec_inn_unknown}.}
   \label{fig:prec_inn_bayesian_observables}
\end{figure}

While weight distributions encode a wealth of information about
the generative model, we do not know from first principles what shape
to expect for well-trained networks.  A way out is to supplement it by
pull distributions, introduced in Eq.\eqref{eq:pulls}, which should
approach a standard Gaussian.  Deterministic generative networks do
not provide us with the necessary information, but a Bayesian
generative network returns a density as well as an uncertainty
estimate on this density~\cite{Bellagente:2021yyh,Butter:2021csz}. We
can then define the mixed ratio
\begin{align}
  t(x_i) 
  = \frac{\mu(x_i) [ 1 - w(x_i)]}{\sigma(x_i)} \; ,
\label{eq:pulls2}
\end{align}
where the mean of the estimated density $\mu(x_i)$ and its uncertainty
$\sigma(x_i)$ are provided by the Bayesian generator, and $w(x_i)$ is
the classifier output.

To extract an error on the likelihood for a specific event, we fix the
network weights to the maximum of their posterior distribution and
generate a dataset.  Next, we use the network as a density estimator
and extract a distribution of likelihoods for each event by sampling
from the network weight distribution. The width of this distribution
should give an estimate of the uncertainty.  However, we have to be
careful in this interpretation, as we cannot treat the event-wise
likelihoods as uncorrelated.

In Fig.~\ref{fig:prec_inn_bayesian} we first look at the correlation
between the classifier weight $w(x)$ and the relative error
$\sigma(\pmd)/\mu(\pmd)$ and the median of the Bayesian INN estimate as a
function of $w$. For events with one and two jets there is a clear
correlation, while for $Z+3$~jet events missing features start to
dominate the classifier, and the two uncertainty estimates lose their
correlation. In the lower panels of Fig.~\ref{fig:prec_inn_bayesian}
we show the pull distributions, normalizing the deviation of the
generated from the true density (encoded in the classifier) by the
uncertainty from the generative network. While we obtain a roughly
Gaussian shape, its width is much smaller than we would expect. The
reason for this is the problem of assigning an uncertainty to
individual phase space points without taking their correlation into
account.

We can understand the conservative uncertainty estimate of the
Bayesian network from the kinematic observables.  We use the same
distribution of likelihoods as for reweighted distributions.  Turning
each distribution into a histogram taking the bin-wise means and
standard deviations, we can also define an error bar for each
histogram bin. In Fig.~\ref{fig:prec_inn_bayesian_observables} we
first show the same four distributions as in
Fig.~\ref{fig:prec_inn_improved_clustering}, but now with a Bayesian
network uncertainty.  For the smooth rapidity and momentum
distributions the event reweighting only has a minor effect,
corresponding to the observation that events with anomalous weights do
not cluster in these distributions.  The BNN uncertainty estimate is
over-conservative in that it easily covers the deviation of the model
from the truth and also the effect of event reweighting.

The situation changes for the $Z$-peak, where the network does well,
the reweighting does not lead to a significant improvement of the
sharp mass peak, but the uncertainty estimate there is too small. For
$\Delta R_{jj}$ we see what happens if the (Bayesian) generative
network ignores a feature altogether --- in this case the missing
collinear enhancement is not accounted for in density estimation and
also not in the uncertainty estimation for the density. This suggests
that the implicit bias of the generative networks does not allow it to
capture the structure. On the other hand, the classifier identifies
this failure mode, and the reweighted distribution reproduces the
truth with high precision.

Finally, in the lower two panels of
Fig.~\ref{fig:prec_inn_bayesian_observables} we show a Bayesian
uncertainty estimate for the challenging cases from
Fig.~\ref{fig:prec_inn_unknown}. We already know that the classifier
identifies the problematic phase space region correctly, and the
reweighted events reproduce the truth distribution. The question is
where this problem comes from.  The Bayesian network output tells us
if this problem is related to a lack of training data or to the
network structure. We see that similar to the first two distributions
the Bayesian uncertainty estimate easily covers the difference between
generated events and truth, as well as the difference between
generated and reweighted events. This clearly points towards a
limitation in the training data, most likely just the size of the
3-jet dataset. As a side remark,
Fig.~\ref{fig:prec_inn_bayesian_observables} would not even have
flagged these two distributions as problematic, this potentially
crucial piece of information requires a dedicated study of events with
anomalous weights.

\section{Conclusions}

Generative networks play an important role in the ML-transformation of 
LHC physics. They can be used for many tasks in event generation, simulation, 
and advanced analysis. This comes with the requirement to control their 
precision in the density estimation over phase space systematically. 
A classifier is perfectly suited to control generative models, as 
motivated by established classifier reweighting. In addition to a 
single AUC value
it provides us with a wealth of information on the strengths, 
weaknesses, and failure modes of the generative model.

We have applied a performance test based on the classifier weight distributions for three different generative tasks\footnote[4]{We collected the datasets used to train the classifiers in a Github repository which is publicly available at \href{https://github.com/heidelberg-hepml/discriminator-metric}{https://github.com/heidelberg-hepml/discriminator-metric.}}. First, we have studied a not very realistic, but challenging modification of generated jet configurations, to find that these modifications can be identified and even corrected for by looking at jets with anomalous weights. Our second case were calorimeter showers, where the weight distributions identified types of showers that the generative model did not learn well, pointing us towards possible improvements of the generative setup. Finally, we have looked at an event generator for $Z$+jets events, for which the classifier weights again allow us to identify and understand the problems of the generative network training. For this case we also showed how our diagnostic can be embedded in a comprehensive precision and uncertainty framework for generative events.

Some standard failure modes appearing in our three applications and diagnosed by the weight distributions are: (i) missing features or missing tails in the generated events, leading to a tail of large weights $w \gg 1$ clustered in phase space; (ii) wrongly learned phase space boundaries or sharp cliffs, leading to a tail towards small weights $w \ll 1$, clustered in phase space; (iii) sharp features learned with reduced resolution, leading to a shift of peak of the weight distribution to values $w<1$ and a compensating enhancement at finite $w>1$, related to the amount of missing resolution and also clustered in phase space. 

The clustering of anomalous weights in the interpretable phase space has, in all cases, allowed us to identify the physics reason behind the poorly performing generative network. Moreover, reweighting the events with the classifier weights over phase space allows us to improve the network and make sure that the weighted events do reproduce all key features. 

Our study shows that a trained classifier can and should be used to analyze the performance of generative networks; the weight distributions not only tests the performance of the generator, it also allows us to identify failure modes, correct for shortcomings, and defines a key ingredient to the development of precision generators for particle physics.

\subsection*{Acknowledgements}

We would like to thank Anja Butter and Ramon Winterhalder for many useful discussions and for the close coordination with Ref.~\cite{Nachman:2023clf}. 
CK and TP would like to thank the Baden-W\"urttemberg-Stiftung for financing through the program \textsl{Internationale Spitzenforschung}, pro\-ject \textsl{Uncertainties – Teaching AI its Limits} (BWST\_IF2020-010). RD and DS are supported by the U.S.~Department of Energy under Award Number DOE-SC0010008.  
TH is funded by the Carl-Zeiss-Stiftung through the project \textsl{Model-Based AI: Physical Models and Deep Learning for Imaging and Cancer Treatment}. 
This research is supported by the Deutsche Forschungsgemeinschaft (DFG, German Research Foundation) under grant 396021762 -- TRR~257: \textsl{Particle Physics Phenomenology after the Higgs Discovery} and through Germany's Excellence Strategy EXC~2181/1 -- 390900948 (the \textsl{Heidelberg STRUCTURES Excellence Cluster}). 

\appendix

\clearpage
\section{Classifier calibration}
\label{app:calibration}

To gauge whether the classifiers used in our study have been well-trained (not overfitted, reasonably close to optimal), one important check is to inspect their calibration curves. The idea of the calibration curve is that a properly learned and optimal classifier $C(x)$ should return the probability that $x$ is class~1, and $1-C(x)$ the probability that $x$ is class~0. Therefore, if we took all events $x$ in the training data (assumed to be balanced) for which $C(x)=C$, a fraction $C$ of them should be class~1. The differential way to write this is
\begin{align}
\frac{\dfrac{\mathrm{d}N_1}{dC}} {\dfrac{\mathrm{d}N_1}{dC} + \dfrac{\mathrm{d}N_0}{\mathrm{d}C}} = C \; .
\label{eq:base_cal}
\end{align}
As in the main body of this paper, we will look at calibration curves in terms of the weights $w$. 
Using Eq.\eqref{eq:wformula}, we can turn Eq.\eqref{eq:base_cal} into a statement about the weights,
\begin{align}\label{eq:weights_cal}
\frac{\mathrm{d}N_1}{\mathrm{d}w}=\frac{\mathrm{d}N_0}{\mathrm{d}w}\,w \; .
\end{align}

\begin{figure}[h!]
    \includegraphics[width=0.33\linewidth]
    {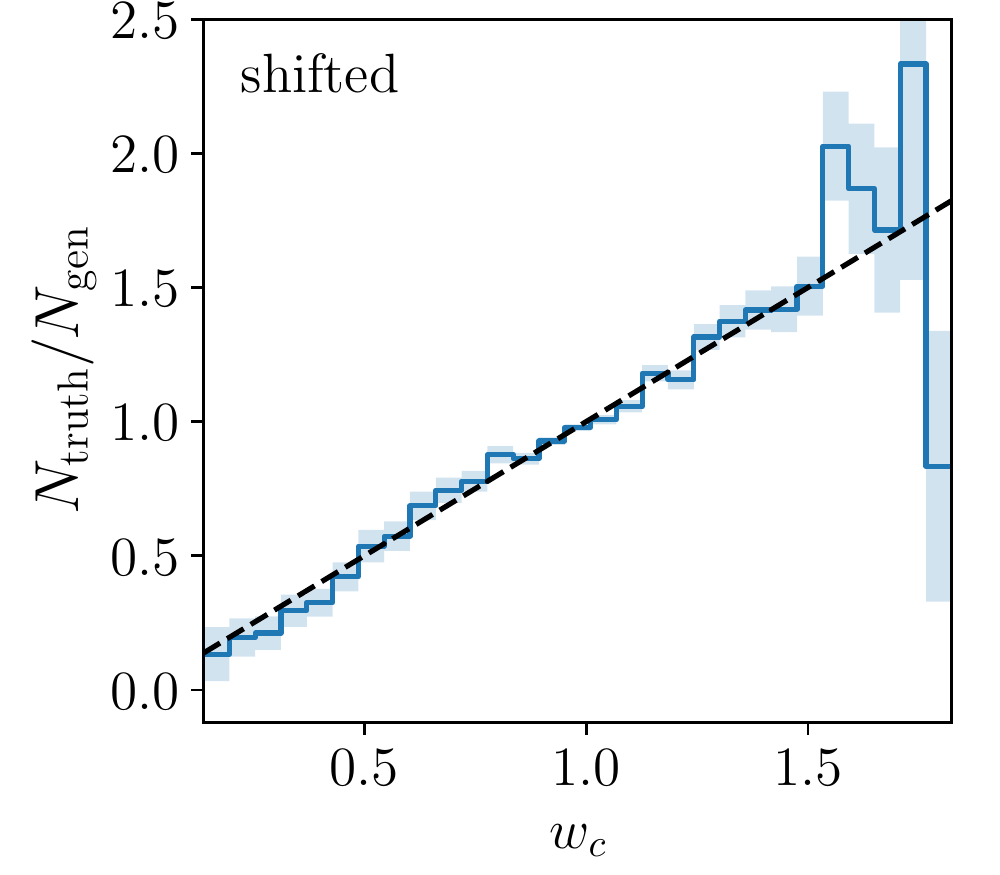}
    \includegraphics[width=0.33\linewidth]
    {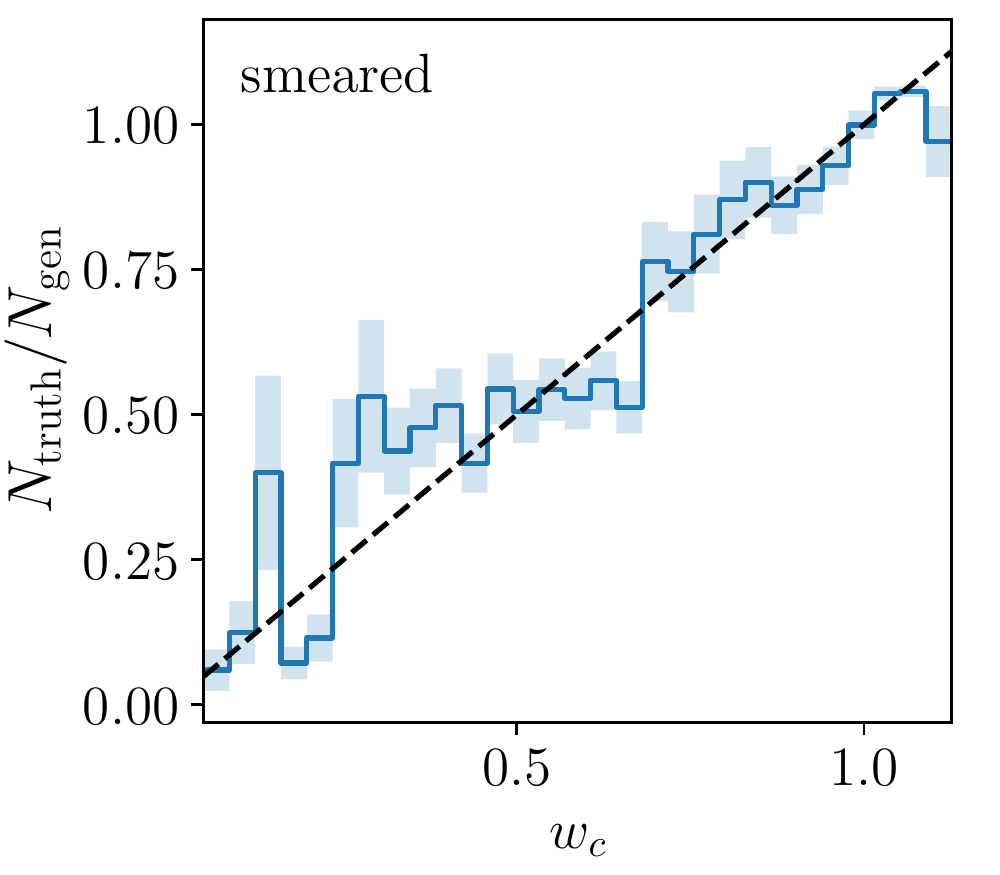}
    \includegraphics[width=0.33\linewidth]
    {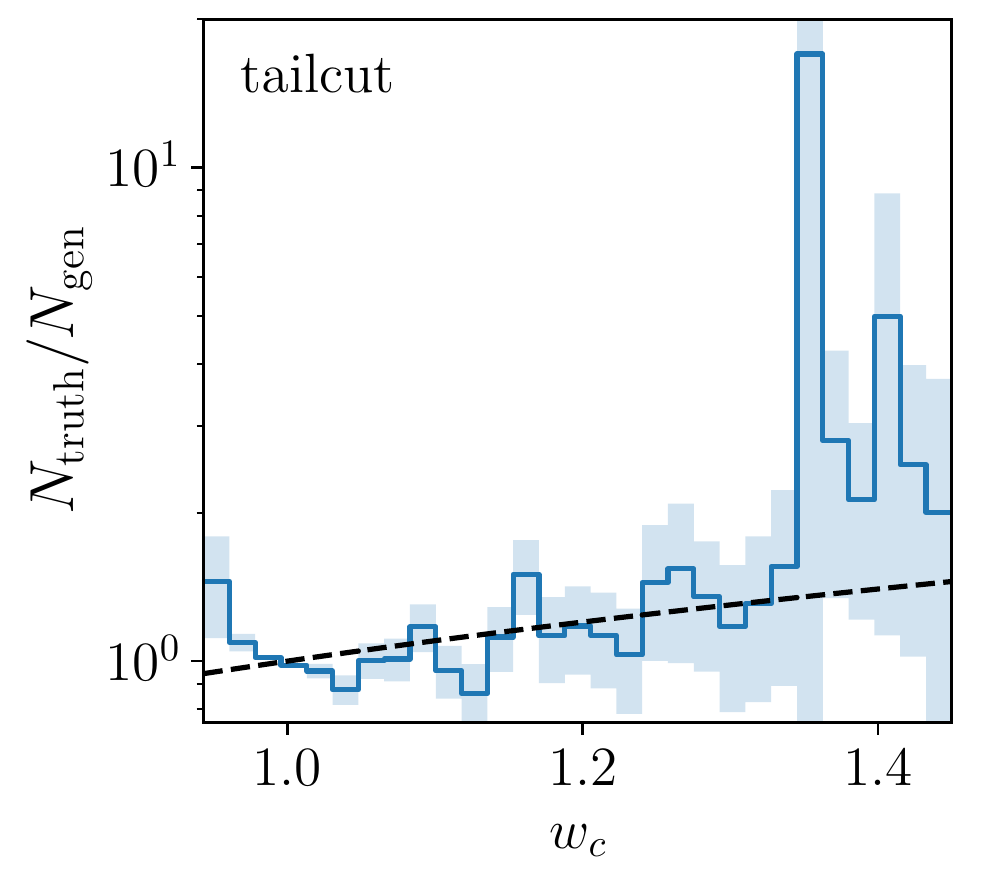}\\
    \includegraphics[width=0.33\linewidth, page=3]
    {figs/caloinn_plots/calibration_plots/cal_curve_unnorm_e}
    \includegraphics[width=0.33\linewidth, page=3]
    {figs/caloinn_plots/calibration_plots/cal_curve_unnorm_g}
    \includegraphics[width=0.33\linewidth, page=3]
    {figs/caloinn_plots/calibration_plots/cal_curve_unnorm_pi}\\
    \includegraphics[width=0.33\textwidth,page=3]
    {/prec_inn_improved_generator/calibration_z1j}
    \includegraphics[width=0.33\textwidth,page=3]
    {/prec_inn_improved_generator/calibration_z2j}
    \includegraphics[width=0.33\textwidth,page=3]
    {/prec_inn_improved_generator/calibration_z3j}\\
    \caption{
        Calibration plots in weight space for the different discriminator
        models. Top to bottom: (i) jets with shifted, smeared and tailcut
        distortions; (ii) normalized showers for $e^+$, $\gamma$, and $\pi^+$;
        (iii) $Z+\{1,2,3\}$~jets using the state-of-the-art generator.
    }
    \label{fig:calibration}
\end{figure}

\begin{figure}[h!]
    \includegraphics[width=0.325\textwidth,page= 1]{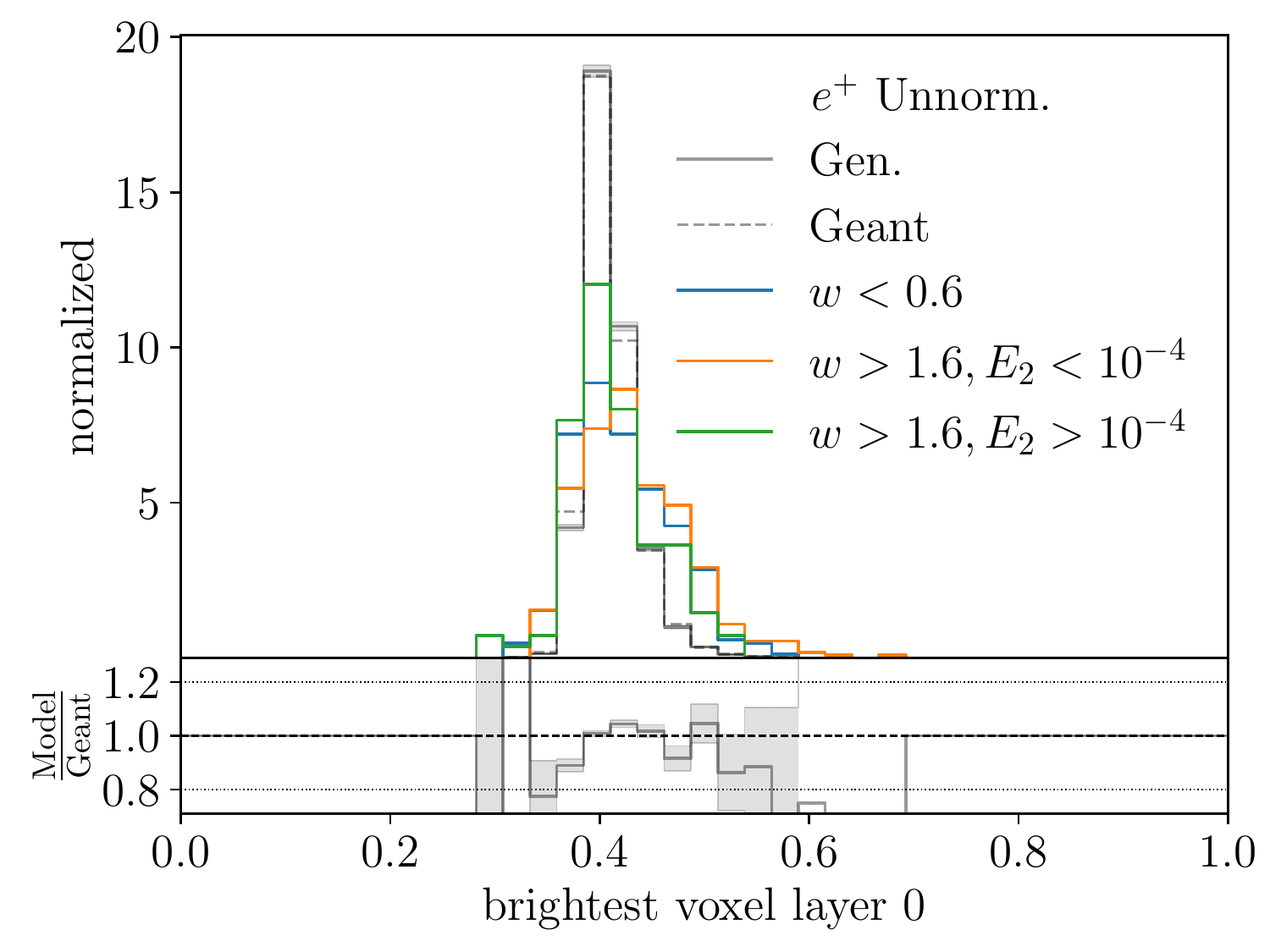}
    \includegraphics[width=0.325\textwidth,page= 2]{figs/caloinn_plots/clustering_unnorm_e.pdf}
    \includegraphics[width=0.325\textwidth,page= 3]{figs/caloinn_plots/clustering_unnorm_e.pdf}
    \includegraphics[width=0.325\textwidth,page= 4]{figs/caloinn_plots/clustering_unnorm_e.pdf}
    \includegraphics[width=0.325\textwidth,page= 5]{figs/caloinn_plots/clustering_unnorm_e.pdf}
    \includegraphics[width=0.325\textwidth,page= 6]{figs/caloinn_plots/clustering_unnorm_e.pdf}
    \includegraphics[width=0.325\textwidth,page= 14]{figs/caloinn_plots/clustering_unnorm_e.pdf}
    \includegraphics[width=0.325\textwidth,page= 15]{figs/caloinn_plots/clustering_unnorm_e.pdf}
    \includegraphics[width=0.325\textwidth,page= 16]
    {figs/caloinn_plots/clustering_unnorm_e.pdf}
\caption{Clustering plots for $e^{+}$: similar pattern of $\gamma$ showers, expected given the similar physics and data structure.}
\label{fig:obs_eplus}
\end{figure}

Equation~\eqref{eq:weights_cal} implies an equivalent way of plotting a calibration curve in
weight space: divide the combined weight distribution in bins and calculate the
ratio $N_\text{truth}/N_\text{gen}$ for each bin. According to Eq.\eqref{eq:weights_cal}, for a well-calibrated classifier these should agree.
We show calibration curves, calculated following this method, for our 
classifiers in Fig.~\ref{fig:calibration}. We see that the classifiers are for the most part very well-calibrated. One possible exception is for $e^+$ at lower weights, but one should keep in mind this is one of the better generative models considered in this work (AUC=0.536), so nearly all the events are in the well-calibrated part of the calibration curve (with $w\approx 1$). Also, as we see in the discussion in Sec.~\ref{sec:calo_cluster} and in Fig.~\ref{fig:obs_eplus}, even if the tails of the classifier are mis-calibrated, it can still extract poorly modeled regions
of phase space and assign, if anything, too extreme weights to them. However, attention is needed when using them for reweighting.

\begin{figure}[b!]
    \includegraphics[width=0.33\textwidth,page=1]{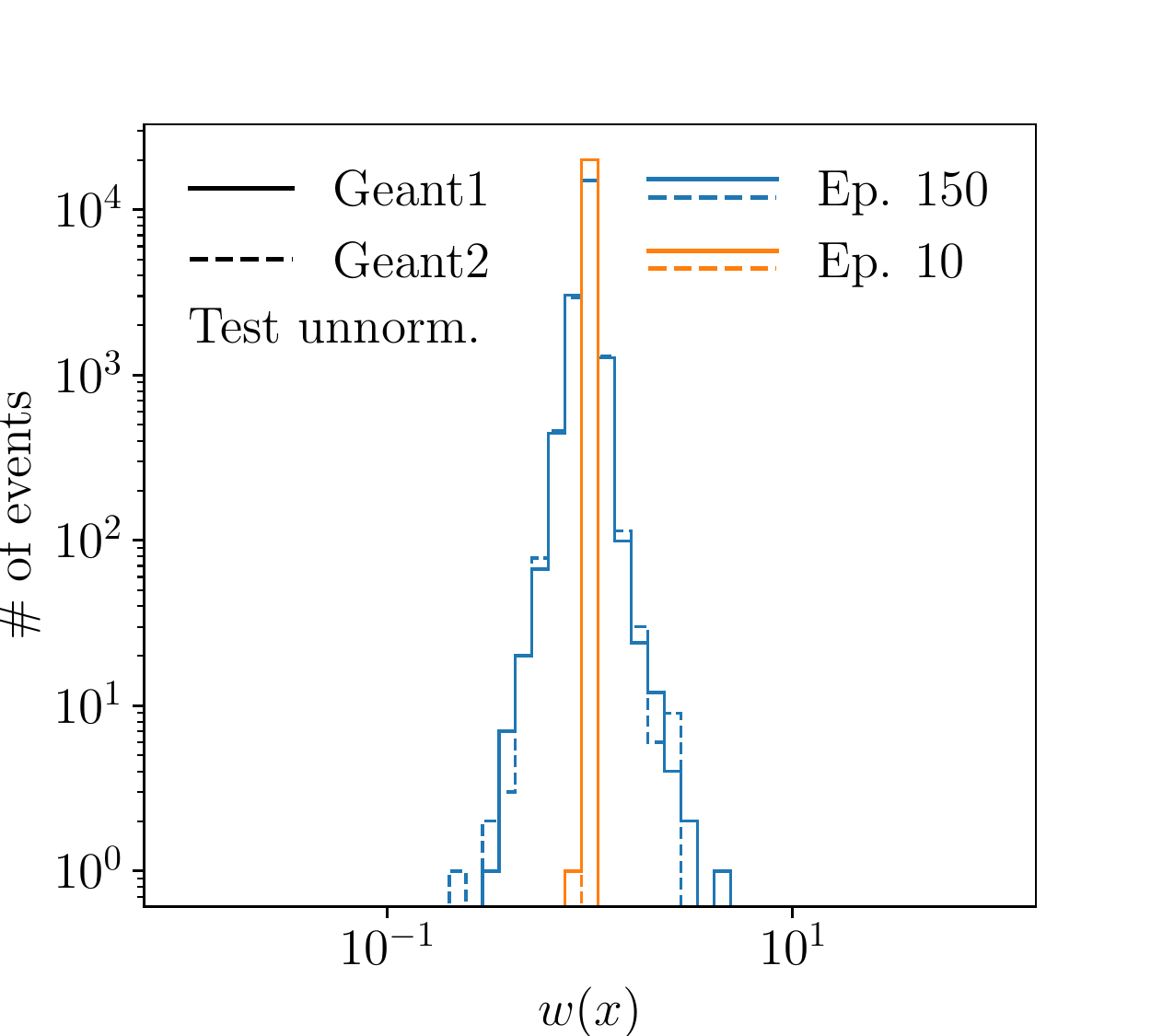}
    \includegraphics[width=0.33\textwidth,page=1]
    {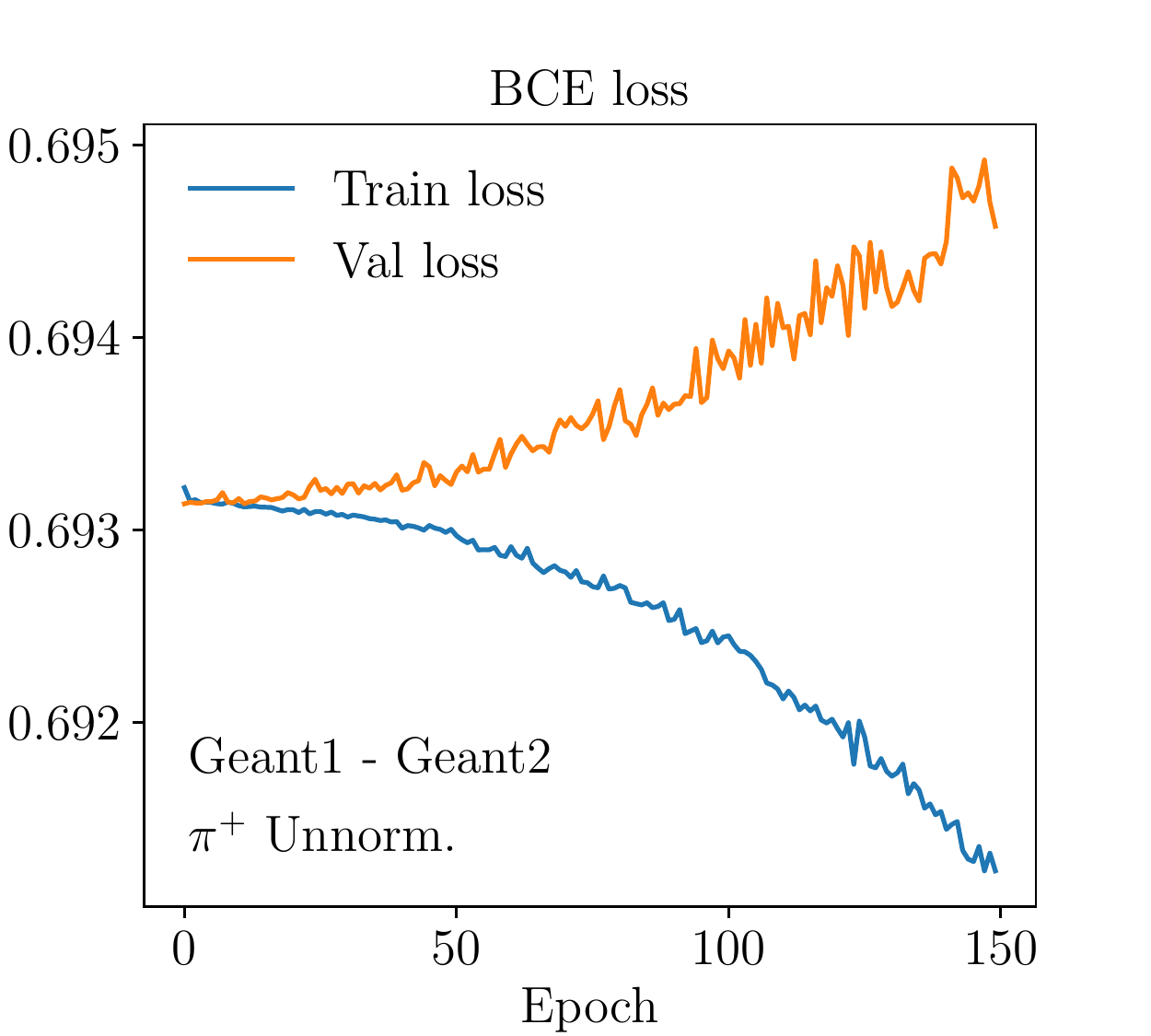}
    \includegraphics[width=0.31\textwidth,page=3]{figs/caloinn_plots/calibration_plots/cal_curve_unnorm_gg.pdf}
    \caption{Left: weight distributions of test
      set for a classifier trained on two different \textsc{Geant} samples for
      pion showers. Center: BCE loss function for training and validation. Right: calibration curve in weight space at epoch 150.}
    \label{fig:calo_gg}
\end{figure}

To confirm that our calibration curves in Fig.~\ref{fig:calibration} are indeed reasonably well-calibrated, we consider the case of an {\it overfitted} classifier obtained by training on two statistically identical \textsc{Geant} samples. In Fig.~\ref{fig:calo_gg} (left) we show the weight distribution for the test dataset after epochs 10 and 150. The middle panel
shows the training and validation losses when training without
learning rate scheduler. The right panel shows the calibration curve for epoch 150. The network learns to distinguish between the
two \textsc{Geant} samples and reweights one noise into the other.  This
guarantees that the weight distribution is symmetric around the
maximum at $w=1$.  However, the weight distributions broaden during
training.  Because the difference between the two datasets is
feature-less, this broadening is the same on the classifier training
and test datasets.
At the same time, Fig.~\ref{fig:calo_gg} and
Eq.\eqref{eq:auc_from_weights} illustrate the benefit of studying the
weight distributions: the AUC evaluated on the test dataset is stable
at 0.5 during the entire training, but the weight distribution shows
that the classifier is moving away from optimality. All in all, we see how all three diagnostics --- weight distribution, training/validation losses, and calibration curve --- indicate a poorly trained classifier, in stark contrast to the classifiers considered in this paper.

\clearpage
\section{Additional kinematic distributions}
\label{app:calo}

\begin{figure}[h!]
    \includegraphics[width=0.325\textwidth,page= 1]{figs/caloinn_plots/clustering_unnorm_g.pdf} 
    \includegraphics[width=0.325\textwidth,page= 2]{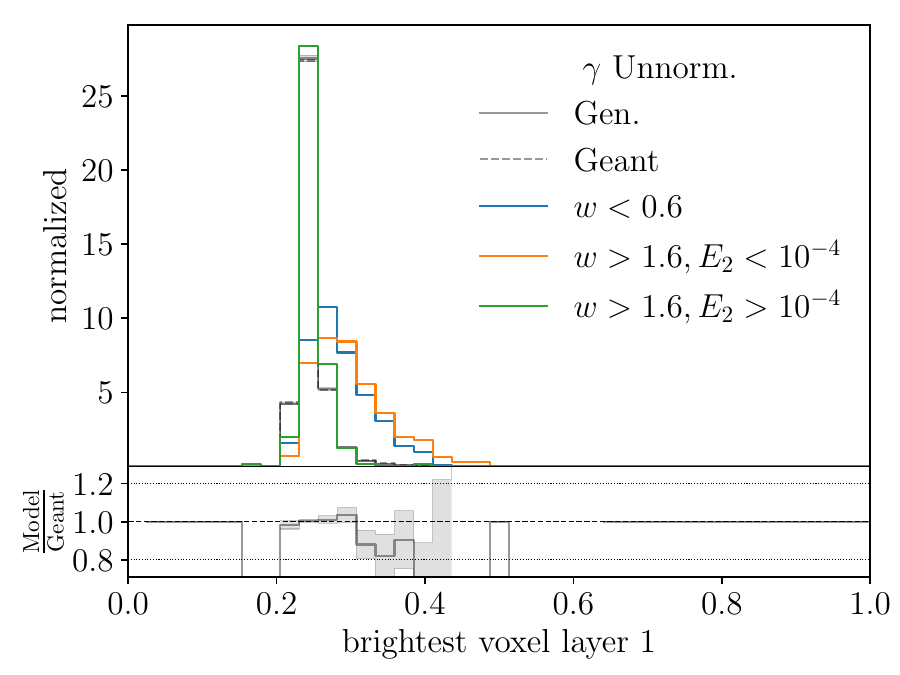}
    \includegraphics[width=0.325\textwidth,page= 3]{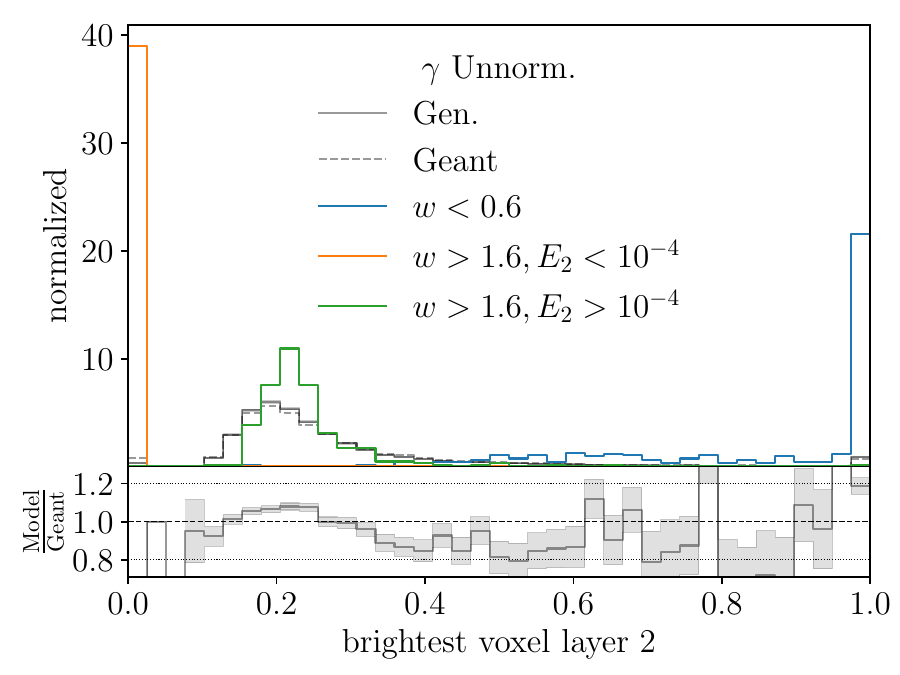}
    \includegraphics[width=0.325\textwidth,page= 4]{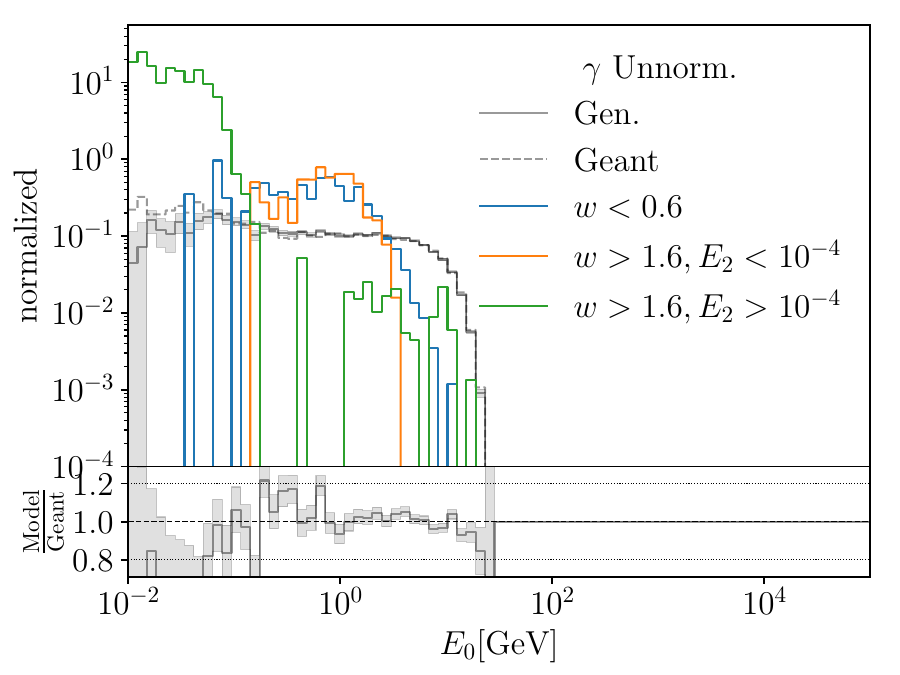}
    \includegraphics[width=0.325\textwidth,page= 5]{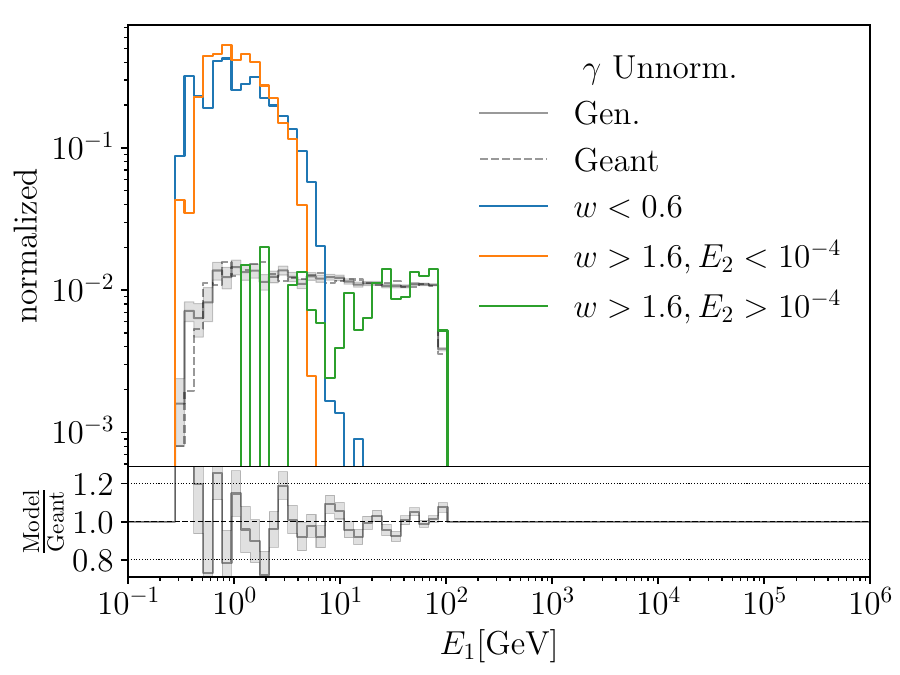}
    \includegraphics[width=0.325\textwidth,page= 6]{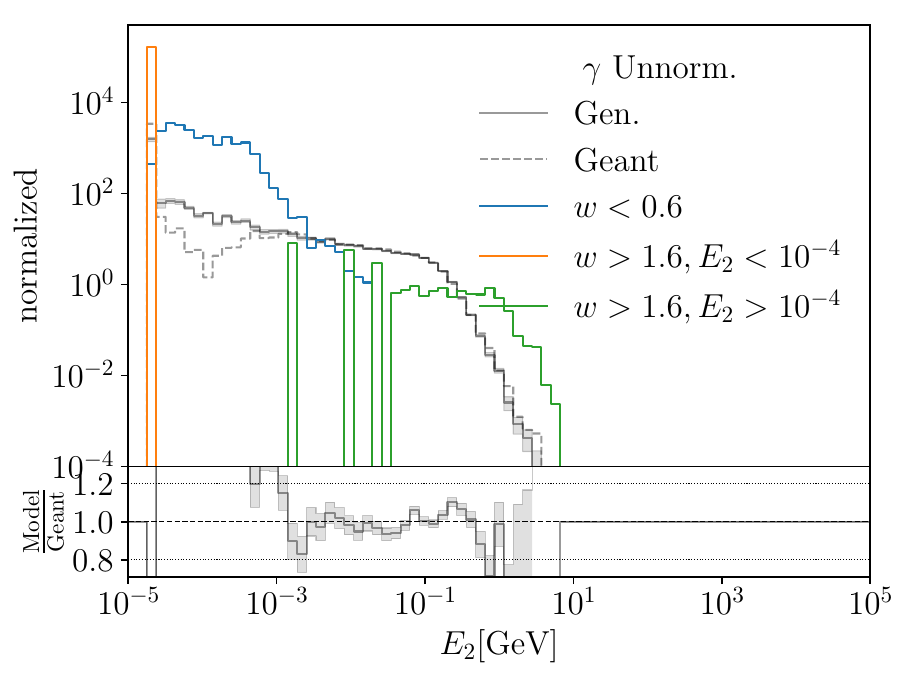}
    \includegraphics[width=0.325\textwidth,page= 14]{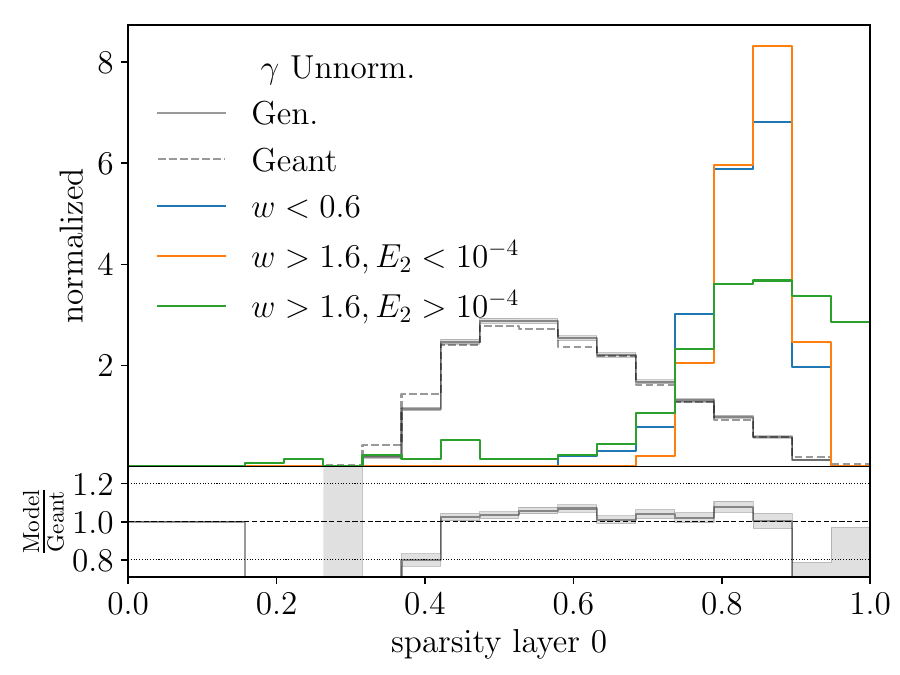} 
    \includegraphics[width=0.325\textwidth,page= 15]{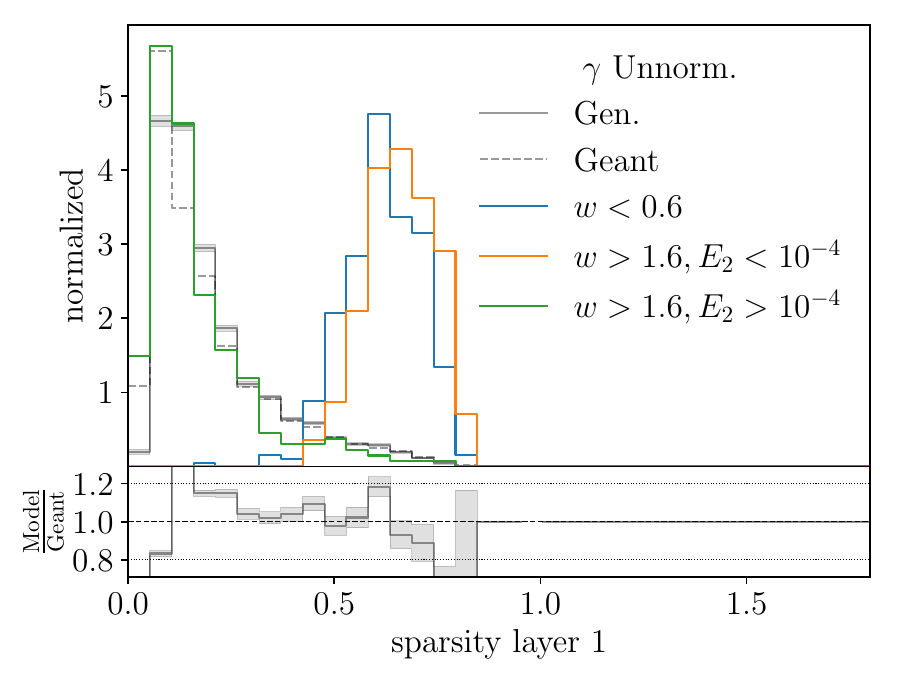}
    \includegraphics[width=0.325\textwidth,page= 16]{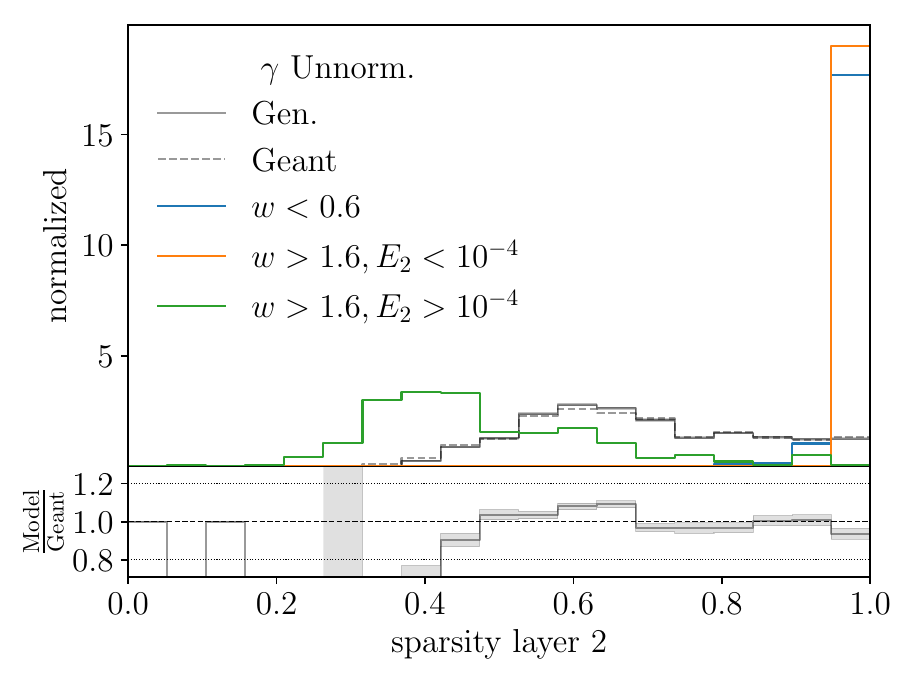}
\caption{Clustering plots for $\gamma$: (i) reweighting is required at low energies, but the pattern is not just the energy; (ii) (orange) under-sampling of soft showers
with zero energy deposition in layer-2; (iii) (blue) induced over-sampling of soft showers in layer-2; (iv) (green) under-sampling of delayed showers, low energy deposition in layer-0.}
\label{fig:obs_gamma}
\end{figure}

\begin{figure}[h!]
    \includegraphics[width=0.325\textwidth,page= 1]{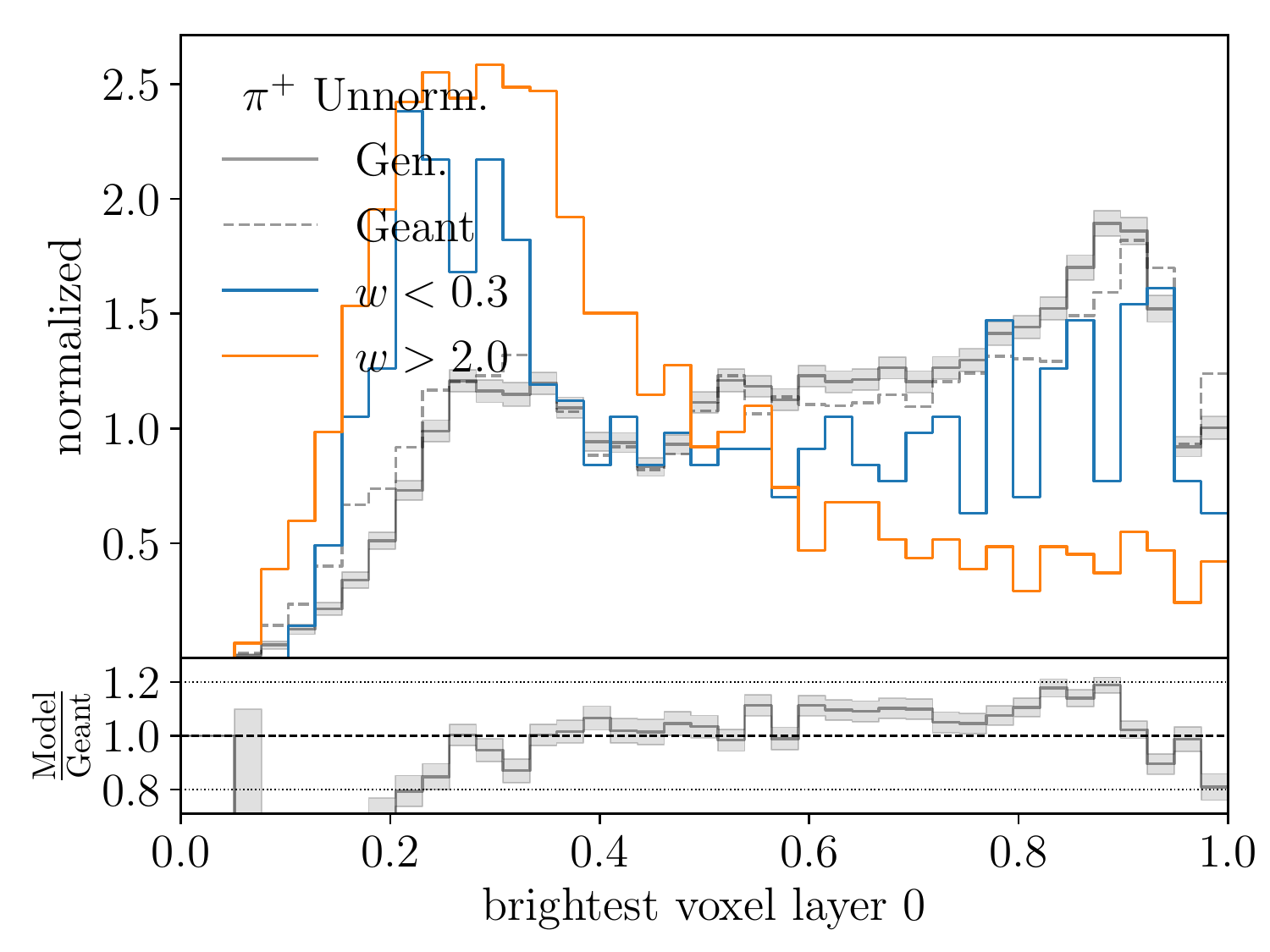} 
    \includegraphics[width=0.325\textwidth,page= 2]{figs/caloinn_plots/clustering_unnorm_pi.pdf}
    \includegraphics[width=0.325\textwidth,page= 3]{figs/caloinn_plots/clustering_unnorm_pi.pdf}
    \includegraphics[width=0.325\textwidth,page= 4]{figs/caloinn_plots/clustering_unnorm_pi.pdf} 
    \includegraphics[width=0.325\textwidth,page= 5]{figs/caloinn_plots/clustering_unnorm_pi.pdf}
    \includegraphics[width=0.325\textwidth,page= 6]{figs/caloinn_plots/clustering_unnorm_pi.pdf}
    \includegraphics[width=0.325\textwidth,page= 14]{figs/caloinn_plots/clustering_unnorm_pi.pdf} 
    \includegraphics[width=0.325\textwidth,page= 15]{figs/caloinn_plots/clustering_unnorm_pi.pdf}
    \includegraphics[width=0.325\textwidth,page= 16]
    {figs/caloinn_plots/clustering_unnorm_pi.pdf}
\caption{Clustering plots for $\pi^{+}$: (i) for the different energies the INN finds all features, but the balance between feature and continuum is not perfect; (ii) in both tails corrections at all energies are applied; (iii) the generator over-samples showers with no energy deposition in layer-1 and layer-2; (iv) large sparsity values are underestimated by the INN.}
\label{fig:obs_piplus}
\end{figure}

\clearpage

\begin{figure}[h]
    \includegraphics[width=0.325\textwidth,page= 1]{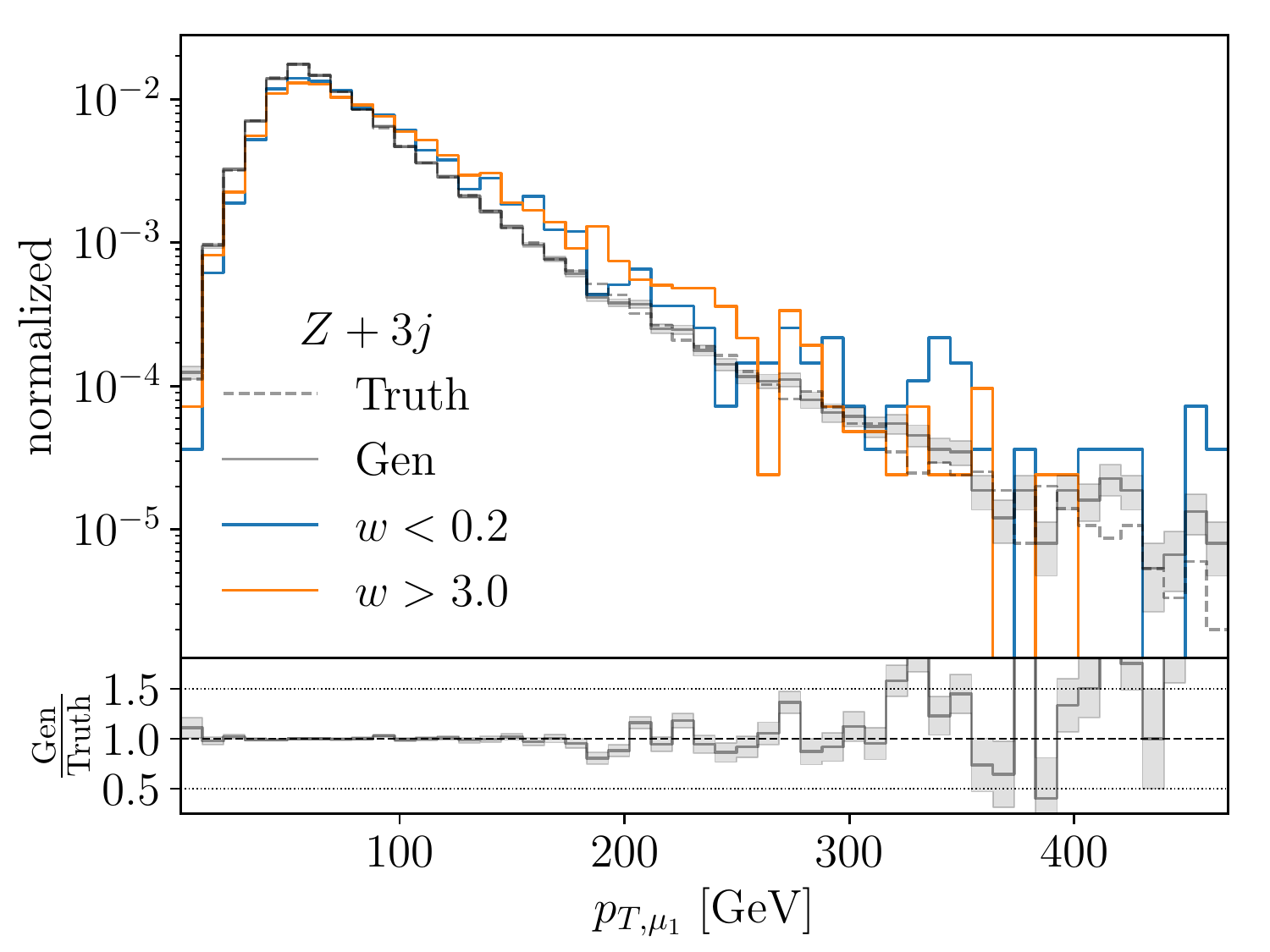}
    \includegraphics[width=0.325\textwidth,page= 2]{/prec_inn_improved_generator/clustering_z3j} \\
    \includegraphics[width=0.325\textwidth,page= 3]{/prec_inn_improved_generator/clustering_z3j}
    \includegraphics[width=0.325\textwidth,page= 4]{/prec_inn_improved_generator/clustering_z3j}
    \includegraphics[width=0.325\textwidth,page=14]{/prec_inn_improved_generator/clustering_z3j} \\
    \includegraphics[width=0.325\textwidth,page= 5]{/prec_inn_improved_generator/clustering_z3j}
    \includegraphics[width=0.325\textwidth,page= 6]{/prec_inn_improved_generator/clustering_z3j}
    \includegraphics[width=0.325\textwidth,page= 7]{/prec_inn_improved_generator/clustering_z3j}
    \includegraphics[width=0.325\textwidth,page= 8]{/prec_inn_improved_generator/clustering_z3j}
    \includegraphics[width=0.325\textwidth,page= 9]{/prec_inn_improved_generator/clustering_z3j}
    \includegraphics[width=0.325\textwidth,page=10]{/prec_inn_improved_generator/clustering_z3j}
    \includegraphics[width=0.325\textwidth,page=11]{/prec_inn_improved_generator/clustering_z3j}
    \includegraphics[width=0.325\textwidth,page=12]{/prec_inn_improved_generator/clustering_z3j}
    \includegraphics[width=0.325\textwidth,page=13]{/prec_inn_improved_generator/clustering_z3j}
    \includegraphics[width=0.325\textwidth,page=15]{/prec_inn_improved_generator/clustering_z3j}
    \includegraphics[width=0.325\textwidth,page=18]{/prec_inn_improved_generator/clustering_z3j}
    \includegraphics[width=0.325\textwidth,page=21]{/prec_inn_improved_generator/clustering_z3j}
    \caption{Set of kinematic distributions for $Z+3$~jets events from the
      state-of-the-art generator in different weight ranges.  The 
      events with small weights are taken from the generated 
      distribution, the events with large weights are taken from the 
      truth distribution.}
    \label{fig:prec_inn_improved_all}
\end{figure}

\clearpage
\bibliographystyle{SciPost-bibstyle-arxiv}
\bibliography{paper}

\end{document}